\documentclass{pasj01}
\usepackage{natbib} 
\usepackage{lscape}
\usepackage{url}
\Received{$\langle$02-Jun-2018$\rangle$}
\Accepted{$\langle$08-Apr-2019$\rangle$}
\Published{$\langle$publication date$\rangle$}

\newcommand{\oiii}{[O{\sc iii}]}
\newcommand{\cii}{[C{\sc ii}]}
\newcommand{\ciii}{C{\sc iii}]}
\newcommand{\lya}{Ly$\alpha$}

\newcommand{\muv}{$M_{\rm UV}$}
\newcommand{\dv}{$\Delta v_{\rm Ly\alpha}$}
\newcommand{\name}{B14-65666}
\newcommand{\ew}{EW$_{\rm 0}$(Ly$\alpha$)}
\newcommand{\lsun}{$L_{\rm \odot}$}
\newcommand{\msun}{$M_{\rm \odot}$}
\newcommand{\ltir}{$L_{\rm TIR}$}
\newcommand{\td}{$T_{\rm d}$}
\newcommand{\bd}{$\beta_{\rm d}$}
\newcommand{\md}{$M_{\rm d}$}
\newcommand{\nhi}{$N_{\rm HI}$}
\newcommand{\zlya}{$z_{\rm Ly\alpha}$}
\newcommand{\zsys}{$z_{\rm sys}$}

\newcommand{\mdyn}{$M_{\rm dyn}$}

\begin{document}

\title{
``Big Three Dragons'': a $z=7.15$ Lyman Break Galaxy Detected in 
\oiii\ $88$ \micron, \cii\ $158$ \micron, and Dust Continuum with ALMA
}
\author{
Takuya Hashimoto\altaffilmark{1,2,3} 
Akio K. Inoue\altaffilmark{1,2}, 
Ken Mawatari\altaffilmark{2,4}, 
Yoichi Tamura\altaffilmark{5},
Hiroshi Matsuo\altaffilmark{3,6}, 
Hisanori Furusawa\altaffilmark{3}, 
Yuichi Harikane\altaffilmark{4,7},
Takatoshi Shibuya\altaffilmark{8},
Kirsten K. Knudsen\altaffilmark{9}, 
Kotaro Kohno\altaffilmark{10,11}, 
Yoshiaki Ono\altaffilmark{4}, 
Erik Zackrisson\altaffilmark{12}, 
Takashi Okamoto\altaffilmark{13}, 
Nobunari Kashikawa\altaffilmark{3,6,7}, 
Pascal A. Oesch\altaffilmark{14}, 
Masami Ouchi\altaffilmark{4,15}, 
Kazuaki Ota\altaffilmark{16}, 
Ikkoh Shimizu\altaffilmark{17},  
Yoshiaki Taniguchi\altaffilmark{18},
Hideki Umehata\altaffilmark{18,19}, 
and 
Darach Watson\altaffilmark{20}. 
}
\altaffiltext{1}{Research Institute for Science and Engineering, Waseda University, Tokyo 169-8555, Japan}
\altaffiltext{2}{Department of Environmental Science and Technology, Faculty
of Design Technology, Osaka Sangyo University, 3-1-1, Nagaito,
Daito, Osaka 574-8530, Japan}
\altaffiltext{3}{National Astronomical Observatory of Japan, 2-21-1 Osawa, Mitaka, Tokyo 181-8588, Japan}
\altaffiltext{4}{Institute for Cosmic Ray Research, The University of Tokyo, Kashiwa, Chiba 277-8582, Japan}
\altaffiltext{5}{Division of Particle and Astrophysical Science, Graduate School of Science, Nagoya} 
\altaffiltext{6}{Department of Astronomical Science, School of Physical Sciences,
The Graduate University for Advanced Studies (SOKENDAI),
2-21-1, Osawa, Mitaka, Tokyo 181-8588, Japan}
\altaffiltext{7}{Department of Physics, Graduate School of Science, 
The University of Tokyo, 7-3-1 Hongo, Bunkyo, Tokyo, 113-0033, Japan}
\altaffiltext{8}{Department of Computer Science, Kitami Institute of Technology, 165 Koen-cho, Kitami, Hokkaido 090-8507, Japan}
\altaffiltext{9}{Department of Space, Earth and Environment, Chalmers University of Technology, Onsala Space Observatory, SE-439 92 Onsala, Sweden}
\altaffiltext{10}{Institute of Astronomy, Graduate School of Science,
The University of Tokyo, 2-21-1 Osawa, Mitaka, Tokyo 181-0015, Japan}
\altaffiltext{11}{Research Center for the Early Universe, Graduate
School of Science, The University of Tokyo, 7-3-1 Hongo, Bunkyo-ku,
Tokyo 113-0033, Japan}
\altaffiltext{12}{Observational Astrophysics, Department of Physics and Astronomy, Uppsala University, Box 516, SE-751 20 Uppsala, Sweden}
\altaffiltext{13}{Department of Cosmosciences, Graduates School of Science, Hokakido University, N10 W8, Kitaku, Sapporo 060-0810, Japan}
\altaffiltext{14}{Geneva Observatory, University of Geneva, Ch. des Maillettes, 51, 1290 Versoix, Switzerland}
\altaffiltext{15}{Kavli Institute for the Physics and Mathematics of the Universe (WPI), Todai Institutes for Advanced Study, The University of Tokyo, 5-1-5 Kashiwanoha, Kashiwa, Chiba 277-8583, Japan}
\altaffiltext{16}{Kyoto University Research Administration Office, Yoshida-Honmachi, Sakyo-ku, Kyoto 606-8501 Japan}
\altaffiltext{17}{Theoretical Astrophysics, Department of Earth \& Space Science, Osaka University, 1-1 Machikaneyama, Toyonaka, Osaka 560-0043, Japan}
\altaffiltext{18}{The Open University of Japan, 2-11 Wakaba, Mihama-ku, Chiba 261-8586, Japan}
\altaffiltext{19}{RIKEN Cluster for Pioneering Research, 2-1 Hirosawa, Wako-shi, Saitama 351-0198, Japan}
\altaffiltext{20}{Dark Cosmology Centre, Niels Bohr Institute, University of Copenhagen, Denmark}
\email{t.hashimoto8@kurenai.waseda.jp}


\maketitle

\begin{abstract}
We present new ALMA observations and physical properties of a Lyman Break Galaxy at $z=7.15$. 
Our target, \name, has a bright ultra-violet (UV) absolute magnitude, $M_{\rm UV} \approx -22.4$, and has been spectroscopically identified in \lya\ with a small rest-frame equivalent width of $\approx4$ \AA. Previous {\it HST} image has shown that the target is comprised of two spatially separated clumps in the rest-frame UV.
With ALMA, we have newly detected spatially resolved \oiii\ 88 \micron, \cii\ 158 \micron, and their underlying dust continuum emission. In the whole system of \name, the \oiii\ and \cii\ lines have consistent redshifts of $7.1520\pm0.0003$, and the \oiii\ luminosity, \textcolor{black}{$(34.4\pm4.1)\ \times 10^{8}$ \lsun}, is \textcolor{black}{about three times} higher than the \cii\ luminosity, $(11.0\pm1.4)\ \times 10^{8}$ \lsun. 
\textcolor{black}{With our two continuum flux densities, the dust temperature is constrained to be \td\ $\approx50-60$ K under the assumption of the dust emissivity index of $\beta_{\rm d} = 2.0-1.5$, leading to a large total infrared luminosity of \ltir\ $\approx 1\times10^{12}$ \lsun.}
Owing to our high spatial resolution data, 
\textcolor{black}{we show that the \oiii\ and \cii\ emission can be spatially decomposed into two clumps associated with the two rest-frame UV clumps whose spectra are kinematically separated by $\approx200$ km s$^{-1}$. We also find these two clumps have comparable UV, infrared, \oiii, and \cii\ luminosities. Based on these results, we argue that \name\ is a starburst galaxy induced by a major-merger. The merger interpretation is also supported by the large specific star-formation rate (defined as the star-formation rate per unit stellar mass), sSFR $= 260^{+119}_{-57}$ Gyr$^{-1}$, inferred from our SED fitting. Probably, a strong UV radiation field caused by intense star formation contributes to its high dust temperature and the \oiii-to-\cii\ luminosity ratio. 
}
\end{abstract}

\section{Introduction}
\label{sec:intro}

Understanding properties of galaxies during reionization, at redshift $z\gtrsim6-7$,  is important.
While a large number of galaxy candidates are selected 
with a dropout technique at $z\gtrsim7$ (e.g., \citealt{ellis2013, bouwens2014.z9, oesch2018}), 
the spectroscopic identifications at $z\gtrsim7$ remain difficult (e.g., \citealt{stark2017} 
and references therein). 
This is mainly due to the fact that the most prominent hydrogen \lya\ line 
is significantly attenuated by the intergalactic medium (IGM).

With the advent of the Atacama Large Millimeter/Submillimeter Array (ALMA) telescope, 
it has become possible to detect rest-frame far-infrared (FIR) fine structure lines in star-forming galaxies at $z>5$ (e.g., \citealt{capak2015, maiolino2015}). 
A most commonly used line is \cii\ 158 \micron, 
which is one of the brightest lines in local galaxies (e.g., \citealt{malhotra1997, brauher2008}). 
To date, more than 21 \cii\ detections are reported at $5\lesssim z \lesssim 7$ (\citealt{carniani2018b} and references therein; \citealt{pentericci2016, matthee2017, smit2018}).

However, based on a compiled sample with \cii\ observations at $z\gtrsim5$, 
\cite{harikane2018b} and \cite{carniani2018b} have revealed that 
\cii\ may be weak for galaxies with strong \lya\ emission, so-called \lya\ emitters 
(LAEs; rest-frame \lya\ equivalent widths \ew\ $\gtrsim20-30$ \AA). 
\cite{harikane2018b} have interpreted the trend with photoionization models 
of {\tt CLOUDY}  (\citealt{ferland2013}) 
implemented in 
spectral energy distribution (SED) models of BEAGLE (\citealt{chevallard2016}). 
The authors show that low metallicity or high ionization states 
in LAEs lead to weak \cii. 
\textcolor{black}{Theoretical studies also show that such ISM conditions lead to the decrease in the \cii\ luminosity (\citealt{vallini2015, vallini2017, olsen2017, lagache2018}).}
If we assume that $z\gtrsim7$ galaxies in general have low metallicity or high ionization states, 
\cii\ may not be the best line to spectroscopically confirm $z\gtrsim7$ galaxies. 
Indeed, a number of null-detections of \cii\ are reported 
at $z\gtrsim7$ (e.g., \citealt{ota2014, schaerer2015, maiolino2015, inoue2016}). 

In fact, based on {\it Herschel} spectroscopy for local dwarf galaxies, 
\cite{cormier2015} have demonstrated that 
\oiii\ 88 \micron\ becomes brighter than \cii\  at low metallicity (see also \citealt{malhotra2001}). 
Based on calculations of {\tt CLOUDY}, \cite{inoue2014alma} also theoretically predict that 
the \oiii\ line at high-$z$ should be bright enough to be detected with ALMA.

Motivated by these backgrounds, 
we are conducting follow-up observations of \textcolor{black}{the} \oiii\ \textcolor{black}{88 \micron\ line} for $z>6$ galaxies with ALMA. 
After the first detection of \oiii\ in the reionization epoch in \cite{inoue2016} at $z=7.21$,
the number of \oiii\ detection\textcolor{black}{s} is rapidly increasing. 
There are currently \textcolor{black}{ten} objects with \oiii\ detections at $z\approx6-9$  
(\citealt{carniani2017a, laporte2017, marrone2018, hashimoto2018a, tamura2018}, \textcolor{black}{\citealt{hashimoto2018c, walter2018}}). Remarkably, \cite{hashimoto2018a} have detected \oiii\ in a $z=9.11$ galaxy 
with a high significance level of $7.4\sigma$. 
Importantly, \oiii\ is detected from all the targeted galaxies \textcolor{black}{(six detections out of six objects) from our team}, i.e., the success rate is currently 100\% . 
These results clearly demonstrate that \oiii\ is a powerful tool to confirm $z>6$ galaxies. 

\cite{inoue2016} have also investigated the FIR line ratio at $z>7$. 
In a combination with the null detection of \cii, 
the authors have shown that their $z=7.21$ LAE has a line ratio of \oiii/\cii\ $>12$ ($3\sigma$). 
The line ratio would give us invaluable information on properties of the interstellar medium (ISM). 
Given the fact that \oiii\ originates only from HII regions whereas 
\cii\ originates both from HII regions and photo-dissociated regions (PDRs), 
\cite{inoue2016} have interpreted the high line ratio as the $z=7.21$ LAE 
having highly ionized HII regions but less PDRs. 
Such properties would lead to a high escape fraction of ionizing photons, 
which is a key parameter to understand reionization. 
Therefore, it is of interest to understand if a high line ratio is common in high-$z$ galaxies (\citealt{inoue2016}).

In this study, we present new ALMA observations and 
physical properties of an Lyman Break Galaxy (LBG) at $z=7.15$. 
Our target, \name, is a very ultra-violet (UV) bright LBG 
with an absolute magnitude, \textcolor{black}{$M_{\rm UV} \approx -22.4$} (\citealt{bowler2014, bowler2017, bowler2018}). 
With the Faint Object Camera and Spectrograph (FOCAS) on the Subaru telescope, 
\cite{furusawa2016} have spectroscopically detected \lya\ 
at the significance level of $5.5\sigma$. 
The authors find that \name\ has a small \ew\ of $3.7^{+1.7}_{-1.1}$ \AA. 
In addition, based on observations with the {\it Hubble Space Telescope} ({\it HST}) 
Wide Field Camera 3 (WFC3) F140W band image, 
\cite{bowler2017} have revealed that \name\  is comprised of two components 
in the rest-frame UV with a projected separation of $\approx2-4$ kpc. 
At high-$z$, such a complicated structure is often interpreted in terms of a merger or clumpy star formation. 
The authors have argued that the large star-formation rate (SFR) inferred from the UV luminosity 
could be naturally explained if the system is a merger-induced starburst. 
More recently, with ALMA Band 6 observations, 
\cite{bowler2018} have detected dust continuum emission at the peak significance level of $5.2\sigma$, 
which is the third detection of dust emission in normal star-forming galaxies at $z>7$ 
(cf., \citealt{watson2015, laporte2017} see also \citealt{knudsen2017}).

In ALMA Cycle 4, we have performed high spatial resolution follow-up observations of \name\ with beam size of $\approx 0''.3 \times 0''.2$ ($0''.3 \times 0''.3$) in Band 6 (Band 8). \textcolor{black}{In ALMA Cycle 5, we have also obtained deeper Band 8 data with a slightly larger beam size of $\approx 0''.4 \times 0''.4$.} 
We successfully detect spatially resolved \cii, \oiii, and dust continuum emission \textcolor{black}{in two bands}, making \name\ the first object at $z\gtrsim6$ with a complete set of these 
three features\footnote{``Big Three Dragons'' is a hand in a {\it Mahjong} game with triplets or quads of all three dragons.}.
The spatially resolved data enable us to investigate the velocity gradients of the \cii\ and \oiii\ lines. 
These emission lines also allow us to investigate the \lya\ velocity offset with respect to the systemic, \dv, which is an important parameter to understand reionization (e.g., \citealt{choudhury2015, mason2017, mason2018}). 
The dust continuum emission also offers us invaluable information on the ISM properties of \name. 
We also derive physical quantities such as the stellar age, the stellar mass ($M_{\rm *}$), 
and the SFR. 
With these quantities, we will discuss kinematical, morphological, and ISM properties of \name.

This paper is organized as follows. 
In \S \ref{sec:data}, we describe our data.
In \S \ref{sec:lines},  we measure  \cii\ and \oiii\ quantities. 
Dust properties are presented in \S \ref{sec:dust}, 
followed by results on luminosity ratios in \S \ref{sec:luminosity_ratio}. 
In \S \ref{sec:sed_fit}, we perform spectral energy distribution (SED) fitting.
In \S \ref{sec:lya}, we derive \dv\ in \name, 
and statistically examine \dv\ at $z\approx6-8$. 
Discussions in the context of (i) properties of \name\ and (ii) reionization  
are presented in \S \ref{sec:discussion}, 
followed by our conclusions  in \S \ref{sec:conclusion}. 
Throughout this paper, magnitudes are given in the AB system
\citep{oke1983}, and we assume a $\Lambda$CDM cosmology 
with $\Omega_{\small m} = 0.272$, $\Omega_{\small b} = 0.045$, $\Omega_{\small \Lambda} = 0.728$
and $H_{\small 0} = 70.4$ km s$^{-1}$ Mpc$^{-1}$ (\citealt{komatsu2011}). 
\textcolor{black}{The solar luminosity, \lsun, is $3.839\times10^{33}$ erg s$^{-1}$.}

\begin{table*}
\tbl{Summary of ALMA Observations. \label{tab:obs}}
{
\begin{tabular}{ccccccc}
\hline
Date & Baseline lengths  & $N_{\rm ant}$ & Central frequencies of SPWs & Integration time & PWV \\ 
(YYYY-MM-DD) & (m) & & (GHz) & (min.) & (mm) \\ 
(1) & (2) & (3) & (4) & (5) & (6)  \\ 
\hline 
\bf{Band 6 \textcolor{black}{(Cycle 4)}} \\ 
\hline
2017-07-09 & $16 - 2647$ & 40 & 218.78, 216.98, 232.66, 234.48 & 37.80 & 0.45  \\ 
2017-07-09 & $16 - 2647$ & 40 & 218.78, 216.98, 232.66, 234.48 & 37.80 & 0.44  \\ 
2017-07-09 & $16 - 2647$ & 40 & 218.78, 216.98, 232.66, 234.48 & 37.80 & 0.42  \\
\hline 
\bf{Band 8 \textcolor{black}{(Cycles 4 + 5)}}  \\ 
\hline 
2016-11-14 & $15 - 918$ & 40 & 405.43, 403.61, 415.47, 417.23 & 14.45 & 0.58  \\ 
2016-11-15 & $15 - 918$ & 40 & 405.43, 403.61, 415.47, 417.23 & 39.72 & 0.63 \\
2018-04-06 & $15 - 483$ & 43 & 404.09, 405.98, 416.02, 417.60 & 47.60 & 0.86 \\
2018-05-01 & $15 - 500$ & 43 & 404.09, 405.98, 416.02, 417.60 & 45.20 & 0.95 \\
2018-05-01 & $15 - 500$ & 43 & 404.09, 405.98, 416.02, 417.60 & 45.22 & 0.83 \\
2018-05-04 & $15 - 500$ & 43 & 404.09, 405.98, 416.02, 417.60 & 45.22 & 0.69 \\
2018-05-05 & $15 - 500$ & 43 & 404.09, 405.98, 416.02, 417.60 & 45.17 & 0.81 \\
\hline
\end{tabular}
}
\tabnote{Note. (1) The observation date; (2) the ALMA's baseline length; (3) the number of antenna used in the observation; (4) the central frequencies of the four spectral windows (SPWs); (5) the on-source integration time; (6) the precipitable water vapor.}
\end{table*}

\section{Observations and Data}
\label{sec:data}
 
\subsection{ALMA Band 6 Observations}
\label{subsec:obs}
 
As summarized in Table \ref{tab:obs}, we observed \name\ with ALMA in Band 6 targeting \cii\ 158 \micron\ in Cycle 4 (ID 2016.1.00954.S, PI: A. K. Inoue). The antenna configuration was C40-6, and the on-source exposure times was 114 minutes. We used four spectral windows (SPWs) with 1.875 GHz bandwidths in the Frequency Division Mode (FDM), totaling the band width of 7.5 GHz. Two SPWs with a 7.813 MHz resolution was used to target the line. One of the two SPWs was centered on the  \lya\ frequency and the other was centered at a higher frequency (i.e, a shorter wavelength) with a small overlap in frequency, taking into account the possible redward velocity offset of the \lya\ line with respect to the systemic redshift (e.g., \citealt{steidel2010, hashimoto2013}).  The remaining two SPWs with a 31.25 MHz resolution were used to observe dust continuum emission at $\approx 163$ \micron. 
A quasar, J0948+0022 (J1058+0133) was used for phase (bandpass) calibrations, and two quasars, J1058+0133 and J0854+2006, were used for flux calibrations (\textcolor{black}{Appendix 2 Table \ref{tab:appendix}}). The flux calibration uncertainty was estimated to be $\lesssim$ 10\%. 
The data were reduced and calibrated using the Common Astronomy Software Application (CASA; \textcolor{black}{\citealt{McMullin2007}}) pipeline version 4.7.2. \textcolor{black}{Using the {\tt CLEAN} task,  we produced two images and cubes with different weighting: (1) The natural weighting to maximize point-source sensitivity on which we perform photometry and (2) the Briggs weighting with the robust parameter of 0.3 to investigate morphological properties\footnote{
For the description of the Briggs \textcolor{black}{weighting} and the robust parameter, 
https://casa.nrao.edu/Release4.1.0/doc/UserMan/UserMansu262.html}.} To create a pure dust continuum image, we collapsed all off-line channels. To create a pure line image, we subtracted continuum using the off-line channels in the line cube with the CASA task {\tt uvcontsub}.
In Table \ref{tab:data}, we summarize the r.m.s. levels, the spatial resolutions, and the beam position angles for the continuum images \textcolor{black}{with two weighting}.

\subsection{ALMA Band 8 Observations}
\label{subsec:obs}

\textcolor{black}{In Cycles 4 and 5, we also observed \name\ with ALMA in Band 8 targeting \oiii\ 88 \micron\ (IDs 2016.1.00954.S and 2017.1.00190.S; PIs: A. K. Inoue; Table \ref{tab:obs}). The antenna configuration was C40-4 (C43-3) for the Cycle 4 (5) observations, and the total on-source exposure times was 282 minutes. The observation strategy was the same as that used in Band 6. 
The combinations of two quasars, (J0948+0022 and J1028-0236), (J1058+0133 and J1229+0203), and (J1058+0133 and J1229+0203) were used for phase, bandpass, and flux calibrations, respectively  (Appendix 2 Table \ref{tab:appendix}). The flux calibration uncertainty was estimated to be $\lesssim$ 10\%. 
The Cycles 4 and 5 datasets were first reduced and calibrated with the CASA pipeline versions 4.7.0 and 5.1.1, respectively, and then combined into a single measurement set with the CASA task {\tt concat}. We  created images and cubes with the natural weighting using the {\tt CLEAN} task. In Table \ref{tab:data}, we summarize the r.m.s. level, the beam size, and the beam position angle for the continuum image.}

\begin{table}
\tbl{Summary of ALMA data. \label{tab:data}}
{
\begin{tabular}{cccc}
\hline
Data & $\sigma_{\rm cont}$  & Beam FWHMs & PA \\ 
& ($\mu$Jy beam$^{-1}$) & (arcsec $\times$ arcsec) & (deg.) \\ 
(1) & (2) & (3) & (4) \\
\hline
Band 6 (natural) & 9.5 &  $0.29\times0.23$ & $-60$ \\
\hline
Band 6 (Briggs) & 11.0  & $0.23\times0.12$ & $-70$ \\
\hline 
Band 8 (natural) & \textcolor{black}{29.4}  & \textcolor{black}{$0.39\times0.37$} & \textcolor{black}{$+62$}\\
\hline
\end{tabular}
}
\tabnote{Note. (1) The ALMA Band used. Weighting is specified in the parenthesis; (2) the $1\sigma$ r.m.s. level of the continuum image; (3) the ALMA's beam FWHM in units of arcsec $\times$ arcsec; (5) the ALMA's beam position angle in units of degree.}
\end{table}

\section{\cii\ 158 \micron\ and \oiii\ 88 \micron\ lines}
\label{sec:lines}

\begin{figure*}[]
\includegraphics[width=17cm]{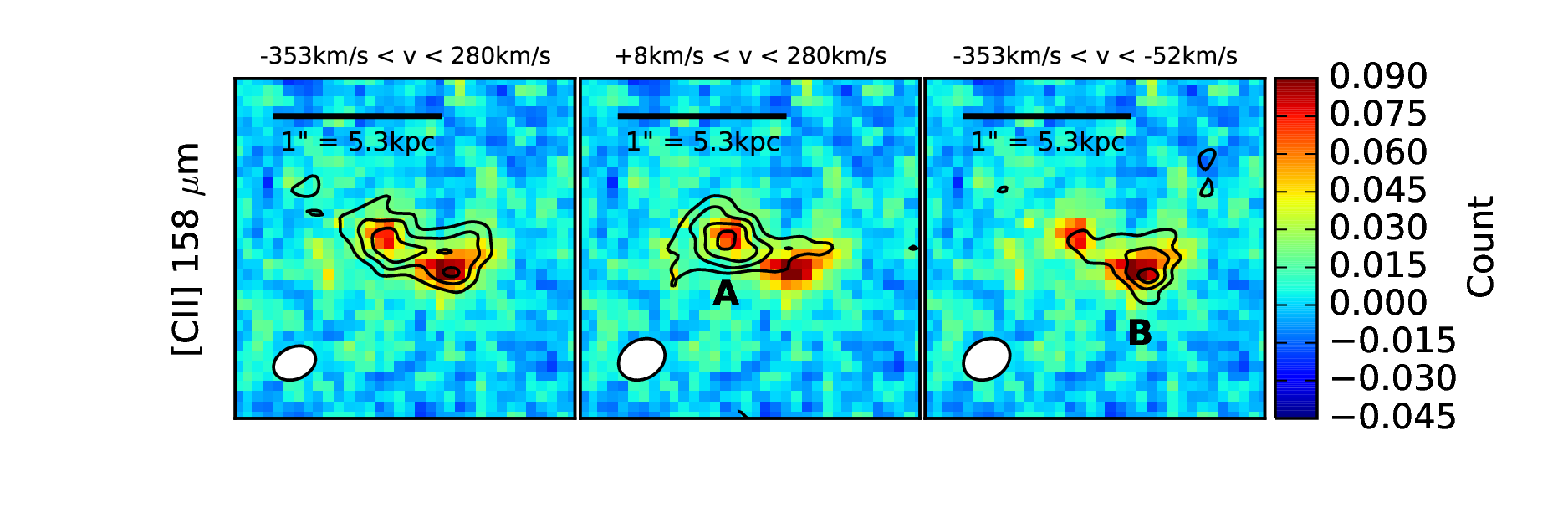}
\includegraphics[width=17cm]{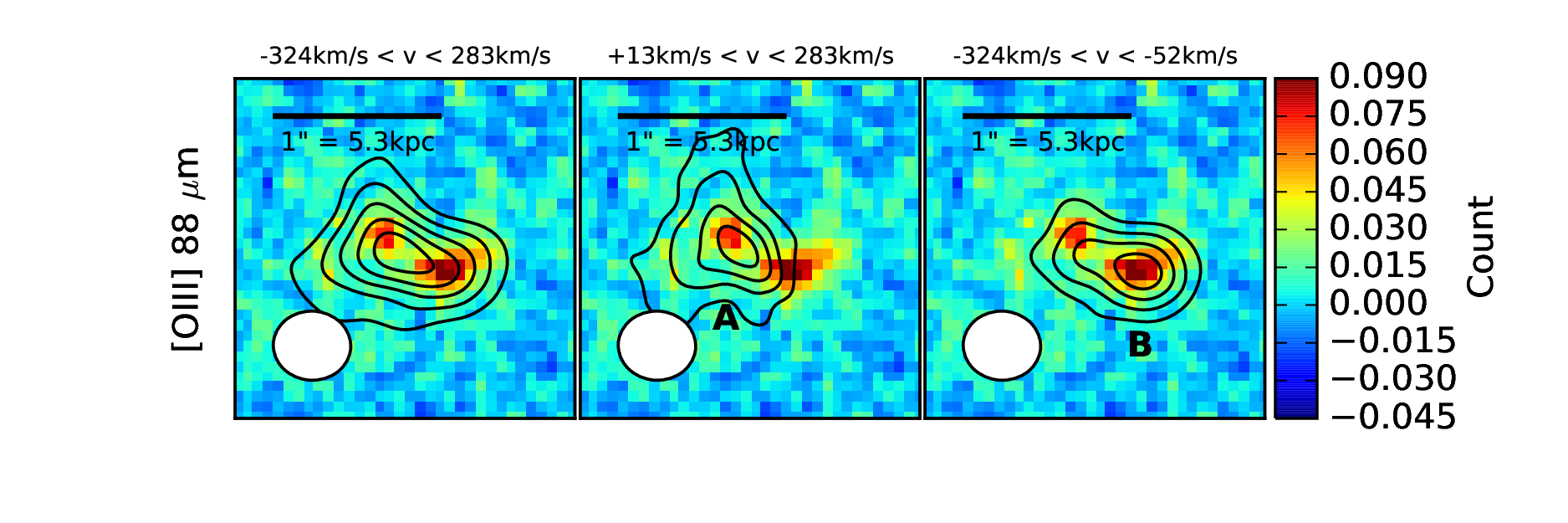}
\caption
{
\textcolor{black}{Top and bottom panels show \cii\ and \oiii\ contours overlaid on the $2''.0\times2''.0$ cutout image of HST F140W, respectively. The left, middle, and right panels correspond to the line image for the whole system, clump A, and clump B, respectively. The velocity range used to extract the images are indicated above the panels, where the velocity zero point is defined as the systemic redshift, \zsys\ $=$ 7.1520. 
({\it Top}) \cii\ line contours 
drawn at ($3$, $5$, $7$, $9$)$\times \sigma$, 
where $\sigma \approx$ 19, 12, and 13 mJy beam$^{-1}$ km s$^{-1}$ for the left, middle and right panel, respectively. 
({\it Bottom}) \oiii\ line contours 
drawn at ($3$, $5$, $7$, $9$, $11$)$\times \sigma$, 
where $\sigma \approx$ 48, 31, and 31 mJy beam$^{-1}$ km s$^{-1}$ for the left, middle and right panel, respectively. 
In each panel, contours are shown by the solid lines and the ellipse at lower left indicates the synthesized beam size of ALMA.}
}
\label{fig:contours}
\end{figure*}

\subsection{Measurements for the Whole System}
\label{subsec:whole}

In Band 6 (8) data, to search for an emission line, we have created a data cube by binning three (six) native channels, resulting in a velocity resolution of $\approx 30 \ (33)$ km s$^{-1}$. At the \name\ position \textcolor{black}{determined in the HST image}, we find a supposed \cii\ (\oiii) feature at around 233.12 (416.27) GHz. This frequency region is free from atmospheric absorption features. In Band 6 (8), we have then created a velocity-integrated intensity image between 232.9 and 233.4 GHz (\textcolor{black}{415.8 and 416.7 GHz}) corresponding to $\approx$ 600 (\textcolor{black}{600}) km s$^{-1}$. 

The top left and bottom left panels of Figure \ref{fig:contours} show \cii\ and \oiii\ contours  overlaid on the HST F140W images, respectively, \textcolor{black}{whose detailed astrometry analyses are presented in Appendix 1, and our measurements are summarized in Table \ref{tab:results}.} 
With our spatial resolution, \cii\ and \oiii\ are spatially resolved. Assuming a 2D Gaussian profile for the velocity-integrated intensity, we measure the beam-deconvolved size of \cii\ to be $(0''.84\pm0''.12) \times (0''.27\pm0''.05)$, where the first and second values represent the FWHMs of the major and minor-axis, respectively, \textcolor{black}{with a positional angle (PA) of $74^{\circ}\pm4^{\circ}$.} \textcolor{black}{At $z=7.15$, the physical size corresponds to $(4.5 \pm 0.6) \times (1.4 \pm 0.3)$ kpc$^{2}$.} Likewise, we obtain the beam-deconvolved size of \oiii\ to be  $(0''.71\pm0''.10) \times (0''.41\pm0''.11)$, \textcolor{black}{corresponding to $(3.8 \pm 0.5) \times (2.2 \pm 0.6)$ kpc$^{2}$ at $z=7.15$, with PA = $76^{\circ}\pm12^{\circ}$. The size and PA values of \cii\ and \oiii\ are consistent with each other.}

\begin{table*}
\tbl{Summary of observational results of B14-65666. \label{tab:results}}
{
\begin{tabular}{cccc}
\hline
Parameters & Total & Clump A & Clump B \\
\hline
R.A. &  10:01:40.69 & 10:01:40.70 & 10:01:40.67 \\ 
Dec. & +01:54:52.42 & +01:54:52.64 & +01:54:52.47 \\ 
$M_{\rm 1500}$ [AB mag.] & -22.4 & -21.5 & -21.8 \\ 
$L_{\rm UV}$ [$10^{11}$ \lsun] & 2.0 & 0.9 & 1.1 \\ 
\hline 
$z_{\rm [OIII]}$ & $7.1521\pm0.0004$ & $7.1523\pm0.0004$ & $7.1488\pm0.0004$ \\
$z_{\rm [CII]}$ & $7.1521\pm0.0004$ & $7.1536\pm0.0004$ & $7.1478\pm0.0005$ \\
$z_{\rm sys.}$ $^{a}$ & $7.1521\pm0.0003$ & $7.1530\pm0.0003$ & $7.1482\pm0.0003$ \\ 
$z_{\rm Ly\alpha}$$^{b}$ & $7.1730\pm0.0012$ & - & - \\
\dv\ [km s$^{-1}$]   & $772\pm45\pm100$ & - & - \\ 
\hline
\oiii\ integrated flux [Jy km s$^{-1}$]& $1.50\pm0.18$  & $ 0.92\pm0.14$ & $0.57\pm0.09$ \\ 
\cii\ integrated flux [Jy km s$^{-1}$] & $0.87\pm0.11$  & $ 0.47\pm0.07$ & $ 0.38\pm0.06$\\ 
FWHM(\oiii) [km s$^{-1}$] & $429\pm37$ & $325\pm32$ & $267\pm34$\\ 
FWHM(\cii) [km s$^{-1}$] & $349\pm31$ & $347\pm29$ & $284\pm40$\\
\oiii\ deconvolved size$^{c}$ [kpc$^{2}$] & $(3.8 \pm 0.5) \times (2.2 \pm 0.6)$ & $(3.8 \pm 0.7) \times (3.0 \pm 0.6)$ & $(3.1 \pm 0.6) \times (1.1 \pm 0.7)$ \\
\cii\ deconvolved size$^{c}$ [kpc$^{2}$] & $(4.5 \pm 0.6) \times (1.4 \pm 0.3)$ & $(3.3 \pm 0.5) \times (1.5 \pm 0.3)$ & $(2.5 \pm 0.6) \times (1.4 \pm 0.5)$ \\ 
\mdyn$^{d}$ [$10^{10}$ \msun] & $8.8\pm1.9$  & $5.7\pm1.6$ & $3.1\pm1.1$ \\ 
\hline
\oiii\ luminosity [$10^{8}$ \lsun] & $34.4\pm4.1$ & $21.1\pm3.2$ & $13.0\pm2.1$ \\
\cii\ luminosity [$10^{8}$ \lsun] & $11.0\pm1.4$ & $6.0\pm0.9$ & $4.9\pm0.8$ \\
\lya\ luminosity [$10^{8}$ \lsun] & $6.8\pm1.3$  & - & - \\ 
\oiii-to-\cii\ luminosity ratio & $3.1\pm0.6$  & $3.5\pm0.8$ & $2.7\pm0.6$ \\ 
\hline
$S_{\rm \nu,90}$ [$\mu$Jy] & $470\pm128$ & $208\pm83$ & $246\pm73$ \\
$S_{\rm \nu,163}$ [$\mu$Jy]& $130\pm25$ & $41\pm23$ & $87\pm26$ \\ 
dust deconvolved size [kpc$^{2}$] & $(3.8 \pm 1.1) \times (0.8 \pm 0.5)$ & $< 1.6 \times1.2$$^{e}$ &  $< 1.6 \times1.2$$^{e}$ \\
\hline
\ltir\ (\td$=48$K, \bd$=$2.0) [$10^{11}$ \lsun] & $9.1\pm1.8$ & $2.9\pm1.6$ & $6.1\pm1.8$ \\
\ltir\ (\td$=54$K, \bd$=$1.75) [$10^{11}$ \lsun] & $10.5\pm2.0$ & $3.3\pm1.9$ & $7.0\pm2.1$ \\
\ltir\ (\td$=61$K, \bd$=$1.5) [$10^{11}$ \lsun] & $12.0\pm2.3$ & $3.8\pm2.1$ & $8.0\pm2.4$ \\
\hline
\md\ (\td$=48$K, \bd$=$2.0) [$10^{6}$ \msun] & $11.1\pm2.1$ & $3.5\pm2.0$ & $7.4\pm2.2$  \\ 
\md\ (\td$=54$K, \bd$=$1.75) [$10^{6}$ \msun] & $9.4\pm1.8$ & $3.0\pm1.7$ & $6.3\pm1.9$ \\ 
\md\ (\td$=61$K, \bd$=$1.5) [$10^{6}$ \msun] & $8.1\pm1.6$  & $2.6\pm1.4$ & $5.4\pm1.6$  \\ 
\hline
\end{tabular}
}
\tabnote{Note. 
$^{a}$ The systemic redshift, $z_{\rm sys.}$, is calculated as the 
$S/N$-weighted mean redshift of $z_{\rm [OIII]}$ and $z_{\rm [CII]}$. \\
$^{b}$ The value is different from the original value in \cite{furusawa2016}, \zlya $=7.168$, 
to take into account air refraction and the motion of the observatory (see \S \ref{sec:lya}).\\
$^{c}$ The values represent major and semi-axis FWHM values of a 2D Gaussian profile. \\
\textcolor{black}{$^{d}$ The dynamical mass of individual clumps is estimated based on the virial theorem assuming the random motion (see \S \ref{subsec:dynamical_mass}). The total dynamical mass of the system is assumed to be the summation of the dynamical masses of the clumps.} \\
\textcolor{black}{$^{e}$We present the beam size of Band 6, i.e., higher angular resolution image, as the upper limit because the emission is unresolved in the individual clumps.}\\
\textcolor{black}{The total infrared luminosity, \ltir, is estimated by integrating the modified-black body radiation at $8-1000$ \micron. For the dust temperature and the emissivity index values, we assume the three combinations of ($T_{\rm d}$ [K], $\beta_{\rm d}$) = (48, 2.0), (54, 1.75), and (61, 1.5) (see \S \ref{sec:dust} for the choices of these values.} The dust mass, \md, is estimated with a dust mass absorption coefficient $\kappa = \kappa_{\rm 0} (\mu/\nu_{\rm 0})^{\beta_{\rm d}}$, where we assume $\kappa_{\rm 0} = 10$ cm$^{2}$ g$^{-1}$ at 250 \micron\ (\citealt{hildebrand1983}).
}
\end{table*}

\begin{figure*}[]
\includegraphics[width=8cm]{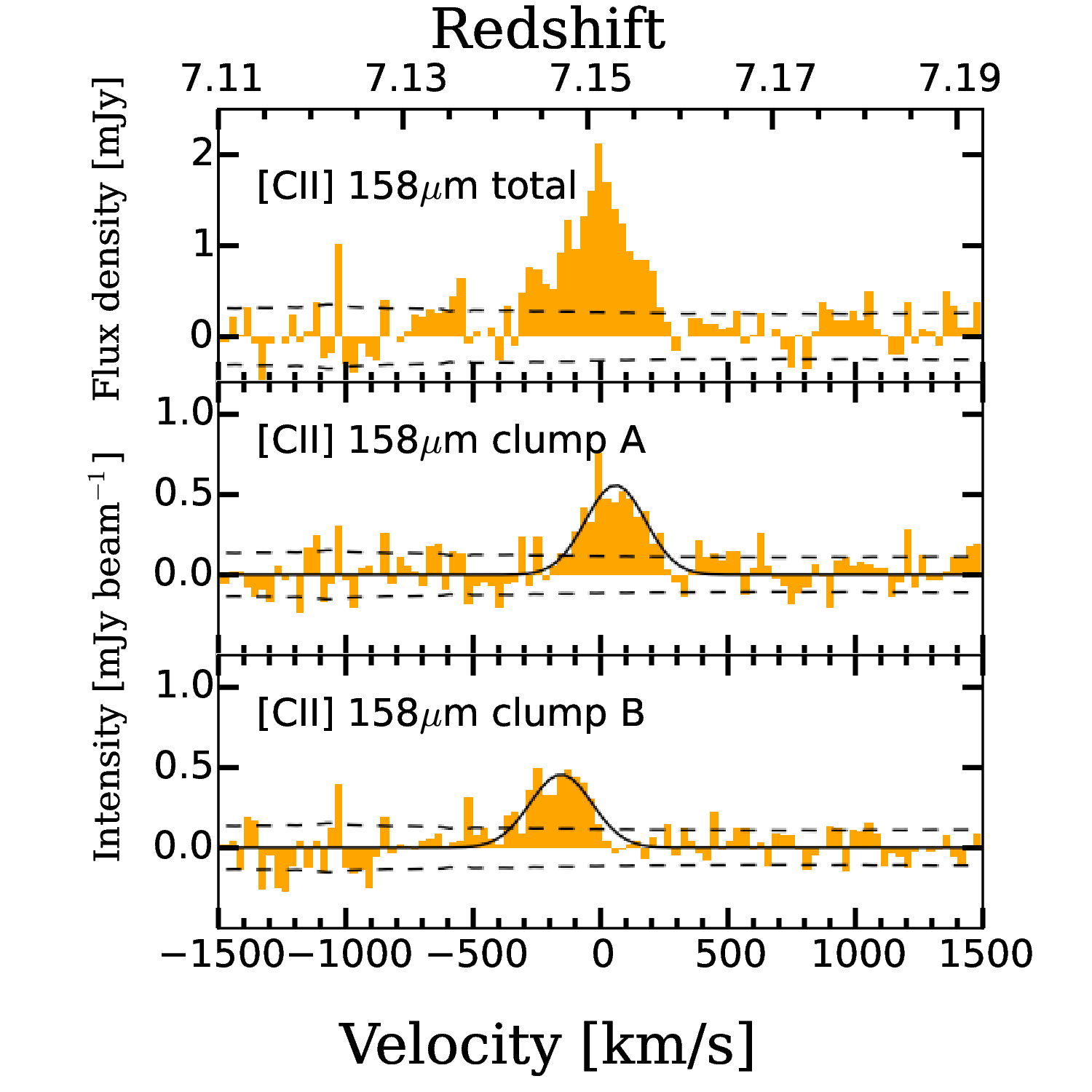}
\includegraphics[width=8cm]{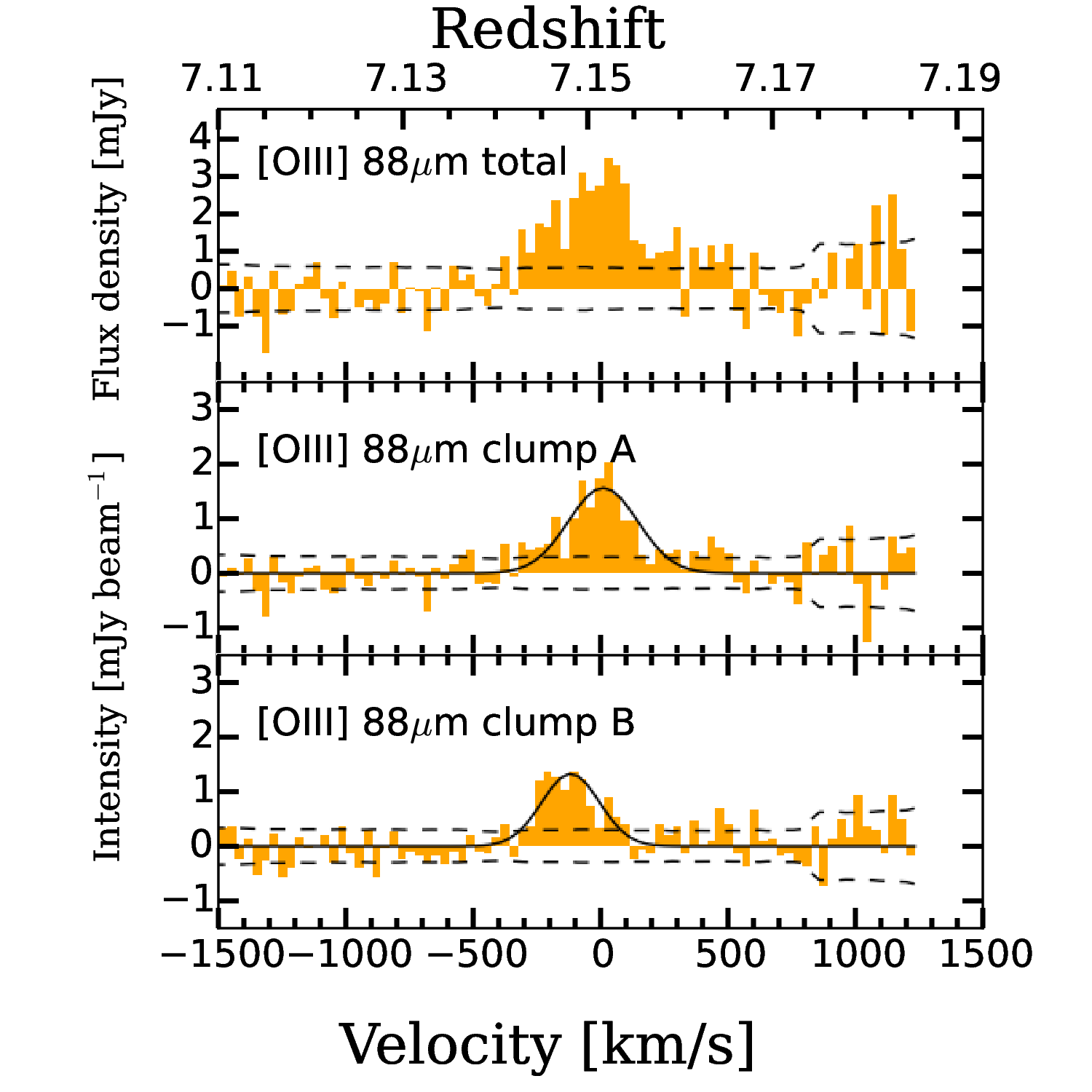}
\caption
{
({\it Left}) Top panel shows the \cii\ spectrum in units of mJy 
extracted from the region with $>3\sigma$ detections in the velocity-integrated intensity image shown in Figure \ref{fig:contours}. 
Middle and bottom panels show the \cii\ spectra  in units of mJy beam$^{-1}$
extracted at the positions of clumps A and B, respectively. 
The black line denotes the best-fit Gaussian for the line, and the black dashed line shows the noise spectra. 
({\it Right}) The same as the left panel, but for \oiii. 
}
\label{fig:1d_spec}
\end{figure*}

We spatially integrate the image with the CASA task {\tt imfit} assuming a 2D Gaussian profile for the velocity-integrated intensity. The velocity-integrated line flux of \cii\ (\oiii) is $0.87\pm0.11$ ($1.50\pm0.18$) Jy km s$^{-1}$. 

To obtain the redshift and FWHM of the lines, we extract spectra from the \cii\ and \oiii\ regions with $>3\sigma$ detections in the velocity-integrated intensity images. The spectra of \cii\ and \oiii\ are shown in the top left and top right panels of Figure \ref{fig:1d_spec}, respectively. Applying a Gaussian line profile and the rest-frame \cii\ (\oiii) frequency\footnote{http://www.cv.nrao.edu/php/splat/} of 1900.5369 (3393.006244) GHz, we obtain the \cii\  (\oiii) redshift of $z=7.1521\pm0.0004$ ($7.1521\pm0.0004$) and the FWHM value of $349\pm31$ ($429\pm37$) km s$^{-1}$. The two redshift and FWHM values are consistent within $\approx 1\sigma$ uncertainties. \textcolor{black}{The $S/N$-weighted mean redshift, \zsys, is $7.1521\pm0.0003$.}

To derive the line luminosity, we use the following relation 
\begin{equation}
L_{\rm line} = 1.04 \times 10^{-3} \times \left( \frac{S_{\rm line} \Delta v}{{\rm Jy\ km\ s^{-1}}} \right) \left( \frac{D_{\rm L}}{{\rm Mpc}} \right)^{2} \left( \frac{\nu_{\rm obs}}{{\rm GHz}} \right) L_{\rm \odot} 
\end{equation}
(\citealt{carilli_walter2013}), 
where $S_{\rm line} \Delta v$ is the velocity-integrated flux, 
$D_{\rm L}$ is the luminosity distance, 
and $\nu_{\rm obs}$ is the observed frequency. 
We obtain $(11.0\pm1.4) \times 10^{8}$ \lsun\ and $(34.4\pm4.1) \times 10^{8}$ \lsun\ for the \cii\ and \oiii\ luminosity, respectively.

\subsection{Measurements for Individual Clumps}
\label{subsec:multi_comp}

Recent ALMA studies of high-$z$ galaxies show spatially separated multiple \cii\ components with projected distances of $\approx 3-7$ kpc (e.g., \citealt{matthee2017, carniani2018b}). As can be seen from Figure \ref{fig:contours}, the HST F140W image of \name\ shows two UV clumps with a projected distance of $\approx3$ kpc, which we refer as clumps A and B. 
\textcolor{black}{Motivated by these,  we decompose the \cii\ and \oiii\  emission into the two clumps using velocity information following \cite{matthee2017} and \cite{carniani2018b}. Our measurements for the individual clumps are also summarized in Table \ref{tab:results}. The top middle and top right panels of Figure \ref{fig:contours} show the \cii\ emission extracted from the velocity range of [+8 km s$^{-1}$, +280 km s$^{-1}$] and [-353 km s$^{-1}$, -52 km s$^{-1}$], respectively, where the velocity zero point is defined as \zsys. Likewise, the decomposed \oiii\ emission are shown in the bottom middle and bottom right panels of  Figure \ref{fig:contours}. The flux centroids of the decomposed \cii\ and \oiii\ emission are consistent with the positions of the two UV clumps, demonstrating the successful decomposition.} 

\textcolor{black}{We perform photometry on individual clumps as in \S \ref{subsec:whole}. The clump A has \cii\ (\oiii) velocity-integrated flux of $0.47\pm0.07$ ($0.92\pm0.14$) Jy km s$^{-1}$. The clump B has \cii\ (\oiii) velocity-integrated flux of $0.38\pm0.06$ ($0.57\pm0.09$) Jy km s$^{-1}$. We then extract line spectra of individual clumps to obtain redshift and line FWHM values as in \S \ref{subsec:whole}. In Figure \ref{fig:1d_spec}, the middle and bottom panels show the spectra extracted at the position of the clump A and B, respectively.  In clump A, we obtain the \cii\ (\oiii) redshift of $7.1536\pm0.0004$ ($7.1523\pm0.004$) and  FWHM of $347\pm29$ ($325\pm32$) km s$^{-1}$. The $S/N$-weighted mean redshift is $7.1530\pm0.0003$. Likewise, in clump B, we obtain the \cii\ (\oiii) redshift of $7.1478\pm0.0005$ ($7.1488\pm0.004$) and  FWHM of $284\pm40$ ($267\pm34$) km s$^{-1}$. The $S/N$-weighted mean redshift is $7.1482\pm0.0003$. Based on these velocity-integrated flux and redshift values, the clump A has \cii\ (\oiii) luminosity of $6.0\pm0.9 \times 10^{8}$  ($21.1\pm3.2 \times 10^{8}$) \lsun. Likewise, the clump B has \cii\ (\oiii) luminosity of $4.9\pm0.8 \times 10^{8}$ ($13.0\pm2.1 \times 10^{8}$) \lsun.}

\begin{figure*}[thbp]
\hspace{+1.5cm}
\includegraphics[width=14cm]{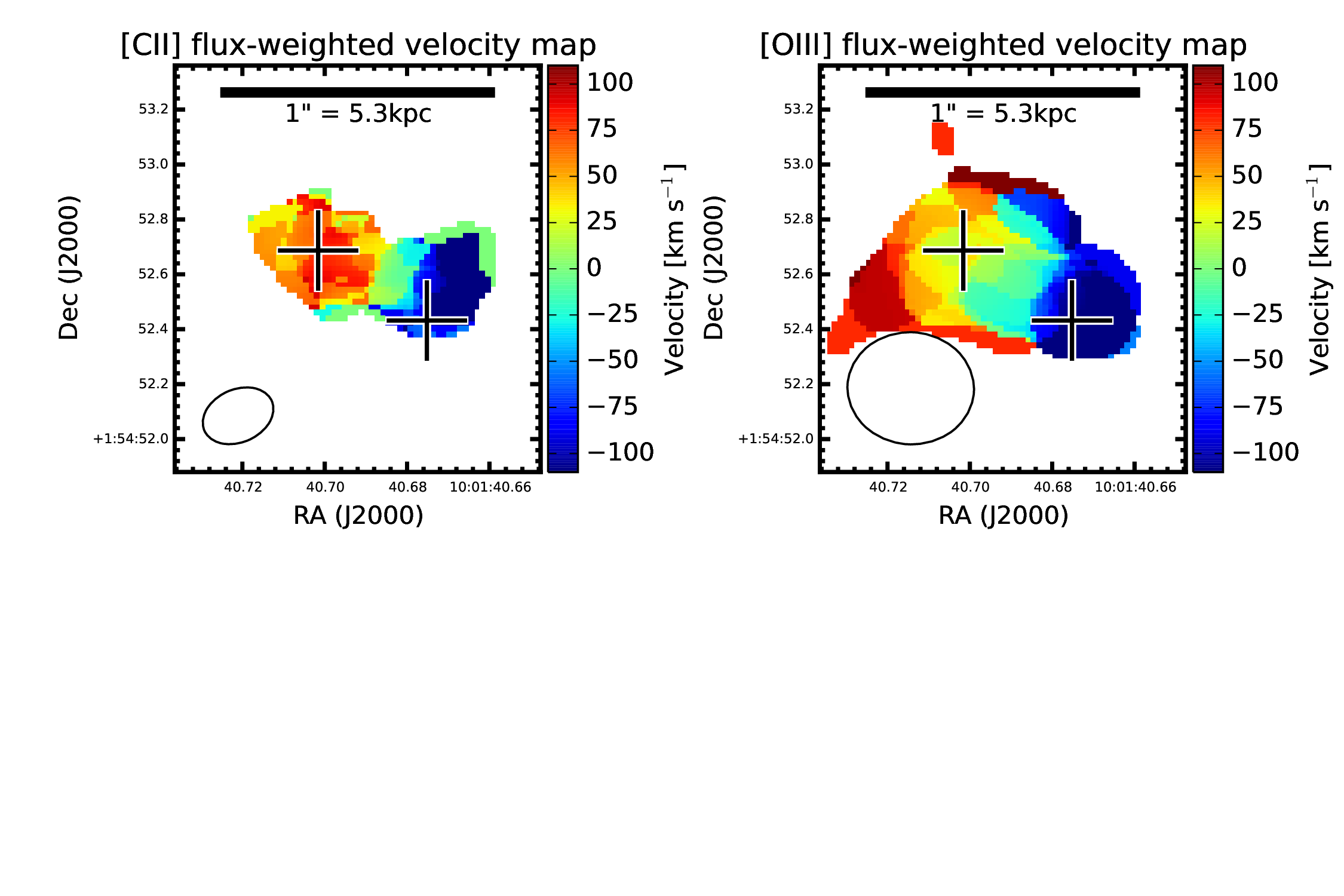}
\vspace{-2.5cm}
\caption
{
Left and right panels show flux-weighted velocity (i.e., Moment 1) maps 
of \cii\ and \oiii\ lines, respectively. 
The velocity zero point is defined as the systemic redshift, \zsys\ $=$ 7.1520.
Only pixels with detections above $3\sigma$ are used to create the maps. 
The flux peak positions of the UV clumps are indicated by the black crosses. 
\textcolor{black}{In each panel, the ellipse at lower left indicates the synthesized beam size of ALMA.} 
}
\label{fig:velocity_map}
\end{figure*}

Based on the $S/N$-weighted mean redshifts, the velocity offset of the two clumps is $177\pm16$ km s$^{-1}$. \textcolor{black}{To better understand the velocity field of \name,} we also create a flux-weighted velocity (i.e., Moment 1) map of \cii\ \textcolor{black}{and \oiii\ } with the CASA task {\tt immoments}. In this procedure, we only include pixels above $3\sigma$ in the velocity-integrated intensity image (c.f., \citealt{jones2017a}). The  left and right panels of Figure \ref{fig:velocity_map} demonstrate that \cii\ and \oiii\ shows an $\approx 200$ km s$^{-1}$ velocity gradient, respectively. 

\begin{figure}[]
\includegraphics[width=7cm]{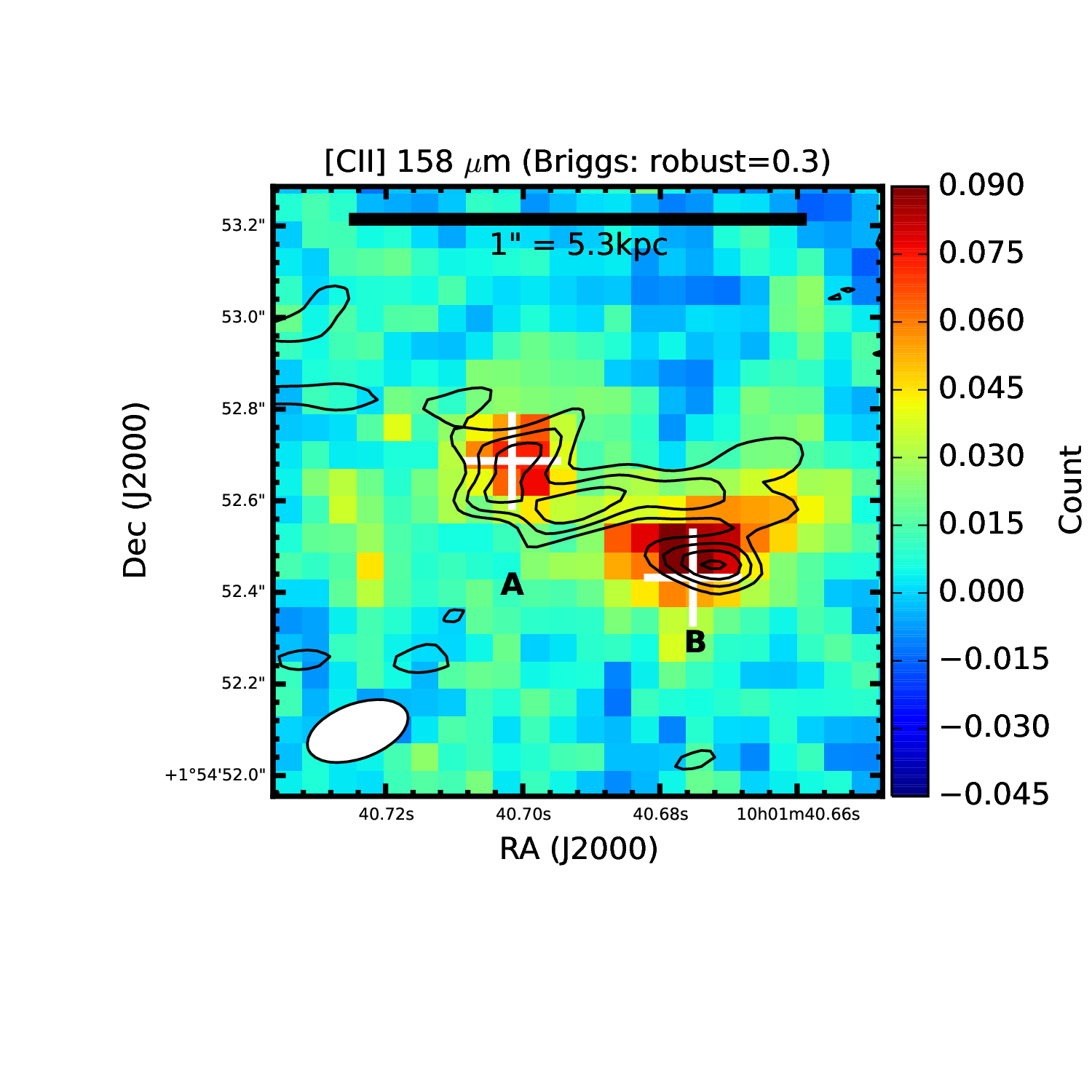}
\vspace{-1cm}
\caption
{
Zoomed-in \cii\ line image with a Briggs weighting (robust $=$ 0.3). 
Contours are drawn at ($-3$, $2$, $3$, $4$, $5$)$\times \sigma$, 
where $\sigma = 26$ mJy beam$^{-1}$ km s$^{-1}$.
White crosses show \textcolor{black}{the positions of clumps A and B.
The ellipse at lower left indicates the synthesized beam size of ALMA.} 
}
\label{fig:CII_briggs}
\end{figure}

Finally, \textcolor{black}{Figure \ref{fig:CII_briggs} shows a higher spatial resolution image of \cii\ with Briggs weighting and a robust parameter $0.3$ (Table \ref{tab:data})}
\footnote{Based on aperture photometry, we find that the higher spatial resolution image recovers $\approx86$\%\ of the total \cii\ flux, implying that the ``resolved-out'' effect is insignificant.}. 
The clump A has extended disturbed morphology, while the clump B has compact morphology. 

\textcolor{black}{To summarize, \name\ has two clumps in UV, \cii, and \oiii\ whose positions are consistent with each other. The spectra of \cii\ and \oiii\ can be decomposed into two Gaussians kinematically separated by $\approx200$ km s$^{-1}$. The velocity field is not smooth, implying that the velocity field may not be due to a rotational disk (see similar discussion in \citealt{jones2017b}). These results indicate that \name\ is a merger system, as first claimed by \cite{bowler2017} (see \S \ref{sec:intro}). A further discussion in terms of merger is presented in \S \ref{subsec:name}. 
Finally, we note that even the individual clumps have the highest \cii\ and \oiii\ luminosities among star forming galaxies at $z>6$ (\cii: \citealt{carniani2018b} and references therein, \oiii: \citealt{inoue2016, laporte2017, carniani2017a, hashimoto2018a, tamura2018}). 
}

\subsection{Dynamical Mass of the Individual Clumps}
\label{subsec:dynamical_mass}

\textcolor{black}{
Assuming the virial theorem, we can derive the \mdyn\ value of the two individual clumps as 
\begin{equation}
M_{\rm dyn.} = C \times \frac{r_{\rm 1/2} \sigma_{\rm line}^{2}}{G}, 
\end{equation}
where $r_{\rm 1/2}$ is the half-light radius, $\sigma_{\rm line}$ is the line velocity dispersion, and $G$ is the gravitational constant. The factor $C$ depends on various effects such as the galaxy's mass distribution, the velocity field along the line of sight, and relative contributions from random or rotational motions. For example, \cite{binney.tremaine2008} shows that $C = $ 2.25 is an average value of known galactic mass distribution models. \cite{erb2006b} have used $C=$ 3.4 under the assumption of a disk geometry taking into account an average inclination angle of the disk. In the case of dispersion-dominated system, \cite{forster-schreiber2009} have proposed that $C=$ 6.7 is appropriate for a variety of galaxy mass distributions. 
Because we do not see a clear velocity field in each clump (Figure \ref{fig:velocity_map}), we use $C =$ 6.7. Adopting the major semi-axis of the 2D Gaussian fit for lines as $r_{\rm 1/2}$ (Table \ref{tab:results}), we obtain \mdyn\ $= (5.7\pm1.6) \times 10^{10}$ and ($3.1\pm1.1) \times 10^{10}$ \msun\ for the clump A and B, respectively. Under the assumption that \mdyn\ of the whole system is the summation of the individual \mdyn\ values, we obtain the total dynamical mass of ($8.8\pm1.9) \times 10^{10}$ \msun. Note that our dynamical mass estimate is uncertain at least by a factor of three due to the uncertainty in $C$. Furthermore, given the nature of merger in \name, the virial theorem may not be applicable for the individual clumps (and the whole system). Thus, our \mdyn\ values should be treated with caution.  In \S \ref{sec:sed_fit}, we compare the dynamical mass with the stellar mass derived from the SED fitting. 
}




\section{Dust}
\label{sec:dust}

\begin{figure*}[th]
\hspace{+2.5cm}
\includegraphics[width=13cm]{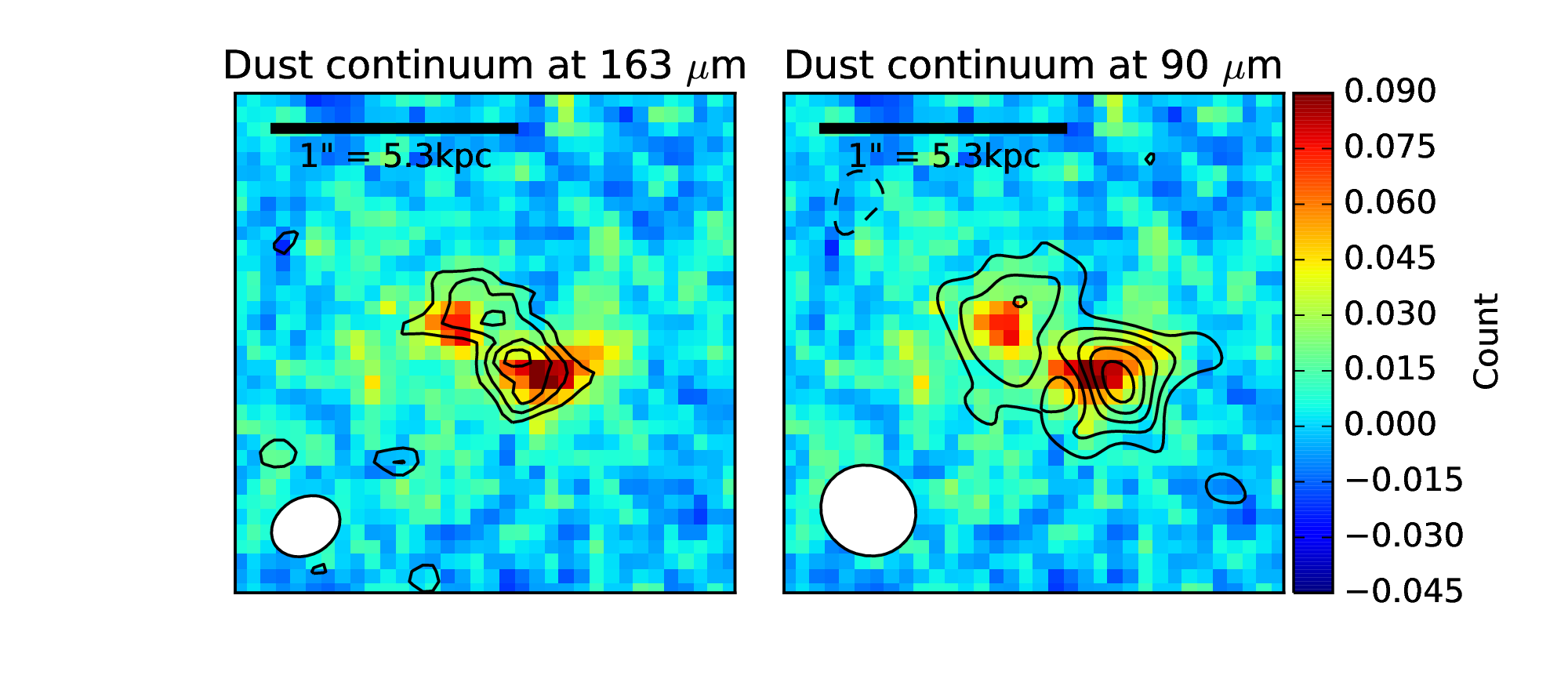}
\caption
{
Left and right panels show ALMA dust continuum images overlaid on the $2''.0\times2''.0$ cutout image of HST F140W, respectively.
({\it Left}) Dust continuum contours at $\approx 163$ $\mu$m 
drawn at ($-3$, $2$, $3$, $4$, $5$)$\times \sigma$, 
where $\sigma = 9.5$ $\mu$Jy beam$^{-1}$.
({\it Right}) Dust continuum contours 
at $\approx 90$ $\mu$m drawn at (-3, 2, 3, 4, 5, 6)$\times \sigma$, 
where $\sigma = 29.4$ $\mu$Jy beam$^{-1}$. 
In each panel, 
negative and positive contours are shown by the dashed and solid lines, respectively, 
and the ellipse at lower left indicates the synthesized beam size of ALMA. 
}
\label{fig:dust}
\end{figure*}

We search for dust thermal emission in the two continuum images at around 163 and 90 \micron. Hereafter, we refer to these two images as \textit{dust163} and \textit{dust90}, respectively. The left and right panels of Figure \ref{fig:dust} show contours of \textit{dust163} and \textit{dust90} overlaid on the HST F140W images, respectively, \textcolor{black}{and our measurements are summarized in Table \ref{tab:results}.} We have detected a signal at the F140W position \textcolor{black}{in the two images. We stress that \name\ is the second star-forming galaxy at $z>7$ after A1689$\_$zD1 (\citealt{watson2015, knudsen2017}) with dust continuum detections in multiple wavelengths. To obtain the continuum flux density of the individual clumps,} we spatially integrate the image with the CASA task {\tt imfit} assuming \textcolor{black}{two-component} 2D Gaussian profiles for the flux density. \textcolor{black}{In this procedure, the positions of the 2D Gaussian components are fixed at the clump positions.} 

\textcolor{black}{
The clump A (B) has flux densities of $S_{\nu, 163\mu m} = 41 \pm 23$ ($87 \pm 26$) $\mu$Jy and $S_{\nu, 90\mu m} = 208 \pm 83$ ($246 \pm 73$) $\mu$Jy. Because the continuum emitting regions of the individual clumps are not spatially resolved, we adopt the beam size of \textit{dust163}, i.e., the higher-angular resolution continuum image, as the upper limits. At $z=7.1520$, the upper limit of $0.29 \times 0.23$ arcsec$^{2}$corresponds to $1.6\times1.2$ kpc$^{2}$.
} 

\textcolor{black}{
For the whole system,
} 
the derived flux densities are $S_{\nu, 163\mu m} = 130 \pm 25$ $\mu$Jy and $S_{\nu, 90\mu m} = 470 \pm 128$ $\mu$Jy. \textcolor{black}{The beam-deconvolved size of \textit{dust163} is $(0''.72\pm0''.20) \times (0''.15\pm0''.10)$, corresponding to $(3.8 \pm 1.1) \times (0.8 \pm 0.5)$ kpc$^{2}$ at $z=7.15$, with a positional angle (PA) of $46^{\circ} \pm 11^{\circ}$. Likewise, the  beam-deconvolved size of \textit{dust90} is $(0''.80\pm0''.25) \times (0''.40\pm0''.16)$, corresponding to $(4.3 \pm 1.4) \times (2.2 \pm 0.9)$ kpc$^{2}$ at $z=7.15$, with PA $= 57^{\circ} \pm 32^{\circ}$. These two size and PA values are consistent with each other within $1\sigma$ uncertainties.  In Table \ref{tab:results}, we only present the size of \textit{dust163} because it has a higher-angular resolution than \textit{dust90}. At the current angular resolution of our data, it is possible that the continuum flux density of each clump is contaminated by the other clump. For accurate estimates, higher angular resolution data are required.} 


Our $S_{\nu, 163\mu m}$ \textcolor{black}{for the whole system} is well consistent with the flux density at 158 $\mu$m presented by \cite{bowler2018}, $168\pm56$ $\mu$Jy, taken at a lower angular resolution, \textcolor{black}{further supporting our dust continuum detections.  \cite{bowler2018} have reported a spatial offset of $\Delta_{\rm tot.} = 0''.60$ predominantly in the North-South direction ($\Delta_{\rm R.A.} = 0''.17$, $\Delta_{\rm Dec.} = 0''.57$) between their Band 6 dust continuum and F140W positions (see Figure 6 in \citealt{bowler2018}). However, as shown in Figure \ref{fig:dust}, we do not find any significant spatial offset between our ALMA dust continuum and F140W positions although our F140W image has consistent astrometry with that of F140W image used in \cite{bowler2018} (Appendix 1). Instead, based on comparisons of our and Bowler et al's Band 6 data, we find a marginal ($2.3\sigma$) spatial offset of $\Delta_{\rm tot.} = 0''.44$ between the two continuum images at the rest-frame wavelength of $\approx160$ \micron\ again predominantly in the North-South direction ($\Delta_{\rm R.A.} = 0''.05$, $\Delta_{\rm Dec.} = 0''.44$) (see Appendix 2 Figure \ref{fig:appendix2}).  Although the origin of the possible spatial offset between the two ALMA Band 6 data is unclear, it would not be due to the resolved-out effect because the two Band 6 continuum flux densities are consistent within $1\sigma$ uncertainties. }

We estimate the total infrared luminosity, \ltir, by integrating the modified black-body radiation over $8-1000$ \micron. \textcolor{black}{In star forming galaxies at $z>6-7$, all but one previous studies have assumed a dust temperature, $T_{\rm d}$, and a dust emissivity index, $\beta_{\rm d}$, because of none or a single dust continuum detection (e.g., \citealt{ota2014, matthee2017, laporte2017}). The only one exception is A1689$\_$zD1 which has dust continuum detections at two wavelengths (\citealt{watson2015, knudsen2017}).  In A1689$\_$zD1, \cite{knudsen2017} have obtained $T_{\rm d}$ ranging from 36 K to 47 K under the assumption that  $\beta_{\rm d}$ ranges from 2.0 to 1.5 with two dust continuum flux densities.  Following analyses of \cite{knudsen2017}, we attempt to estimate $T_{\rm d}$ with fixed $\beta_{\rm d}$ values. For the whole system, correcting for the cosmic microwave background (CMB) effects (\citealt{da_cunha2013, ota2014}), we obtain the best-fit $T_{\rm d}$ values of 61 K, 54 K, and 48 K for $\beta_{\rm d}=$ 1.5, 1.75, and 2.0, respectively, where the $1\sigma$ temperature uncertainty is about 10 K for each case. Because of loose constrains on $T_{\rm d}$ for the individual clumps, we assume the same combinations of $\beta_{\rm d}$ and $T_{\rm d}$ as in the whole system. We obtain \ltir\ $\approx 1 \times 10^{12}$, $3 \times 10^{11}$, and $7 \times 10^{11}$ \lsun, for the whole system, clump A, and clump B, respectively (Table \ref{tab:results}). Owning to the two dust continuum detections, the \ltir\ value is relatively well constrained in \name.}  

Assuming a dust mass absorption coefficient $\kappa = \kappa_{0} (\nu/\nu_{0})^{\beta_{\rm d}}$, where $\kappa_{0} = 10$ cm$^{2}$ g$^{-1}$ at $250$ $\mu$m (\citealt{hildebrand1983}), \textcolor{black}{we obtain the dust mass, \md\ $\approx 1 \times10^{7}$, $3\times10^{6}$, and $6\times10^{6}$ \msun, for the whole system, clump A, and clump B, respectively (Table \ref{tab:results}). We note that the dust mass estimate is in general highly uncertain due to the unknown $\kappa_{0}$ value. For example,} 
if we use $\kappa_{0} = 0.77$ cm$^{2}$ g$^{-1}$ at $850$ $\mu$m (\citealt{dunne2000}), the dust mass estimates become twice as large. 





\section{Luminosity Ratios}
\label{sec:luminosity_ratio}

\subsection{IR-to-UV luminosity ratio (IRX) and IRX-$\beta$ relation}
\label{subsec:irx}

\begin{table*}
\tbl{The ALMA spectroscopic literature sample: UV continuum slopes\label{tab:irx1}}
{
\begin{tabular}{lccccccc}
\hline
Name & Redshift & Waveband1 & Waveband2 & $\lambda_{\rm rest}$ & $\beta$ & $L_{\rm UV}$ & $\mu$\\
(1) & (2) & (3) & (4) & (5) & (6) & (7) & (8)\\ 
 & &  &  & (\AA) &  & ($10^{10}$\lsun) &  \\
\hline 
MACS1149-JD1 &  9.11  & F140W $=25.88\pm0.02$ & F160W $=25.70\pm0.01$  & $1400-1600$ & $-0.76\pm0.16$ & $(8.6\pm0.2)/\mu$ & 10 \\
A2744$\_$YD4 & 8.38  &  F140W $=26.46\pm0.04$ &  F160W $=26.42\pm0.04$ & $1500-1700$ & $-1.63\pm0.53$ & $(4.5\pm0.2)/\mu$ & $1.8\pm0.3$  \\
MACS0416$\_$Y1 & 8.31 &  F140W $=26.08\pm0.05$ &  F160W $=26.04\pm0.05$ & $1500-1700$ & $-1.72\pm0.50$ & $(6.3\pm0.3)/\mu$ & $1.4$\\ 
A1689$\_$zD1 & 7.5 &  F140W $=24.64\pm0.05$ &  F160W $=24.51\pm0.11$ & $1600-1900$ & $-1.10\pm0.83$ & $(20.5\pm0.9)/\mu$ & 9.3 \\
SXDF-NB1006-2 & 7.21  &  $J =25.46\pm0.18$ &  $H >25.64$ & $1500-1900$ &$< -2.6$ & $9.1\pm1.5$ & $-$ \\ 
B14-65666 & 7.15 & $J=24.7^{+0.2}_{-0.2}$ & $H = 24.6^{+0.3}_{-0.2}$ & $1500-1900$ & $-1.85^{+0.54}_{-0.53}$ $^{a}$ & $19.9\pm3.4$ & $-$\\ 
IOK-1 & 6.96 & F125W $=25.42\pm0.05$ &  F160W $=25.44\pm0.06$  & $1600-2000$ & $-2.07\pm0.26$ & $9.0\pm0.4$  & $-$ \\
\textcolor{black}{SPT0311-58E} & 6.90 & F125W $=25.28\pm0.10$ &  F160W $=24.98\pm0.12$  & $1600-2000$ & $-0.88\pm0.58$ & $(10.1\pm0.9)/\mu$  & $1.3$ \\
COS-3018555981 & 6.85 & F125W $-^{b}$ &  F160W $-^{b}$   & $1600-2000$ & $-1.22\pm0.51$ & $11\pm1^{c}$  & $-$ \\
COS-2987030247 & 6.81 & F125W $-^{b}$  &  F160W $-^{b}$  & $1600-2000$ & $-1.18\pm0.53$ & $13\pm1^{c}$  & $-$ \\
Himiko & 6.60 & F125W $=24.99\pm0.08$ &  F160W $=24.99\pm0.10$  & $1600-2100$ & $-2.00\pm0.48$ & $12.4\pm1.1$  & $-$ \\
\hline
\end{tabular}
}
\tabnote{Note. 
(1) Object Name; (2) Spectroscopic redshift; (3) and (4) Two wavebands and their photometry values to derive the UV continuum slope; 
(5) Rest-frame wavelength range probed by the wavebands; (6) UV spectral slope; (7) UV luminosity at $\approx1500$ \AA\ obtained from the photometry value of the Waveband1; and (8) lensing magnification factor.  
\textcolor{black}{Upper limits represent $3\sigma$.}
\\
Redshift and photometry values are taken from the literature as summarized below. \\
MACS1149-JD1: Redshift (\citealt{hashimoto2018a}); F140W and F160W (\citealt{wei.zheng2017});\\
A2744$\_$YD4: Redshift (\citealt{laporte2017}); F140W and F160W (\citealt{wei.zheng2014});\\
MACS0416$\_$Y1: Redshift (\citealt{tamura2018}); F140W and F160W (\citealt{laporte2015});\\
A1689$\_$zD1: Redshift (\citealt{watson2015}); F125W and F160W (\citealt{watson2015});\\
SXDF-NB1006-2: Redshift (\citealt{inoue2016}); $J$ and $H$ (\citealt{inoue2016});\\
B14-65666: Redshift (This Study); $J$ and $H$ (\citealt{bowler2014});\\
IOK-1: Redshift (\citealt{ota2014}); F125W and F160W (\citealt{jiang2013}; the object No. 62 in their Table 1);\\
\textcolor{black}{SPT0311-058E: Redshift (\citealt{marrone2018}); F125W and F160W (\citealt{marrone2018});} \\ 
\textcolor{black}{COS-3018555981: Redshift (\citealt{smit2018});} \\ 
\textcolor{black}{COS-2987030247: Redshift (\citealt{smit2018});} \\ 
Himiko: Redshift (\citealt{ouchi2013}); F125W and F160W (\citealt{ouchi2013})\\
$^{a}$ The UV spectral slope in \cite{bowler2018} based on the updated photometry values.\\ 
$^{b}$ Not available. \\ 
$^{c}$ Values taken from \cite{smit2018}. \\ 
}
\end{table*}

\begin{table*}
\tbl{The ALMA spectroscopic literature sample: Infrared-excess (IRX)\label{tab:irx2}}
{
\begin{tabular}{lcccccl}
\hline
Name &  $S_{\rm \nu}$ & Rest-Wavelength &  \ltir & IRX  & Ref. \\ 
& ($\mu$Jy) & ($\mu$m)  & ($10^{10}$\lsun) &   & \\ 
(1) & (2) & (3) & (4) & (5) & (6)  \\ 
\hline 
MACS1149-JD1  & $< 53/\mu$  & 90 & $< 11.4/\mu$ &  $<0.12$ & \cite{hashimoto2018a}\\ 
A2744$\_$YD4  & $(175\pm69)/\mu^{a}$ & 90 & $(32.7\pm12.9)/\mu$ & $0.86\pm0.22$ & \cite{laporte2017} \\ 
MACS0416$\_$Y1 & $(137\pm26)/\mu$ & 91 & $(25.7\pm4.9)/\mu$ & $0.61\pm0.09$ & \cite{tamura2018} \\
A1689$\_$zD1  & $(1330\pm140)/\mu$ & 103 & $(255\pm26.8)/\mu$ & $1.09\pm0.05$ & \cite{knudsen2017}\\ 
SXDF-NB1006-2  & $<42$ & 162 & $<19.6$ & $<0.33$ & \cite{inoue2016}\\
B14-65666  & $130\pm25$ & 163 & $61.8\pm11.9$ & $0.49\pm0.12$ & \cite{bowler2018}, This Study\\ 
IOK-1   & $<63$  & 162 & $<28.0$ & $<0.49$ & \cite{ota2014} \\ 
\textcolor{black}{SPT0311-E} & $(1530\pm70)/\mu$$^{b}$ & 159 & $(640\pm30)/\mu$ & $1.80\pm0.04$ & \cite{marrone2018}\\ 
COS-3018555981 & $<87$ & 158 & $<35.4$ & $<0.51$ & \cite{smit2018}\\
COS-2987030247 & $<75$ & 158 & $<30.2$ & $<0.37$ & \cite{smit2018}\\ 
Himiko & $<51$  & 153 & $<18.0$ & $<0.16$ & \cite{ouchi2013}\\
\hline
\end{tabular}
}
\tabnote{Note. 
(1) Object Name; (2) and (3) Dust continuum flux density and its rest-frame wavelength; (4) Total IR luminosities estimated by integrating the modified-black body radiation at $8-1000$ $\mu$m with $T_{\rm d} =$ 50 K and $\beta_{\rm d} = 1.5$; (5) IRX values with $T_{\rm d} =$ 50 K and $\beta_{\rm d} = 1.5$; and (6) Reference. 
Upper limits represent $3\sigma$. 
\\
$^{a}$ Continuum flux density after performing primary beam correction. \\
\textcolor{black}{$^{b}$ Continuum flux density before the lensing correction is estimated from the intrinsic flux density of $1.18\pm0.05$ mJy and $\mu=1.3$ (\citealt{marrone2018}).} \\
}
\end{table*}

The relation between the IR-to-UV luminosity ratio, IRX$\equiv$ log$_{\rm 10}$(\ltir/$L_{\rm UV}$), and the UV continuum slope, $\beta$, is useful to constrain the dust attenuation curve of galaxies (e.g., \citealt{meurer1999}). Local starburst galaxies are known to follow the Calzetti's curve (e.g., \citealt{calzetti2000, takeuchi2012}). Studies have shown that high-$z$ galaxies may favor a steep attenuation curve similar to that of the Small Magellanic Could (SMC) (e.g., \citealt{reddy2006, kusakabe2015}). Based on a stacking analysis of LBGs at $z\approx2-10$, \cite{bouwens2016} show that high-$z$ galaxies have a low IRX value at a given $\beta$, even lower than the SMC curve. This is interpreted as a steep attenuation curve or a high \td\ at high-$z$, the latter being supported from detailed analyses of the IR spectral energy distribution in high-$z$ analogs (\citealt{faisst2017}, see also \citealt{behrens2018}). On the other hand, several studies claim that there is no or little redshift evolution in the IRX-$\beta$ relation at least up to $z\approx5$ (\citealt{fudamoto2017, koprowski2018}). Thus, a consensus is yet to be reached on the high-$z$ IRX-$\beta$ relation. 

At $z\gtrsim6$, little is understood about the IRX-$\beta$ relation due to the small sample with dust continuum detections (\citealt{bowler2018}). Therefore, \name\ would provide us with a clue to understand the IRX-$\beta$ relation at $z > 6$. \cite{bowler2018} have first discussed the position of \name\ in the IRX-$\beta$ relation. The authors have compared \textcolor{black}{the whole system of} \name\ with the $z\approx3-5$ results (\citealt{fudamoto2017, koprowski2018}), a stacking result of $z\approx4-10$ (\citealt{bouwens2016}), and with another $z>7$ galaxy that has a dust continuum detection, A1689$\_$zD1 (\citealt{watson2015, knudsen2017}). 

In this study, we focus on the IRX-$\beta$ relation at $z>6.5$ based on a compiled sample of \textcolor{black}{eleven} spectroscopically confirmed galaxies. 
Tables \ref{tab:irx1} and \ref{tab:irx2} summarize our sample from the literature and this study.  The sample includes \textcolor{black}{five} galaxies with dust continuum detections: A1689$\_$zD1 (\citealt{watson2015, knudsen2017}), A2744$\_$YD4 (\citealt{laporte2017}), MACS0416$\_$Y1 (\citealt{tamura2018}), \textcolor{black}{SPT0311-58E (\citealt{marrone2018})}, and \name\ (see also \citealt{bowler2018}). In addition, the sample includes objects with deep $3\sigma$ upper limits on the IRX obtained with ALMA: Himiko (\citealt{ouchi2013, schaerer2015}), IOK-1 (\citealt{ota2014, schaerer2015}), SXDF-NB1006-2 (\citealt{inoue2016}), COS-301855981, COS-29870300247 (\citealt{smit2018}), and MACS1149-JD1 (\citealt{hashimoto2018a}).

For fair comparisons of the data points, 
we uniformly derive $\beta$ from two photometry values following the equation (1) of \cite{ono2010b}. 
We use the combination of (F125W, F160W) and (F140W, F160W) \textcolor{black}{or ($J$, $H$)} 
at two redshift bins of $z=6.60-7.21$ and $z=7.5-9.11$, respectively. 
These wavebands probe the rest-frame wavelength ranges of 
$\approx 1600-2000$ \AA\ and  $1500-1700$ \AA\ at the two redshift bins. 
Because of the difference in the probed wavelength range, 
the derived $\beta$ values should be treated with caution. 
The estimated $\beta$ values are summarized in Table \ref{tab:irx1}. 
To estimate the UV luminosity, $L_{\rm UV}$, of the sample, we consistently use the rest-frame $\approx1500$ \AA\ magnitude of the waveband1 in Table \ref{tab:irx1}. 
\textcolor{black}{To obtain \ltir\ of the literature sample, we have assumed  \td\ $=50$ K and $\beta_{\rm d} = 1.5$. These assumptions would be reasonable because A1689$\_$zD1 and \name\ have \td\ $\approx40 - 60$ K at $\beta_{\rm d} = 2.0 - 1.5$ (see \S \ref{sec:dust}).}

\textcolor{black}{In \name, because only one HST photometry data point (F140W) is available, we cannot obtain $\beta$ of the individual clumps. Therefore, we investigate the IRX and $\beta$ values of the entire system. We adopt $\beta = -1.85^{+0.54}_{-0.53}$ in \cite{bowler2018} calculated with updated $J$- and $H$-band photometry values. For fair comparisons to other data points, we assume \td\ $=$ 50 K and $\beta_{\rm d} = 1.5$ to obtain \ltir\ $=(6.2\pm1.2) \times 10^{11}$ \lsun, which is about a factor two lower than the values presented in Table \ref{tab:results}. With $L_{\rm UV} = 2.0 \times 10^{11}$ \lsun, we obtain the IRX value of $0.5\pm0.1$. If we instead use the \ltir\ values in Table \ref{tab:results}, we obtain a slightly higher IRX value, $0.7\pm0.1$. Therefore, we adopt IRX $=0.5^{+0.3}_{-0.1}$ as a fiducial value.} 

\begin{figure}[]
\hspace{-1cm}
\includegraphics[width=10cm]{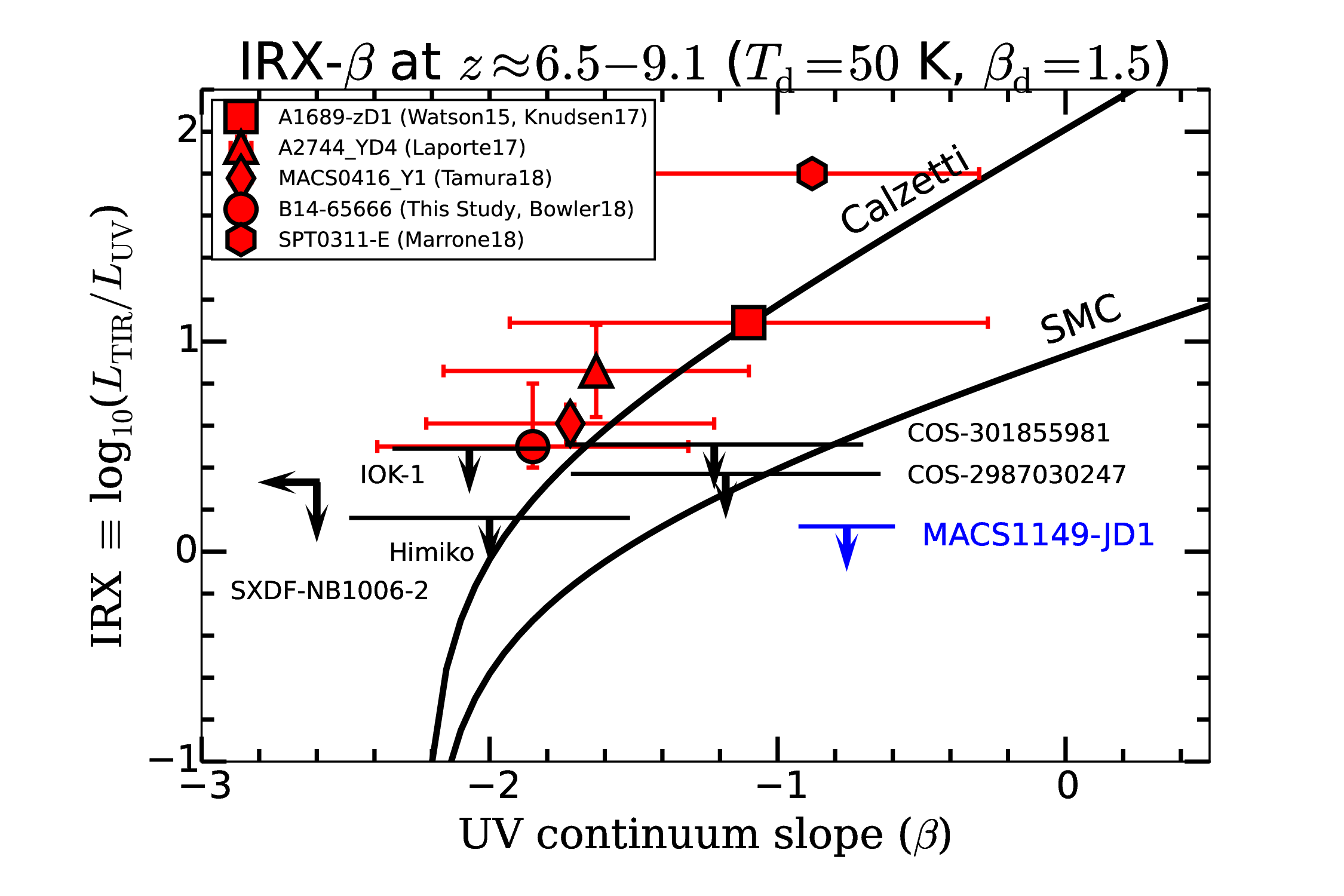}
\caption
{
The IRX, plotted against the UV slope, $\beta$, for \textcolor{black}{eleven} spectroscopically identified galaxies at $z\approx6.5-9.1$.  For fair comparisons of data points from the literature, we have uniformly derived $\beta$ and IRX values (see \S \ref{subsec:irx} for the details). 
We plot the IRX value under the assumption of \td\ $=50$ K and $\beta_{\rm d} = 1.5$.   
\textcolor{black}{The arrows correspond to $3\sigma$ upper limits.} 
In each panel, the two solid black lines indicate the IRX-$\beta$ relation 
based on the Calzetti and SMC dust laws (\citealt{bouwens2016}). 
The \textcolor{black}{five} red symbols denote objects with dust continuum detections; A1689$\_$zD1 (\citealt{watson2015, knudsen2017}), A2744$\_$YD4 (\citealt{laporte2017}), MACS0416$\_$Y1 (\citealt{tamura2018}), SPT0311-58E (\citealt{marrone2018}), and \name. The details of the data are summarized in Tables \ref{tab:irx1} and \ref{tab:irx2}. 
}
\label{fig:irx_beta}
\end{figure}

Figure \ref{fig:irx_beta} shows the $z>6.5$ galaxies in the IRX-$\beta$ relation.
We also plot the IRX-$\beta$ relations based on the Calzetti and SMC dust laws 
assuming the intrinsic $\beta$ value of $-2.2$ (\citealt{bouwens2016}). 
We find that \textcolor{black}{five} LBGs with dust continuum detections are consistent with the Calzetti's curve if we assume \td\ $=50$ K \textcolor{black}{and $\beta_{\rm d} = 1.5$}. 
Among the six null-detections, 
SXDF-NB1006-2 has a very steep UV slope $\beta<-2.6$ ($3\sigma$). 
Such a steep $\beta$ can be reproduced if we assume 
a very young stellar age ($<10$ Myr) or low metallicity 
(e.g., \citealt{schaerer2003, bouwens2010}, see also Figure 10 of \citealt{hashimoto2017a}). 
Likewise, MACS1149-JD1 has a stringent upper limit on the IRX value.  
Although it is possible that MACS1149-JD1 lies below the SMC curve, 
we note that the presented $\beta$ value of MACS1149-JD1 probes the rest-frame 
wavelength range of $1400-1600$ \AA. Deeper $K$ band data would allow us 
to compute $\beta$ in the wavelength range of $\approx1500-2200$ \AA\ 
which is comparable to that probed in previous high-$z$ studies 
(e.g., \citealt{bouwens2009, hashimoto2017b}). 

Although a large and uniform sample with dust continuum detections is needed to understand the typical attenuation curve at $z>6.5$,  our first results show that there is no strong evidence for a steep (i.e., SMC-like) attenuation curve at least for the \textcolor{black}{five} LBGs detected in dust. 


\subsection{\oiii/\cii\ Luminosity Ratio}
\label{subsec:line_ratio}

The line luminosity ratio, \oiii/\cii, would give us invaluable information 
on chemical and ionization properties of galaxies (e.g., \citealt{inoue2016, marrone2018}). 
For example, in local galaxies, 
a number of studies have examined the line ratio (\citealt{malhotra2001, brauher2008, madden2012, cormier2015}). 
These studies have shown that dwarf metal-poor galaxies have high line ratios, \oiii/\cii\ $\approx 2-10$, 
whereas metal-rich galaxies have low line ratios, \oiii/\cii\ $\approx0.5$. 
Alternatively, if the ISM of galaxies is highly ionized, the \cii\ luminosity would be weak 
because \cii\ emission is predominantly emitted from the PDR (e.g., \citealt{vallini2015, katz2017}).

In \name, the line luminosity ratio is \textcolor{black}{\oiii/\cii\ $= 3.1\pm0.6$, $3.5\pm0.8$, $2.7\pm0.6$ for the whole system, clump A, and clump B, respectively (Table \ref{tab:results}). }
We compare our results with those in other high-$z$ galaxies in the literature:
two $z\approx7$ star-forming galaxies 
and \textcolor{black}{two $z\approx 6 - 7$ sub-millimeter galaxies (SMGs).}  
\cite{inoue2016} have detected \oiii\ from a $z=7.21$ LAE 
with the \ew\  value of 33 \AA\ (SXDF-NB1006-2: \citealt{shibuya2012}). 
With the null detection of \cii, the authors have shown that SXDF-NB1006-2 has a total line luminosity ratio of \oiii/\cii\ $>12$ ($3\sigma$). 
\cite{carniani2017a} have reported detections of \oiii\ and \cii\ in a galaxy at $z=7.11$
(BDF-3299: \citealt{vanzella2011, maiolino2015}). 
BDF-3299 has a large \ew\ $= 50$ \AA\ and thus can be categorized into LAEs. 
The galaxy has spatial offsets between \oiii, \cii, and UV emission. 
Under the assumption that both \cii\ and \oiii\ are associated with the UV emission, 
we obtain the total line ratio of $3.7 \pm 0.6$ 
using the \cii\ luminosity ($4.9\pm0.6\times10^{8}$ \lsun) 
and the \oiii\ luminosity ($18\pm2\times10^{8}$ \lsun)
\footnote{
\cite{carniani2017a} have obtained the line ratio at the \oiii\ emitting region without \cii\ emission  
$>$ 8 at $5\sigma$. 
Because the value is obtained in a partial region of the galaxy, 
we have computed the total line luminosity ratio for fair comparisons to other data points.}.
Recently,  \cite{marrone2018} have detected both \oiii\ and \cii\ 
from a lensed SMG at $z=6.90$ comprised of two galaxies (SPT0311-058E and SPT0311-058W). 
The total line luminosity ratio is $1.27\pm0.18$ and $0.56\pm0.17$ for SPT0311-058E and SPT0311-058W, respectively, where the $1\sigma$ values take the uncertainties on magnification factors into account. 
%
\textcolor{black}{Finally, \cite{walter2018} have detected \oiii\ in an SMG at $z=6.08$ located at the projected distance of $\approx61$ kpc from a quasar at the same redshift. In the SMG, J2100-SB, the authors have presented the line luminosity ratio of $1.58\pm0.24$ combining the previous \cii\ detection (\citealt{decarli2017})\footnote{\textcolor{black}{We do not include quasars of \cite{hashimoto2018c} and \cite{walter2018} to focus the sample on normal star-forming galaxies.}}.}

Based on the combined sample of these literature objects with \name, 
we investigate the relation between the line luminosity ratio and the bolometric luminosity 
estimated as $L_{\rm bol} \approx L_{\rm UV} + L_{\rm TIR}$. 
\textcolor{black}{In \name, we obtain $L_{\rm bol} = (12.5\pm1.5) \times 10^{11}$, $(4.2\pm0.5) \times 10^{11}$  and $(8.1\pm1.0) \times 10^{11} L_{\rm \odot}$ for the whole system, clump A, and clump B, respectively (Table \ref{tab:results}).}
For the two LAEs without dust continuum detections, 
we derive the upper limits of $L_{\rm bol}$ as the $L_{\rm UV}$ measurements 
plus the $3\sigma$ upper limits of \ltir, where we assume \td\ $=$ \textcolor{black}{50 K} and $\beta_{\rm d}=1.5$. 
The $L_{\rm UV}$ value is used as lower limits of $L_{\rm bol}$
\footnote{\textcolor{black}{In SXDF-NB1006-2, with the luminosity values in Tables \ref{tab:irx1} and \ref{tab:irx2}, we obtain $9.1\times10^{10}$ and $2.9\times10^{11}$ \lsun\ for the lower and upper limits on $L_{\rm bol}$, respectively. In BDF-3299, based on $L_{\rm UV} = 3.3\times10^{10}$ \lsun\ (Table 2 in \citealt{carniani2018b}) and the $3\sigma$ \ltir\ upper limit of $<0.9\times10^{11}$ \lsun\ (\citealt{carniani2017a}),  we obtain $3.3\times10^{10}$ and $1.6\times10^{11}$ \lsun\ for the lower and upper limits on $L_{\rm bol}$, respecitvely.}
}. 
The \ltir\ value of the two SMGs are well constrained 
from multiple dust continuum detections at different wavelengths 
(see Extended Data Figure 7 in \citealt{marrone2018}). 
Because these SMGs have $L_{\rm UV}$/\ltir\ $\approx0.002-0.02$, 
we assume $L_{\rm bol} \approx$ \ltir\ for these objects. 
\textcolor{black}{We thus adopt $L_{\rm bol} = (4.6\pm1.2) \times 10^{12}$ and $(3.3\pm0.7) \times 10^{13} L_{\rm \odot}$ for SPT0311-58E and SPT0311-58W, respectively. 
Similarly, because J2100-SB is not detected in the rest-frame UV/optical wavelengths, we assume $L_{\rm bol} \approx$ \ltir\ $= (1.9\pm0.03) \times 10^{12}$ \lsun\ (\citealt{walter2018}).}

\begin{figure}[]
\includegraphics[width=8cm]{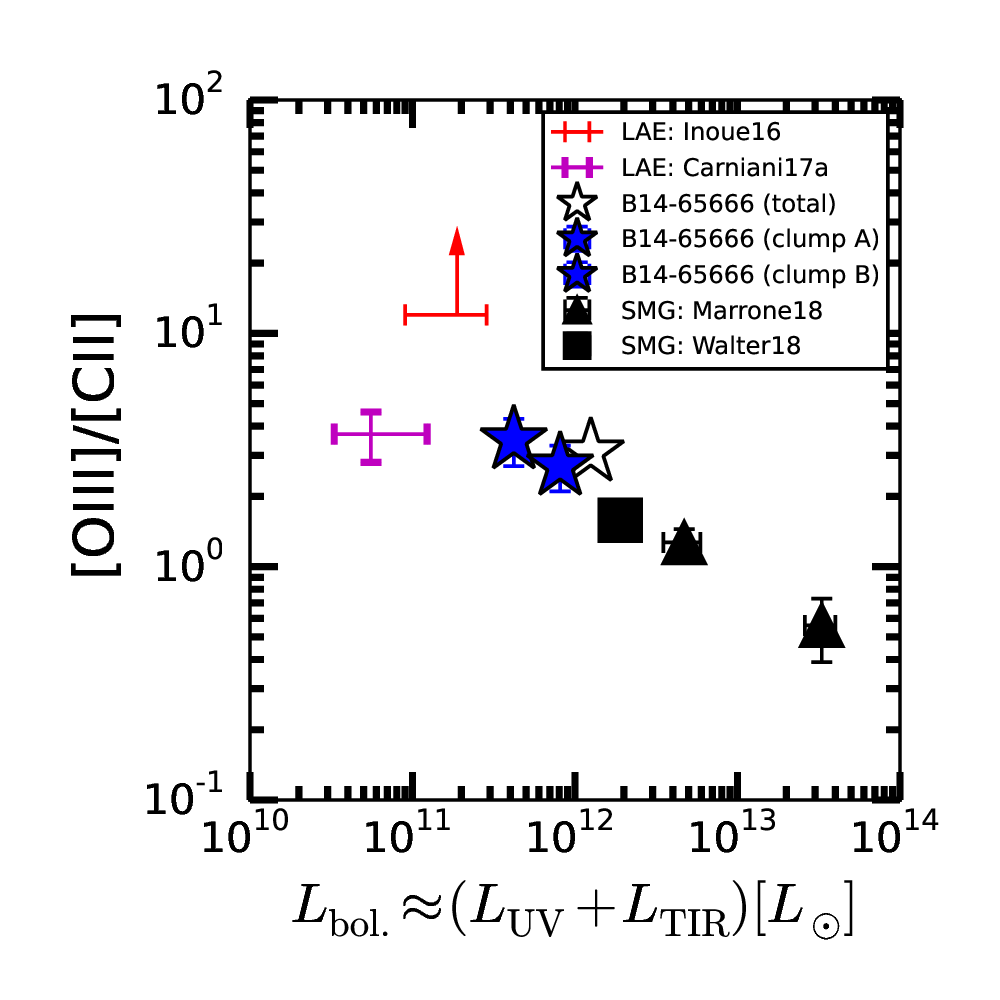}
\caption
{
\oiii-to-\cii\ line luminosity ratio plotted against the bolometric luminosity 
estimated as the summation of the UV and IR luminosities for $z\approx6-7$ objects. 
The red arrow represents the $3\sigma$ lower limit of the line luminosity ratio 
in the LAE of \cite{inoue2016}. 
For the two LAEs without \ltir\ measurements, 
the upper limits of $L_{\rm bol.}$ are estimated as 
the summation of $L_{\rm UV}$ and the $3\sigma$ upper limits on \ltir, 
where we assume \td\ $=$ \textcolor{black}{50} K and $\beta_{\rm d} = 1.5$. 
The lower limits of $L_{\rm bol.}$ for the two LAEs correspond to $L_{\rm UV}$. 
\textcolor{black}{Detailed calculations of these values are presented in \S \ref{subsec:line_ratio}.}
}
\label{fig:lbol_ratio}
\end{figure}

Figure \ref{fig:lbol_ratio} shows a clear anti-correlation, 
although a larger number of galaxies are needed for a definitive conclusion. 
Given that the bolometric luminosity traces the mass scale of a galaxy 
(i.e., the stellar and dark matter halo masses and/or the SFR), 
the possible trend implies that lower mass galaxies having higher luminosity ratios. 
These would in turn indicate that lower mass galaxies have  
lower metallicity \textcolor{black}{and/or} higher ionization states (cf., \citealt{maiolino.manucci2019, nakajima2016}). \textcolor{black}{Because we do not have direct measurements of these parameters in the sample, we leave further discussion to future studies.}




\section{SED fit}
\label{sec:sed_fit}

We perform stellar population synthesis model fitting to \name\ 
to derive the stellar mass ($M_{\rm *}$), dust attenuation ($A_{\rm V}$), 
the stellar age, stellar metallicity ($Z$), and the SFR. 

We use the $Y$, $J$, $H$, and $K$ band data taken by UltraVISTA (\citealt{bowler2014})
and  the deep \textit{Spitzer}/IRAC 3.6 and 4.5 $\mu$m data (\citealt{bowler2017}). 
The clumps A and B are not resolved under the coarse angular resolution of ground based telescopes. Therefore, the photometry values represent the total system of \name. We thus perform SED fitting to the total system. 
In addition, we use our \textcolor{black}{dust continuum flux densities and the \oiii\ flux.}  
We do not use the \cii\ flux. 
This is due to the difficulty in modeling \cii\ which arises both from the HII region and the PDR 
(see \citealt{inoue2014alma}).

The SED fitting code used in this study is the same as that used in 
\cite{hashimoto2018a} \textcolor{black}{and \cite{tamura2018}}. 
For the detailed procedure, we refer the reader to \cite{mawatari2016} and the relevant link
\footnote{
{https://www.astr.tohoku.ac.jp/\textasciitilde mawatari/KENSFIT/KENSFIT.html}
}. 
Briefly, the stellar population synthesis model of GALAXEV (\citealt{bc03}) is used. 
The nebular continuum and emission lines of \cite{inoue2011} are included. 
The \oiii\ line flux is estimated based on metallicity and the SFR 
with semi-empirical models (\citealt{inoue2014alma, inoue2016}).  
A Calzetti's law (\citealt{calzetti2000}) is assumed for dust attenuation, 
\textcolor{black}{which is appropriate for \name\ based on the results of the IRX-$\beta$ relation (\S \ref{subsec:irx}).}
The same attenuation value is used 
for the stellar and nebular components (e.g., \citealt{erb2006b, kashino2013}). 
An empirical dust emission templates of \cite{rieke2009} is adopted. 
The Chabrier initial mass function (\citealt{chabrier2003}) with $0.1-100$ \msun\ is adopted,  
and a mean IGM model of \cite{inoue2014igm} is applied. 
We fix the object's redshift of $7.1520$.
To estimate the best-fit parameters, 
we use the least $\chi^{2}$ formula of \cite{sawicki2012} 
including an analytic treatment of upper limits for non-detections. 
Uncertainties on the parameters are estimated based on a Monte Carlo technique ($N=300$).

\begin{figure}[]
\includegraphics[width=8cm]{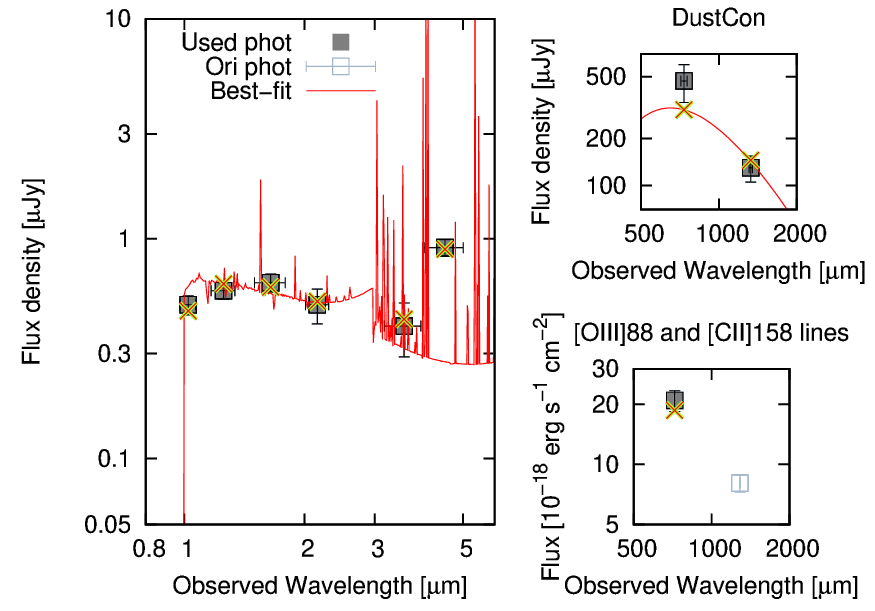}
\caption
{
Best-fit SED \textcolor{black}{(left)} taking into account the dust continuum flux densities \textcolor{black}{(top right)} 
and the \oiii\ flux \textcolor{black}{(bottom right)} for a constant star-formation model. 
In the left panel, black squares show $Y, J, H, K$-band photometry and the IRAC channel 1 and 2 measurements, \textcolor{black}{from left to right. An open-white square ('Ori phot') indicates the photometric data point of \cii\ 158 \micron, 
which is not used in the SED fitting (see the text for the details)}. 
Horizontal and vertical error bars represent the wavelength range of the filters 
and the $1\sigma$ uncertainties, respectively. 
The red solid line indicates the SED model and the corresponding band flux densities 
are shown by crosses. 
In the top right panel, the black squares are \textcolor{black}{the dust continuum flux densities at 90 and 163 \micron\ and their $1\sigma$ uncertainties.}
In the bottom right panel, the black square shows the observed \oiii\ flux and its $1\sigma$ uncertainty, 
while the cross is the model prediction. 
}
\label{fig:sed}
\end{figure}

For simplicity, we assume a constant star formation history (SFH). 
Figure \ref{fig:sed} shows the best-fit SED of \name\ and Table \ref{tab:sed} summarize the estimated physical quantities. 
The strong \oiii\ line flux indicates a very high current SFR, 
\textcolor{black}{$200^{+82}_{-38}$ \msun\ yr$^{-1}$.} 
We note that our stellar mass, 
$M_{\rm *} = 7.7^{+1.0}_{-0.7} \times 10^{8}$ \msun, 
or log($M_{\rm *}/M_{\rm \odot}) = 8.9\pm0.1$, \textcolor{black}{and the dust extinction value, $A_{\rm V} = 0.3^{+0.19}_{-0.10}$,  are consistent with the results of \cite{bowler2014, bowler2018} within $1\sigma$ uncertainties. In \S \ref{subsec:name}, we use the SED-fitting results to discuss the properties of \name.}

\begin{table}
\tbl{Results of SED fit\label{tab:sed}}
{
\begin{tabular}{cc}
\hline
Parameters & Values \\ 
\hline 
$\chi^2$ & 4.30 \\
$\nu$ & 4 \\
A$_{V}$ [mag] & $0.30^{+0.19}_{-0.10}$ \\
Age [Myr] & $3.8^{+1.8}_{-1.3}$ \\
Metallicity & $0.008^{+0.008}_{-0.004}$ \\
Escape fraction & $0.0^{+0.2}_{-0.0}$ \\
Stellar mass (\msun) [$10^{8}$\,M$_\odot$] & $7.7^{+1.0}_{-0.7}$ \\
SFR [M$_\odot$\,yr$^{-1}$] & $200^{+82}_{-38}$ \\
\hline
\end{tabular}
}
\tabnote{Note. 
The stellar mass and SFR values are obtained with the Chabrier IMF with $0.1-100$ \msun. 
\textcolor{black}{The solar metallicity corresponds to 0.02. 
Because we have nine data points (Figure \ref{fig:sed}) and attempt to constrain five parameters, the degree of freedom, $\nu$, is four. 
Note that the SFR value is not a variable parameter because SFR is readily obtained from the stellar age and stellar mass under the assumption of a star formation history.} 
}
\end{table}

\textcolor{black}{To test the validity of our SED-fit results, we examine if the derived stellar and dust masses can be explained in a consistent manner. 
}
Based on a combination of the stellar mass of $\approx 7.7 \times10^{8}$ \msun\ and 
the effective number of supernovae (SN) per unit stellar mass in the Chabrier IMF, 
0.0159 \msun$^{-1}$ (e.g., \citealt{inoue2011dust}), 
we obtain the number of SN $\approx 1.2\times10^{7}$. 
Thus, the dust mass of \textcolor{black}{$\approx 1 \times10^{7}$ \msun} requires the dust yield per SN  \textcolor{black}{$\approx 0.8$ \msun, which can be achieved if the dust destruction is insignificant (e.g., \citealt{michalowski2015}). 
}
The \textcolor{black}{dust-to-stellar mass ratio, log($M_{\rm *}/M_{\rm d}) \approx$ -1.9, is high but within the range observed in local galaxies (see Figure 11 in \citealt{remy-ruyer2015}) and can be } explained by theoretical models (\citealt{popping2017, calura2017}).

\textcolor{black}{
Finally, we compare the stellar mass with the dynamical mass (\S \ref{subsec:dynamical_mass}). 
The derived stellar mass, $M_{\rm *} = 7.7^{+1.0}_{-0.7} \times 10^{8}$ \msun, is well below \mdyn\ $= (8.8\pm1.9) \times 10^{10}$ \msun. The dynamical-to-stellar mass ratio, log($M_{\rm dyn}/M_{\rm *}$) $\approx 2.0$, is high but within the range obtained in star-forming galaxies at $z\approx2-3$ (e.g., \citealt{erb2006b, gnerucci2011}). 
}

\textcolor{black}{
One might think that the stellar age ($\approx4$ Myr) seems too young to reproduce the dust mass. However, we note that the stellar age deduced from the SED fitting indicates the age after the onset of current star formation activity. Given the signature of merger activity, we could infer the existence of star formation activity well before the observing timing at $z=7.15$ which forms two galaxies and a significant fraction of the dust mass (see such an example in \citealt{tamura2018}). 
}

\section{\lya\ velocity offset}
\label{sec:lya}

Because \lya\ is a resonant line, it is known that the \lya\ redshift, \zlya, does not exactly match  the systemic redshift defined by optically-thin nebular emission lines, e.g., \oiii\ and \cii. The discrepancy between the two redshifts provides a valuable probe of the interstellar medium (ISM) and the surrounding intergalactic medium (IGM). For example, based on radiative transfer calculations, theoretical studies (\citealt{dijkstra2006, verhamme2006, verhamme2015, gronke2015}) predict that the \lya\ line is redshifted (blueshifted) with respect to the systemic redshift if a galaxy has an outflowing (inflowing) gas in the ISM. When \lya\ photons enter the IGM, its spectral profile is further altered due to the damping wing of \lya\ absorption by the intergalactic neutral hydrogen, increasing \zlya\ (\citealt{haiman2002, laursen2013}). 

\begin{figure}[]
\includegraphics[width=9cm]{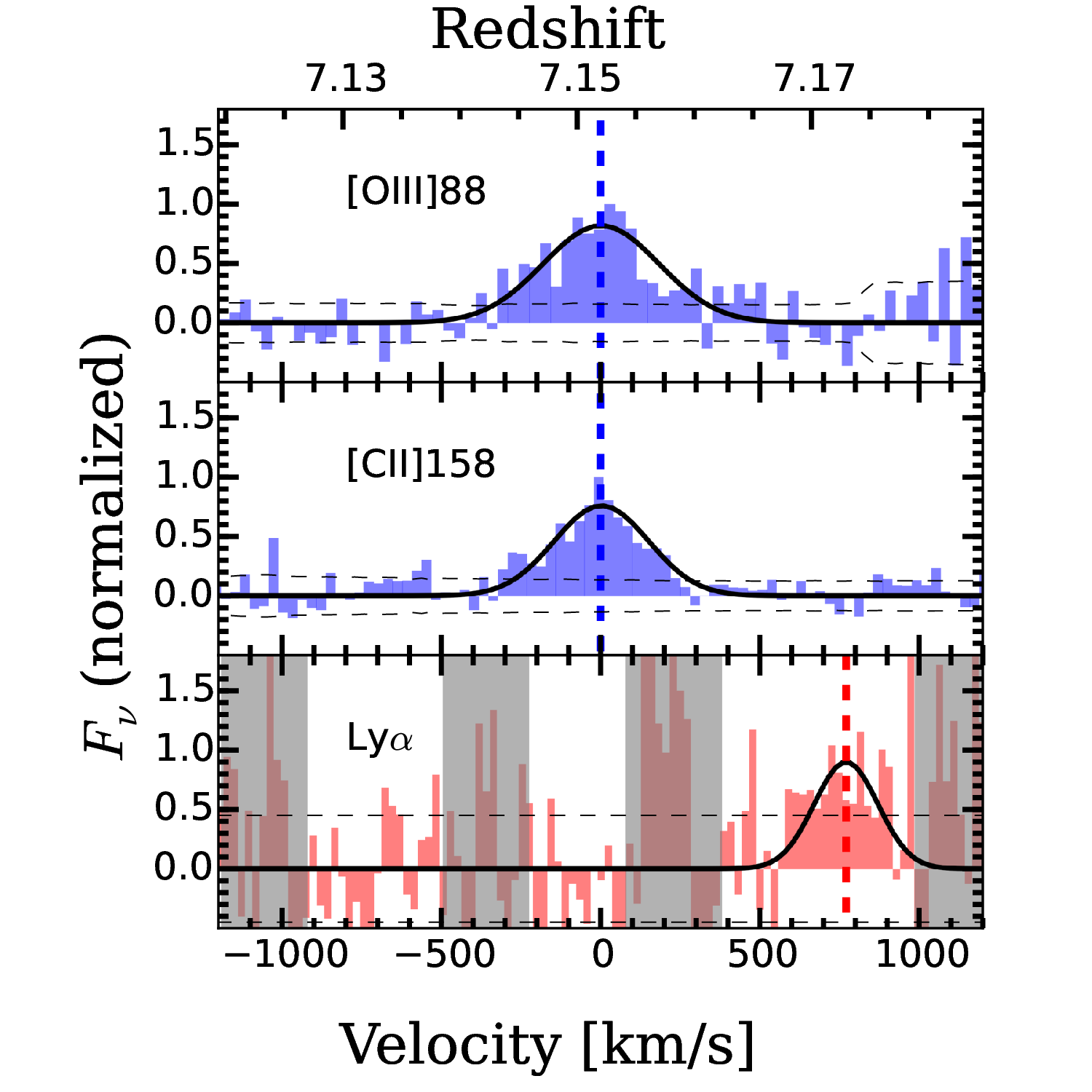}
\caption
{
Top, middle, and bottom panel show ALMA \oiii\ 88 $\mu$m, ALMA \cii\ 158 $\mu$m, 
and Subaru/FOCAS \lya\ spectra in velocity space 
with a resolution of $\sim 30$ km s$^{-1}$,  $33$ km s$^{-1}$, and $25$ km s$^{-1}$, respectively.
The velocity zero point corresponds to the systemic redshift $z = 7.1520$ (blue dashed line) 
and the \lya\ offset is $\simeq 770$ km s$^{-1}$ (red dashed line). 
Grey rectangles show regions contaminated by night sky emission. 
The black \textcolor{black}{dashed} lines indicate the RMS noise level for the velocity resolutions, 
and the black curves are Gaussian fit to the lines. 
}
\label{fig:vel_offset1}
\end{figure}

We measure the velocity offset of the \lya\ line calculated as 
\begin{equation}
\Delta v_{\rm Ly\alpha} = c \times \frac{z_{\rm Ly\alpha} - z_{\rm sys}}{1+z_{\rm sys}}, 
\end{equation}
where $c$ is the speed of light.
We have remeasured \zlya\ in the spectrum of \cite{furusawa2016} \textcolor{black}{by reading the peak wavelength of a Gaussian fit to the line}, taking into account air refraction and the motion of the observatory. With the vacuum rest-frame wavelength of $1215.67$ \AA, we have obtained \zlya $=7.1730\pm0.0012$ in the Solar system barycentric frame (Table \ref{tab:results}). Thus, we obtain \dv\ $= 772\pm45$ km s$^{-1}$ (Figure \ref{fig:vel_offset1}). 
\textcolor{black}{We note that the quality of our FOCAS spectrum is insufficient  to determine the exact spatial position of \lya. Thus, we add a systematic uncertainty of $\pm 100$ km s$^{-1}$ to reflect the fact that \name\ is comprised of two clumps kinematically separated by $\approx 200$ km s$^{-1}$ (Figure \ref{fig:velocity_map}). Hereafter we adopt \dv\ $= 772\pm45\pm100$ km s$^{-1}$.}

We compare the \dv\ value of \name\ with those in the literature. The \lya\ velocity offsets are investigated in hundreds of galaxies at $z\approx2-3$ where both \lya\ and H$\alpha$ or \oiii\ 5007 \AA\ are available (e.g., \citealt{steidel2010, hashimoto2013, hashimoto2015, erb2014, shibuya2014b}). At $z\approx2-3$, galaxies have velocity offsets ranging from 100 to 1000 km s$^{-1}$ with a mean value of $200-400$ km s$^{-1}$ (e.g., \citealt{erb2014, trainor2015, nakajima2018}). 
At $z>6$,  there are 16 galaxies whose \dv\ values are measured. The systemic redshifts of these galaxies are based on either \cii\ 158 \micron\ and \oiii\ 88 \micron\ (\citealt{willott2015, knudsen2016, inoue2016, pentericci2016, bradac2017, carniani2017a, carniani2018a, laporte2017, matthee2017}) or rest-frame UV emission lines such as CIII]1909 \AA\ and OIII]1666 \AA\ (\citealt{stark2015a, stark2017, mainali2017, verhamme2018}). At $z\approx6-8$, velocity offsets of $100-500$ km s$^{-1}$ are reported. We summarize these literature sample at $z\approx6-8$ in Table \ref{tab:compilation}. 
Compared with these literature values, the \dv\ value of \name\ is the largest at $z\approx6-8$, and even larger than the typical \dv\ value at $z\approx2-3$. 

\textcolor{black}{From the point of view of \lya\ radiative transfer in expanding shell models, there are two possible interpretations of a large \dv\ value. The first interpretation is that a galaxy has a large neutral hydrogen column density, \nhi, in the ISM. In the case of large \nhi, the number of resonant scattering of \lya\ photons becomes large, which in turn increases the \dv\ value (e.g., \citealt{verhamme2015}). In this case, \ew\ becomes small because \lya\ photons suffer from more dust attenuation due to a larger optical path length. Such a trend is confirmed at $z\approx2-3$, where larger \dv\ values are found in galaxies with brighter \muv\ (hence larger \nhi; \citealt{garel2012}) and smaller \ew\ (e.g., \citealt{hashimoto2013, shibuya2014b, erb2014}). However, it is unclear if such a trend is also true at $z>6$. 
The second interpretation is that a galaxy has a large outflow velocity, $ v_{\rm out}$. Because \lya\ photons scattered backward of receding gas from the observer preferentially escape from the galaxy, the \lya\ velocity offset is positively correlate with the outflow velocity as  \dv\ $\sim 2\times v_{\rm out}$ (e.g., \citealt{verhamme2006}).}

\begin{figure*}[t]
\hspace{-2cm}
\includegraphics[width=22cm]{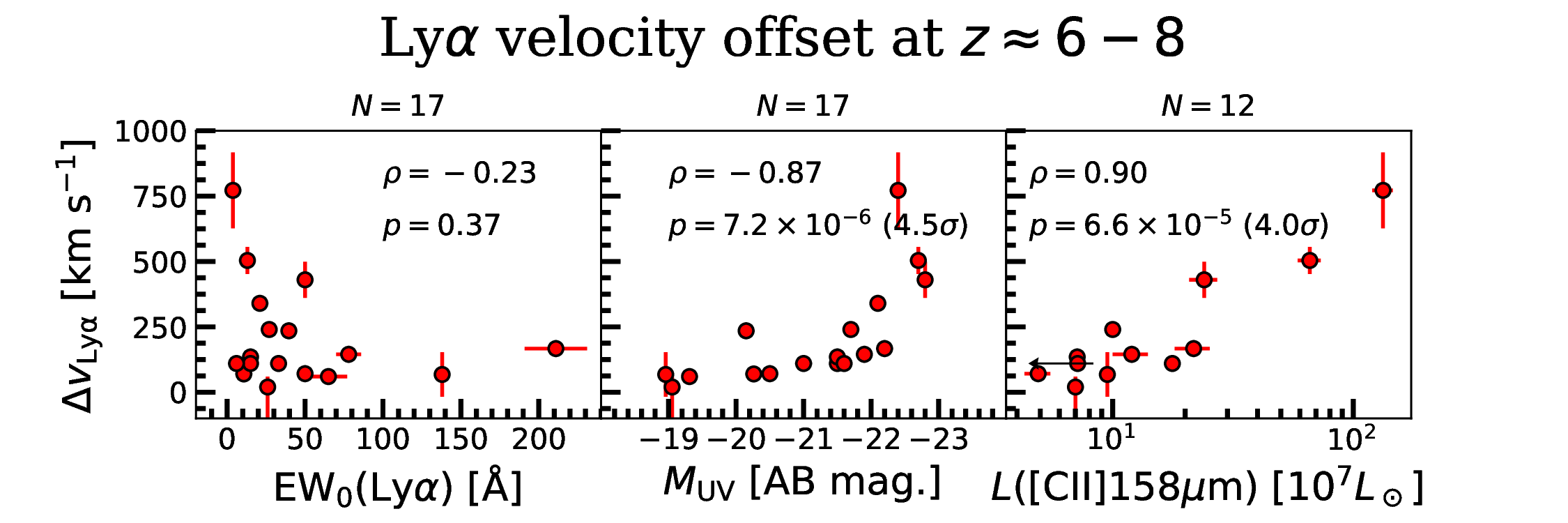}
\caption
{
Compilation of \lya\ velocity offsets at $z\approx6-8$ from this study and the literature (Table \ref{tab:compilation}). 
The \dv\ value is plotted against \ew\ (left panel), \muv\ (middle panel), and the \cii\ 158 \micron\ luminosity (right panel).
Error bars for the literature sample are shown if available. 
In each panel, $N$  shows the number of individual data points. 
In the middle and right panels, the values of \muv\ and \cii\ luminosity are corrected for magnification factors. 
In each panel, $\rho$ indicates the Spearman rank correlation coefficient for the relation, 
and $p$ denotes the probability satisfying the null hypothesis. 
}
\label{fig:vel_offset2}
\end{figure*}

\textcolor{black}{To investigate the two scenarios,} we explore correlations between \dv, \ew, and \muv\ at $z\approx6-8$ based on 17 galaxies in Table \ref{tab:compilation}. In addition, we also investigate the relation between \dv\ and \cii\ luminosities for the first time. Figure \ref{fig:vel_offset2} shows \dv\ values plotted against \ew, \muv, and \cii\ luminosities. To evaluate the significance of the relation, we perform Spearman rank correlation tests. 

\begin{figure}[t]
\hspace{-1cm}
\includegraphics[width=10cm]{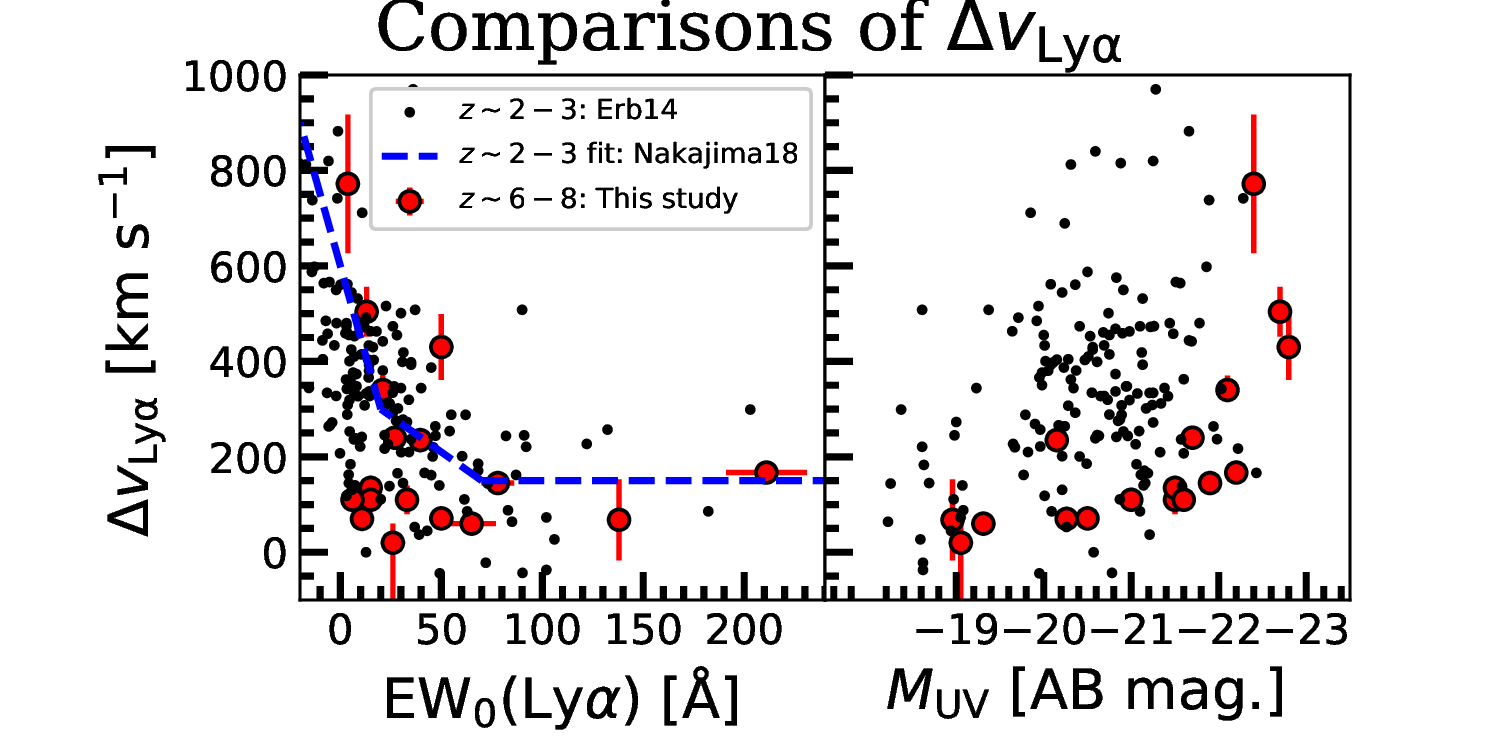}
\caption
{
\textcolor{black}{Comparisons of \dv\ at $z\approx6-8$ and at $z\approx 2-3$ 
as a function of \ew\ and \muv. In each panel, the red circles show the data points at $z\approx6-8$ whereas 
black small dots show data points at $z\approx 2-3$ taken from \cite{erb2014}. 
In the left panel, the blue dashed line indicates the average relation at $z\approx2-3$ 
presented in \cite{nakajima2018}.}
}
\label{fig:vel_offset3}
\end{figure}

In the left panel of Figure \ref{fig:vel_offset2}, the correlation is weak \textcolor{black}{with a Spearman rank correlation coefficient of $p = 0.37$}. \textcolor{black}{In the left panel of Figure \ref{fig:vel_offset3}, we compare our \dv\ values at $z\approx6-8$ with those at $z\approx2-3$ (\citealt{erb2014, nakajima2018}). At $z\approx2-3$, \cite{erb2014} have reported a $7\sigma$ anti-correlation}. It is possible that the correlation at $z\approx6-8$ is diluted because both \dv\ and \ew\ values are affected by the IGM attenuation effect. A large number of objects with \dv\ measurements, \textcolor{black}{particularly those with large \ew\ values,} are needed to conclude if the correlation exists or not at $z\gtrsim6$. Nevertheless, we note that the scatter of \dv\ value becomes larger for smaller \ew\ galaxies.  Such a trend is consistent with results at $z\approx2-3$, as shown in the left panel of Figure \ref{fig:vel_offset3}\footnote{Theoretically, the trend is explained as a secondary effect of \lya\ radiative transfer caused by the viewing angle of galaxy disks (\citealt{zheng2014}).}.  
In the middle panel of Figure \ref{fig:vel_offset2}, we confirm a $4.5 \sigma$ correlation between \muv\ and \dv, indicating that brighter \muv\ objects have larger \dv. Although the trend is consistent with that at $z\approx2-3$ \textcolor{black}{(right panel of Figure \ref{fig:vel_offset3})}, we have identified the trend at $z\approx6-8$ for the first time. 
In the right panel of Figure \ref{fig:vel_offset2}, we identify a positive correlation at the significance level of $4.0\sigma$, indicating that galaxies with higher \cii\ luminosities have larger \dv\ values. 

\textcolor{black}{The correlations in the middle and right panels of Figure \ref{fig:vel_offset2} support the two aforementioned scenarios for the large \dv\ in \name. In the \nhi\ scenario where we interpret \dv\ as \nhi, the middle panel of Figure \ref{fig:vel_offset2} indicates that UV brighter objects have larger \nhi. This is consistent with the results of semi-analytical models implementing the \lya\ radiative transfer calculations (\citealt{garel2012}; see their Figure 12). In this scenario, the right panel of Figure \ref{fig:vel_offset2} indicates that higher \cii\ luminosity objects have larger \nhi.  Such a trend is indeed confirmed in our Galaxy and some nearby galaxies (e.g., \citealt{bock1993, matsuhara1997}). 
}

\textcolor{black}{
In the outflow scenario where we translate \dv\ as the outflow velocity, given the correlation between the SFR and the \cii\ luminosity (e.g., \citealt{de.looze2014, matthee2017, carniani2018b, herrera-camus2018a}) or the UV luminosity, the middle and right panels of Figure \ref{fig:vel_offset2} show that larger SFR objects have stronger outflows. This is consistent with the observational results in the local Universe (e.g., \citealt{martin2005, weiner2009, sugahara2017}). 
}

\textcolor{black}{
In summary, the large \dv\ value in \name\ can be explained as a result of large \nhi\ and/or strong outflow, if the expanding shell models are applicable to \name. To break the degeneracy among the two parameters, it is useful to directly measure the outflow velocity from the blue-shift of the UV metal absorption lines with respect to the systemic redshift (e.g., \citealt{steidel2010, shibuya2014b}), because it is difficult to directly measure the \nhi\ value from observations at high-$z$. 
We note, however, that simplified shell models may not be appropriate for \name\ because of the presence of a merger in \name. It is possible that the turbulent motion due to the merger facilitates the \lya\ escape in spite of the large dust content in \name\ (e.g., \citealt{herenz2015}). Clearly, future spatially-resolved \lya\ data are crucial to understand the exact origin of the large \dv\ value in \name. 
}

\begin{table*}
\tbl{\dv\ literature sample\label{tab:compilation}}
{
\begin{tabular}{llccccccl}
\hline
Name & Lines & \zsys & \dv & \muv & \ew & $L$(\cii) & $\mu$ & Ref. \\ 
& &  & (km s$^{-1}$) & (AB mag.) & (\AA) & ($10^{7}$\lsun) &  &  \\ 
(1) & (2) & (3) & (4) & (5) & (6) & (7) & (8) & (9)  \\ 
\hline 
A2744$\_$YD4 & \oiii & 8.38  & 70 & $-20.9$ + 2.5 log($\mu$) & $10.7\pm2.7$ &  NA$^{a}$ & $1.8\pm0.3$ & L17\\ 
EGS-zs8-1& \ciii1907, 1909 & 7.72  & $340^{+15}_{-30}$ & $-22.1$ & $21\pm4$ & NA$^{a}$ & - & St17 \\ 
SXDF-NB1006-2 & \oiii & 7.21  & $110\pm30$ & $-21.5$ & 33.0 & $ 8.3 < (3\sigma)$ & - & Sh12, I16 \\ 
B14-65666 & \cii, \oiii & 7.15 & \textcolor{black}{$772\pm45\pm100$} & \textcolor{black}{$-22.4$} & $3.7^{+1.7}_{-1.1}$ & $133\pm13$ & - & B14, F16, This Study\\ 
COSMOS13679 & \cii & 7.14  & $135$ & $-21.5$ & 15 & 7.12 & - & P16 \\
BDF-3299 (clump I) & \cii & $7.11$  & $71$ & \textcolor{black}{$-20.5$} & 50 & $4.9\pm0.6$  & - & V11, Maio15, Ca17a \\ 
A1703-zd6 & OIII]1666 &  $7.04$  & $60$  & $-21.1$ + 2.5 log($\mu$) & $65\pm12$  & NA$^{a}$  & $5.2$ & Sc12, St15b  \\ 
RXJ1347.1-1145 & \cii& 6.77  & $20^{+140}_{-40}$ & $-20.8$ + 2.5 log($\mu$) $^{b}$& $26\pm4$ & $7.0^{+1.0}_{-1.5}/\mu$ $^{b}$ & $5.0\pm0.3$ & B17\\
NTTDF6345 & \cii & 6.70  & $110$ & $-21.6$ & 15 & 17.7 & - & P16 \\
UDS16291 & \cii & 6.64  & $110$ & $-21.0$ & 6 & 7.15 & - & P16 \\
COSMOS24108 & \cii & 6.62  & $240$ & $-21.7$ & 27 & 10.0 & - & P16 \\
CR7 (full) & \cii & 6.60  & $167\pm22$ & $-22.2$ & $211\pm20$ & $21.7\pm3.6 ^{c}$  & - & S15, Mat17 \\
Himiko (Total) & \cii & 6.59  & $145\pm15$ & $-21.9$ & $78^{+8}_{-6}$ & $12\pm2$  & - & O13, Ca18 \\ 
CLM1 & \cii & 6.16 & $430\pm69$ & $-22.8$ & 50 &  $24\pm3.2$ & - &  Cu03, W15 \\ 
RXJ2248-ID3 & OIII]1666, CIV & $6.11$ & $235$ & $-22.0$ + 2.5 log($\mu$) $^{d}$ & $39.6\pm5.1$ & NA$^{a}$ & 5.5 &  Main17 \\ 
WMH5 & \cii & 6.07  & $504\pm52$ & $-22.7$ & 13 &  $66\pm7.2$ & - &  W13, W15 \\ 
A383-5.1 & \cii & 6.03 & $68\pm85$ & $-21.6$ + 2.5 log($\mu$) $^{e}$ & 138 & $9.5/\mu$ & 11.4 &  R11, K16 \\ 
\hline
\hline
\end{tabular}
}
\tabnote{Note. Properties of the compiled sample with \dv\ measurements at $z>6$ from the literature and this study. Error values are presented if available. \\
(1) The object name;  (2) the emission line(s) used to measure the systemic redshift, $z_{\rm sys}$; (3) the systemic redshift; (4) the \lya\ velocity offset with respect to the systemic redshift; (5) the UV absolute magnitude in the AB magnitude system; (6) the rest-frame \lya\ equivalent width; (7) the \cii\ luminosity; (8) the lensing magnification factor; (9) references. \\
$^{a}$ `NA' indicates that the \cii\ luminosity is not available. \\
$^{b}$ Values before corrected for magnification are inferred from B17 under the assumption of $\mu=5.0$. \\
$^{c}$ Aperture luminosity in M17 is adopted (see Table 1 of Mat17). \\
$^{d}$ Values before corrected for magnification are inferred from Main17 under the assumption of $\mu=5.5$. \\
$^{e}$ Inferred from $Y=26.15$ \\ 
Reference Cu03: \cite{cuby2003},  R11: \cite{richard2011}, V11: \cite{vanzella2011}, O12: \cite{ono2012}, Sc12: \cite{schenker2012}, Sh12: \cite{shibuya2012}, W13: \cite{willott2013}, Maio15: \cite{maiolino2015}, W15: \cite{willott2015}, So15: \cite{sobral2015}, St15a: \cite{stark2015a}, St15b: \cite{stark2015b}, K16: \cite{knudsen2016}, P16: \cite{pentericci2016}, I16: \cite{inoue2016}, St17: \cite{stark2017} B17: \cite{bradac2017}, Ca17a: \cite{carniani2017a}, Ca18: \cite{carniani2018a}, L17: \cite{laporte2017}, Mat17: \cite{matthee2017}, Main17: \cite{mainali2017}. 
}
\end{table*}

\section{Discussion}
\label{sec:discussion}

\subsection{A consistent picture of \name}
\label{subsec:name}

\name\ is the first \textcolor{black}{star-forming galaxy} with a complete set of \cii, \oiii, and dust continuum emission 
in the reionization epoch. In conjunction with the HST F140W data (\citealt{bowler2017}) and the \lya\ line (\citealt{furusawa2016}), the rich data allow us to discuss the properties of \name\ in detail.  

\textcolor{black}{
In \S \ref{subsec:multi_comp}, we have inferred that \name\ is a merger.  This is based on the fact that (i) the morphology of \name\ shows the two clumps in UV, \cii, and \oiii\ whose positions are consistent with each other (Figure \ref{fig:contours}), (ii) the spectra of \cii\ and \oiii\ can be decomposed into two Gaussians kinematically separated by $\approx200$ km s$^{-1}$ (Figures \ref{fig:1d_spec} and \ref{fig:velocity_map}), and (iii) the velocity field is not smooth as expected in a rotational disk.}
\textcolor{black}{In the same direction, }\cite{jones2017a} have concluded that a galaxy at $z=6.07$, WMH5, would be merger rather than a rotational disk based on two separated \cii\ clumps, the \cii\ velocity gradient, and the \cii\ spectral line composed of multiple Gaussian profiles.

\textcolor{black}{In \name, we note that the clumps A and B have UV, IR, and line luminosities that are consistent within a factor of two (Table \ref{tab:results}), implying that \name\ would represent a major-merger at $z=7.15$. In addition, even the individual clumps have very high luminosities among $z>6$ star-forming galaxies. This suggests that \name\ traces a highly dense region at the early Universe. Although our current data do not show companion objects around \name\ (e.g., \citealt{decarli2017}), future deeper ALMA data could reveal companion galaxies around \name. 
}

A merger event would enhance the star forming activity. 
Based on \textcolor{black}{the results of our SED fitting (Table \ref{tab:sed})}, we calculate the specific SFR, defined as the SFR per unit stellar mass (sSFR $\equiv$ SFR/$M_{\rm *}$). 
The sSFR of \textcolor{black}{$260^{+119}_{-57}$ Gyr$^{-1}$} is larger than 
those for galaxies on the star formation main sequence at $z\approx6-7$ 
(e.g., \citealt{stark2013, speagle2014, santini2017}). 
This suggests that  \name\ is indeed undergoing bursty star-formation (\citealt{rodighiero2011}). 

\textcolor{black}{
Interestingly, the high sSFR value of \name\ is also consistent with its relatively high luminosity-weighted dust temperature, \td\ $\approx50 - 60$ K,  under the assumption of $\beta_{\rm d} = 2.0 - 1.5$ (\S \ref{sec:dust}). Indeed, \cite{faisst2017} have shown that objects with larger sSFR have higher \td\ (see Figure 5 of \citealt{faisst2017}) based on the compiled sample of local galaxies that include dwarf metal-poor galaxies, metal rich galaxies, and (ultra-)luminous infrared galaxies. Probably, a strong UV radiation field driven by intense star-formation activity in \name\ leads to high \td\ as a result of effective dust heating (e.g., \citealt{inoue.kamaya2004}). 
This hypothesis can also explain the high \oiii-to-\cii\ luminosity ratio in \name. The strong UV radiation can efficiently ionize \oiii\ (with ionization potential of 35.1 eV)  against \cii\ (ionization potential of 11.3 eV). Indeed, in the local Universe, there is a positive correlation between the \oiii-to-\cii\ luminosity ratio and the dust temperature (see e.g., Figure 11 in \citealt{herrera-camus2018a}). 
}

\subsection{\lya\ velocity offsets at $z\approx6-8$ and implications for reionization: Enhanced \lya\ visibility for bright galaxies}
\label{subsec:discussion_dv}

In this section, we discuss implications on reionization from the compiled \dv\ measurements 
at $z\approx6-8$. 
The \lya\ velocity offset at $z\approx6-8$ is useful to constrain the reionization process as described below. Based on spectroscopic observations of LAEs and LBGs, previous studies have shown that the fraction of galaxies with strong \lya\ emission increases from $z=2$ to $6$ (e.g., \citealt{cassata2015}), but suddenly drops at $z>6$ (e.g., \citealt{stark2010, pentericci2011, ono2012, schenker2012, schenker2014}). 
This is often interpreted as a rapid increase of the neutral gas in the IGM at $z\approx6$, 
significantly reducing the visibility of \lya. 
\cite{ono2012} have revealed that the amplitude of the drop 
is smaller for UV bright galaxies than for UV faint galaxies. 
More recently, \cite{stark2017} have demonstrated a striking LAE fraction of 100\%\ 
in the sample of most luminous LBGs (\citealt{oesch2015, zitrin2015, roberts-borsani2016}). 
\cite{stark2017} have discussed possible origins of the enhanced \lya\ visibility of these UV luminous galaxies, one of which is that their large \lya\ velocity offsets make \lya\ photons less affected by the IGM attenuation when \lya\ photons enter the IGM.

In the middle and right panels of Figure \ref{fig:vel_offset2}, 
we have statistically demonstrated that 
the \dv\ value becomes larger for galaxies with brighter UV or \cii\ luminosities at $z\approx6-8$. 
This means that the \lya\ visibility is indeed enhanced in brighter galaxies 
(see also \citealt{mainali2017, mason2018}), which would give us a reasonable explanation 
on the high \lya\ fraction in luminous galaxies.

\section{Conclusion}
\label{sec:conclusion}

We have conducted high spatial resolution ALMA observations of an LBG at $z=7.15$. 
Our target, \name, has a bright UV absolute magnitude, $M_{\rm UV} \approx$ \textcolor{black}{-22.4}, 
and has been spectroscopically identified in \lya\ with a small rest-frame 
equivalent width of $\approx4$ \AA. 
\textcolor{black}{Previous {\it HST} image has shown that the target is comprised of two spatially separated clumps in the rest-frame UV, referred to as the clump A (Northern East) and  clump B (Southern West) in this study.}
Based on our ALMA Band 6 and Band 8 observations, 
we have \textcolor{black}{newly} detected spatially resolved \cii\ 158 \micron, \oiii\ 88 \micron, 
and dust continuum emission \textcolor{black}{in the two Bands.}
\name\ is the first object with a complete set of \lya, \oiii, \cii, and dust continuum emission 
which offers us a unique opportunity to investigate detailed kinematical and ISM properties 
of high-$z$ galaxies. 
Our main results are as follows: 
\begin{itemize}
\item \textcolor{black}{
Owing to our high spatial resolution observations, the \cii\ and \oiii\ emission can be spatially decomposed into two clumps whose positions are consistent with those of the two UV clumps revealed by {\it HST} (Figure \ref{fig:contours}). The \cii\ and \oiii\ line spectra extracted at the positions of these clumps also show that the lines are composed of two Gaussian profiles kinematically separated by $\approx200$ km s$^{-1}$ (Figures \ref{fig:1d_spec} and \ref{fig:velocity_map}). These results suggest that \name\ is a merger. }
\item \textcolor{black}{The whole system, clump A, and clump B have the \oiii\ luminosity of $(34.4\pm4.1) \times 10^{8}$, $(21.2\pm3.2) \times 10^{8}$, and  $(13.0\pm2.1) \times 10^{8}$ \lsun, respectively, and the \cii\ luminosity of $(11.0\pm1.4) \times 10^{8}$, $(6.0\pm1.9) \times 10^{8}$, and  $(4.9\pm0.8) \times 10^{8}$ \lsun, respectively. The total line luminosities are the highest so far detected among $z>6$ star-forming galaxies. Even the individual clumps have very high line luminosities. The \oiii-to-\cii\ luminosity ratio is $3.1\pm1.6$, $3.5\pm0.8$, and $2.7\pm0.6$ for the whole system, clump A, and clump B, respectively  (\S \ref{sec:lines}; Table \ref{tab:results}).}
\item \textcolor{black}{In the whole system of \name, the dust continuum flux densities at 90 and 163 \micron\ 
are $S_{\nu, 90\mu m}  = 470\pm128$ and $S_{\nu, 163\mu m} = 130\pm25$ $\mu$Jy, respectively. 
Based on the continuum ratio $S_{\nu, 90\mu m}$/$S_{\nu, 163\mu m}$ 
and assuming the emissivity index in the range of $\beta_{\rm d}=2.0 - 1.5$, 
we have estimated the dust temperature to be \td\ $\approx 50 - 60$K. 
Assuming these \td\ and $\beta_{\rm d}$ values, we have obtained \ltir\  $\approx 1 \times 10^{12}$, 
$3 \times 10^{11}$, and $7 \times 10^{11}$ \lsun\ for the whole system, clump A, and clump B, respectively, 
by integrating the modified black-body radiation over $8-1000$ \micron.
With a typical dust mass absorption coefficient, the dust mass is estimated to be $\approx 1 \times 10^{7}$ , 
$3 \times 10^{6}$, and $6 \times 10^{6}$ \msun\ for the whole system, clump A, and clump B, respectively 
(\S \ref{sec:dust}; Figure \ref{fig:dust}). 
}
\item 
We have investigated the IRX-$\beta$ relation at $z\approx6.5-9.1$ 
based on \textcolor{black}{eleven} spectroscopically identified objects including  
\textcolor{black}{five} LBGs with dust continuum detections. 
For fair comparisons of data points, 
we have uniformly computed the $\beta$ and IRX values for the entire sample. 
We find that the \textcolor{black}{five} LBGs with dust detections are well characterized by 
the Calzetti's dust attenuation curve (\S \ref{subsec:irx}; Figure \ref{fig:irx_beta}). 
\item We have created a combined sample of \textcolor{black}{six} galaxies with \oiii-to-\cii\ luminosity ratios: 
Our object, two literature LAEs, and \textcolor{black}{three literature SMGs}. 
We have found that the luminosity ratio becomes larger for objects with lower 
bolometric luminosities defined as the sum of the UV and IR luminosities. 
The results indicate that galaxies with lower bolometric  luminosities \textcolor{black}{(i.e., lower masses)} have 
either lower metallicities or higher ionization states 
(\S \ref{subsec:line_ratio}; Figure \ref{fig:lbol_ratio}). 
\item 
To estimate the stellar mass, SFR, and the stellar age, 
we have performed SED fitting \textcolor{black}{for the whole system of \name} taking ALMA data into account. 
We have obtained the total stellar mass of $\approx 7.7\times10^{8}$\msun\ 
and the total SFR of $\approx 200$ \msun\ yr$^{-1}$. 
\textcolor{black}{The specific SFR (defined as the SFR per unit stellar mass) is 260 Gyr$^{-1}$, indicating} that \name\ is a \textcolor{black}{starburst galaxy} (\S \ref{sec:sed_fit}; Figure \ref{fig:sed}). 
\item \textcolor{black}{In the whole system of \name,} the \cii\ and \oiii\ lines have consistent redshifts of $z=7.1520\pm0.0003$. 
On the other hand, \lya\ is significantly redshifted with respect to the ALMA lines 
by \dv\ $= 772\pm45\pm100$ km s$^{-1}$, 
which is the largest so far detected among the $z>6$ galaxy population. 
The very large \dv\ would be due to 
\textcolor{black}{
the presence of large amount of neutral gas or large outflow velocity} (\S \ref{sec:lya}; Figure \ref{fig:vel_offset1}). 
\item Based on a compiled sample of 17 galaxies at $z\approx6-8$ with 
\dv\ measurements from this study and the literature, 
we have found a $ 4.5\sigma$ ($4.0\sigma$) correlation between \dv\ and UV magnitudes (\cii\ luminosities) 
in the sense that \dv\ becomes larger for brighter UV magnitudes and brighter \cii\ luminosities. 
These results are in a good agreement with a scenario that 
the \lya\ emissivity during the reionization epoch depends on the galaxy's luminosity 
(\S \ref{subsec:discussion_dv}; Figure \ref{fig:vel_offset2}). 

\end{itemize}

Given the rich data available and spatially extended nature, 
\name\ is one of the best target for follow-up observations 
with ALMA and {\it James Webb Space Telescope}'s NIRSpec IFU mode 
to spatially resolve e.g., gas-phase metallicity, the electron density, and Balmer decrement.

\section*{Acknowledgments}

This paper makes use of the following ALMA data: \textcolor{black}{ADS/JAO.ALMA\#2015.1.00540.S}, ADS/JAO.ALMA\#2016.1.00954.S, and \textcolor{black}{ADS/JAO.ALMA\#2017.1.00190.S.} ALMA is a partnership of ESO (representing its member states), NSF (USA) and NINS (Japan), together with NRC (Canada), NSC and ASIAA (Taiwan), and KASI (Republic of Korea), in cooperation with the Republic of Chile. The Joint ALMA Observatory is operated by ESO, AUI/NRAO and NAOJ.
This work is based in part on data collected at Subaru Telescope, which is operated by the National Astronomical Observatory of Japan. Data analysis were in part carried out on common use data analysis computer system at the Astronomy Data Center, ADC, of the National Astronomical Observatory of Japan. 

T.H. and A.K.I. appreciate support from NAOJ ALMA Scientific Research Grant Number 2016-01A. We are also grateful to KAKENHI grants 26287034 and 17H01114 (K.M. and A.K.I.), 17H06130 (Y.Tamura, K. Kohno), \textcolor{black}{18H04333} (T.O.), 16H02166 (Y.Taniguchi), 17K14252 (H.U.), JP17H01111 (I.S.), 16J03329 (Y.H.), and 15H02064 (M.O.).  
E.Z. acknowledges funding from the Swedish National Space Board. K.O. acknowledges the Kavli Institute Fellowship at the Kavli Institute for Cosmology at the University of Cambridge, supported by the Kavli Foundation. K.Knudsen acknowledges support from the Knut and Alice Wallenberg Foundation. 

We acknowledge Nicolas Laporte, Stefano Carniani, Dan Marrone, and Dawn Erb for providing us with their data. 
\textcolor{black}{We thank Kouichiro Nakanishi, Daisuke Iono, and Bunyo Hatsukade for discussions in ALMA astrometry.}  
We appreciate Fumi Egusa, Kazuya Saigo and Seiji Fujimoto for discussions in handling with ALMA data, and Alcione Mora for help in the GAIA archive data. 
We are grateful to Rebecca A. A. Bowler, \textcolor{black}{Charlotte Mason, Haruka Kusakabe}, Miju Lee, Kenichi Tadaki, \textcolor{black}{Masayuki Umemura}, Hidenobu Yajima, Kazuhiro Shimasaku, Shohei Aoyama, Toru Nagao, Masaru Kajisawa, Kyoko Onishi, Takuji Yamashita, Satoshi Yamanaka, Andrea Ferrara, Tanya Urrutia, Sangeeta Malhotra, and Dan Stark  for helpful discussions. 

\appendix
\label{sec:appendix}

\section{Astrometry of the HST F140W image}
\label{sec:astrometry_hst}

In this study, we compare the spatial position of \name\ in ALMA data with that in the HST F140W band image data (PI: R.A.A. Bowler). Several studies have reported spatial offsets between ALMA-detected objects and their HST-counterparts either due to astrometry uncertainties or physical offsets in different wavelengths (e.g., \citealt{laporte2017, gonzalez_lopez2017, carniani2018b, dunlop2017}).  

\begin{figure}[]
\includegraphics[width=8cm]{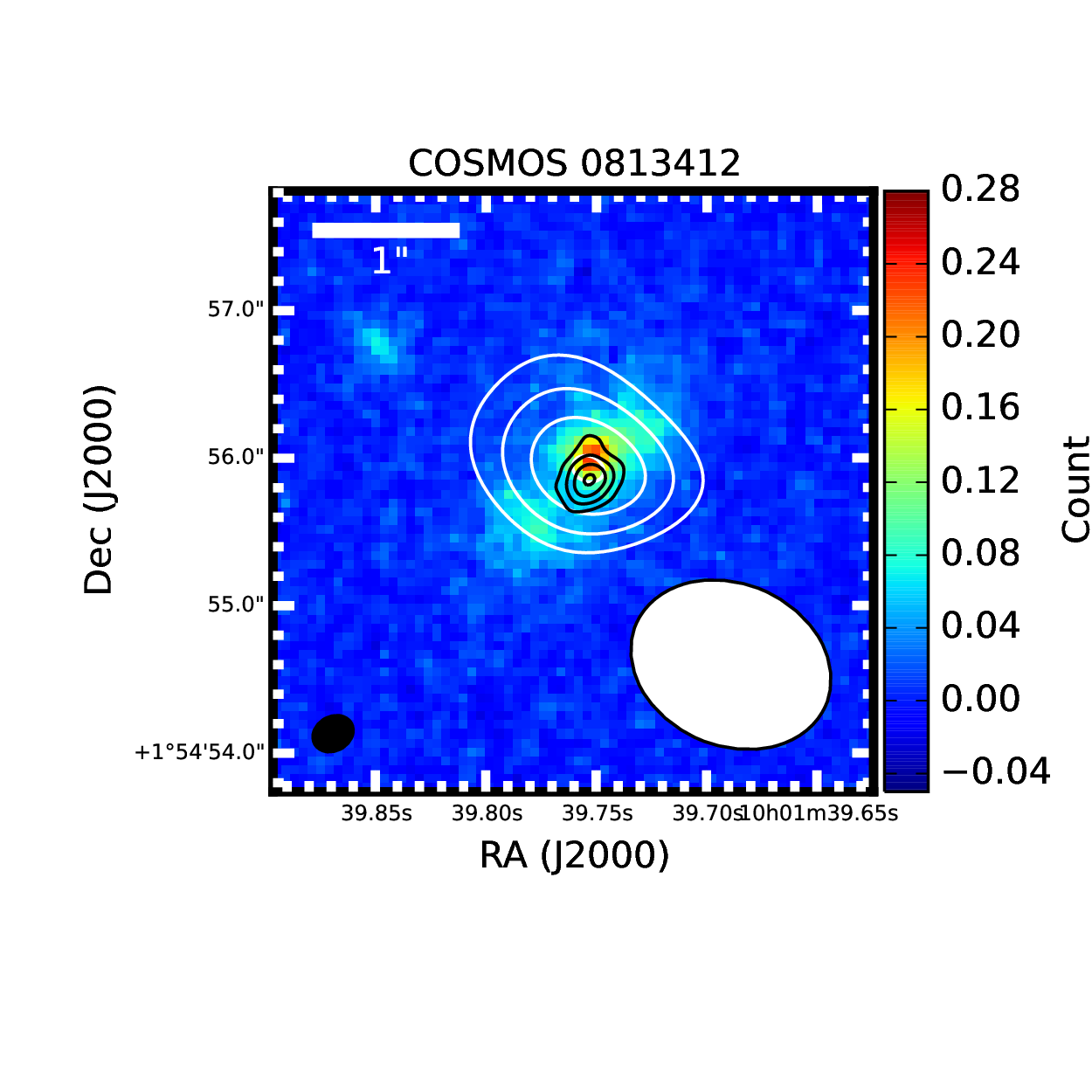}
\caption
{
Black and white contours show the continuum images of a serendipitously detected object at $z=1.93$, 
COSMOS 0813412 (\citealt{schinnerer2007}), in our ALMA Band 6 data and Bowler et al's Band 6 data, respectively, overlaid on its counterpart in the HST F140W image with re-calibrated astrometry. 
Black contours are drawn at ($4$, $8$, $12$, $16$)$\times \sigma$ where $\sigma = 9.5$ $\mu$Jy beam$^{-1}$ in our Band 6 data,  while white contours are drawn at ($4$, $6$, $8$, $10$)$\times \sigma$ where $\sigma = 27.8$ $\mu$Jy beam$^{-1}$ in Bowler et al's Band 6 data. 
The ellipses at the lower left and right corner indicate the synthesized beam sizes of our and Bowler et al's Band 6 data, respectively. 
}
\label{fig:appendix1}
\end{figure}

To better calibrate the archival F140W image, we first search for bright stars in the catalog {\tt gaiadr1.gaia${\_}$source} released in the framework of the GAIA project\footnote{https://gea.esac.esa.int/archive/} (\citealt{gaia2016}). Similar astrometry calibrations are performed in \cite{carniani2018b}. Because of the small sky area, $\approx 2'.3 \times 2'.0$, covered by the Bowler et al.'s archival HST data, only a single star is matched to the catalog. Thus, we instead use the catalog {\tt gaiadr1.sdssdr9${\_}$original$_{\_}$valid} from the same project. The latter catalog includes a larger number of objects while its astrometry is originally taken from the SDSS project.
Using 17 objects uniformly distributed in the field-of-view, we have performed {\tt IRAF} tasks {\tt ccmap} and {\tt ccsetwcs} to calibrate astrometry. The applied shift around \name\ is $\approx0''.23$ \textcolor{black}{($\Delta_{\rm R.A.} = 0''.14$ toward the West-East direction and $\Delta_{\rm Dec.} = 0''.18$ toward the North-South direction)} for the archival HST image. To check the accuracy of our astrometry, we make use of a serendipitous continuum detection (16$\sigma$) in Band 6 from a galaxy at $z=1.93$, COSMOS 0813412 (\citealt{schinnerer2007}). Figure \ref{fig:appendix1} shows that the centroids of ALMA continuum and its HST F140W counterpart are consistent with each other within $\approx0''.15$ uncertainties, demonstrating successful astrometry calibration. \textcolor{black}{We have also confirmed that the re-calibrated F140W image has astrometry consistent with that in \cite{bowler2018} (in private communication with R. A. A. Bowler).}   

\section{Astrometry of Three ALMA Datasets}
\label{sec:appendix_astrometry_alam}

\textcolor{black}{
We present detailed analyses on astrometry of three ALMA datasets: Our ALMA Band 6 and 8 data plus Bowler et al's Band 6 data. For this purpose, we have re-analyzed the Bowler et al's  Band 6 data using CASA ver 4.5.1. We obtain the beam size and the $1\sigma$ noise level well consistent with those presented in \cite{bowler2018}. We confirm their dust continuum detection, although the peak significance level is slightly lower, $4.6\sigma$, than the value reported in \cite{bowler2018}. 
}

\textcolor{black}{
As shown in Table \ref{tab:appendix}, the same phase and bandpass calibrators are used in the three datasets.  In addition, the same flux calibrator is used in our ALMA Band 6 and 8 data. Therefore, we firstly analyze the sky coordinates of three calibrators, J0948+0022, J1058+0133, and J0854+2006, to examine if there exists astrometry uncertainty arising from calibrators. We have confirmed that the centroids (i.e., flux peak positions) of these calibrators are consistent with each other to the order of $< 0''.01$, indicating that the coordinates of the three datasets are well aligned. 
}

\textcolor{black}{
Secondly, we compare the positions of COSMOS 0813412 in our and Bowler et al's Band 6 data. Figure \ref{fig:appendix1} shows the dust continuum contours of COSMOS 0813412 overlaid on the HST F140W image with re-calibrated astrometry. We obtain the centroids of (RA, Dec) = (10:01:39.753, +01.54.55.860) and (10:01:39.754, +01.54.55.920) in our and Bowler et al's Band 6 data, respectively. This corresponds to the spatial offset of $0''.06$ toward the North-South direction. To evaluate its significance, we calculate the positional uncertainties, $\Delta p$, with the equation\footnote{ https://help.almascience.org/index.php?/Knowledgebase/Article/View/319/0/what-is-the-astrometric-accuracy-of-alma}  
\begin{equation}
\Delta p \ {\rm [mas]} = 60 \times \frac{100 \ {\rm GHz}}{{\rm FREQ}} \times \frac{10 \ {\rm km}}{{\rm BSL}} \times \frac{1}{{\rm SNR}}, 
\end{equation}
where FREQ is the observing frequency in GHz, BSL is the maximum baseline length in km, and the SNR corresponds to the peak significance level. Based on FREQ $=$  225 GHz (233 GHz), BSL $=$ 2.65 km (0.33 km), and SNR $\approx$16 (10), $\Delta p$ $= 0''.01$ ($0''.08$) in our (Bowler et al's) Band 6 data. Thus, the spatial offset of $0''.06$ is within the $1\sigma$ positional uncertainties, confirming that the two ALMA Band 6 data have consistent astrometry. 
}

\begin{figure}[]
\includegraphics[width=7cm]{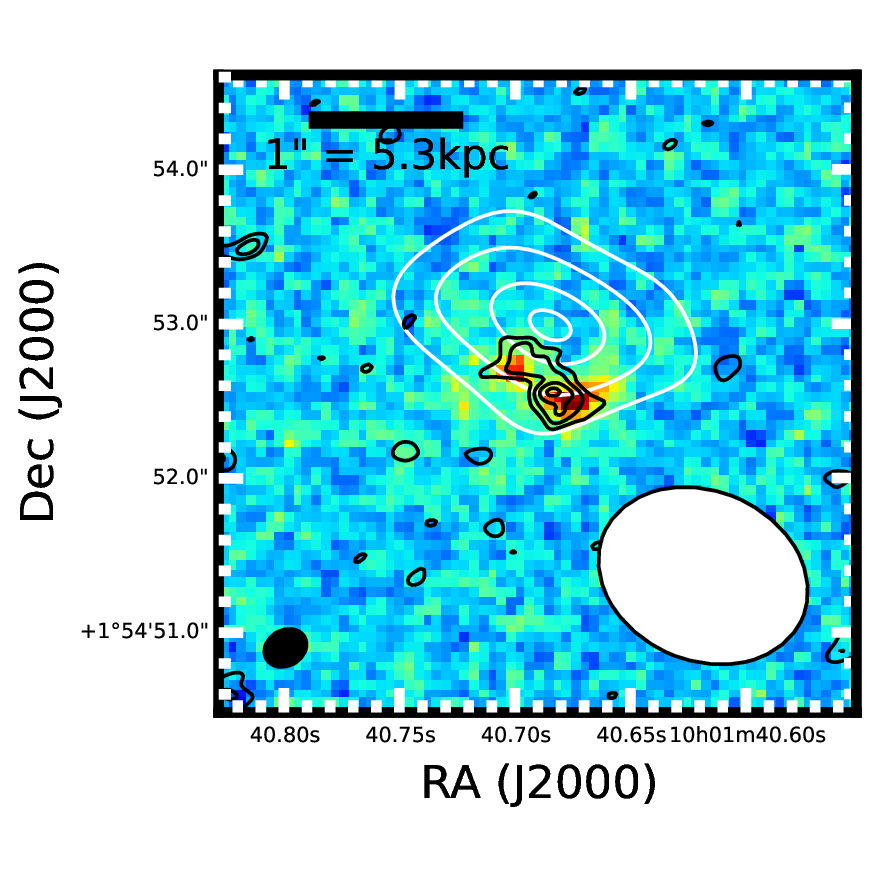}
\caption
{
Black and white contours show the continuum images of \name\ in our ALMA Band 6 data and Bowler et al's Band 6 data, respectively, overlaid on its counterpart in the HST F140W image with re-calibrated astrometry. 
Black contours are drawn at (2, 3, 4, 5)$\times \sigma$ where $\sigma = 9.5$ $\mu$Jy beam$^{-1}$ in our Band 6 data,  while white contours are drawn at (2, 3, 4, 4.5)$\times \sigma$ where $\sigma = 27.8$ $\mu$Jy beam$^{-1}$ in Bowler et al's Band 6 data. 
The ellipses at the lower left and right corner indicate the synthesized beam sizes of our and Bowler et al's Band 6 data, respectively. 

}
\label{fig:appendix2}
\end{figure}

\textcolor{black}{
Finally, we examine the spatial positions of \name\ using our and Bowler et al's Band 6 data. We do not attempt to compare the positions of \name\ between Band 6 and 8 data for astrometry analyses because they probe dust continuum emission at different wavelengths. Figure \ref{fig:appendix2} shows the two dust continuum contours overlaid on the $4''.0 \times 4''.0$ cutout image of F140W. The dust continuum centroids are obtained as (RA, Dec) =  (10:01:40.683, +01.54.52.56) and (10:01:40.686, +01.54.53.00) in our and Bowler et al's Band 6 data, respectively, corresponding to a possible spatial offset of $\Delta_{\rm tot.} = 0''.44$ predominantly in the North-South direction ($\Delta_{\rm R.A.} = 0''.05$, $\Delta_{\rm Dec.} = 0''.44$). 
As in COSMOS 0813412, we calculate the positional uncertainties to be $0''.02$ ($0''.17$) in our (Bowler et al's) Band 6 data based on the peak significance of $5.3\sigma$ ($4.6\sigma$). Therefore, the spatial offset corresponds to $2.3\sigma$ ($=0.44/(0.17+0.02)$), indicating that the offset is marginal (Table \ref{tab:appendix2}). 
}



\begin{table*}
\tbl{Calibrators for ALMA Observations. \label{tab:appendix}}
{
\begin{tabular}{cccc}
\hline
Data & phase calibrators & bandpass calibrators & flux calibrators\\ 
(1) & (2) & (3) & (4)   \\ 
\hline 
\bf{This Study} \\ 
\hline
Band 6 (Cycle 4) & J0948+0022 & J1058+0133 & J1058+0133, J0854+2006 \\ 
Band 8 (Cycle 4) & J0948+0022 & J1058+0133 & J1058+0133, J0854+2006 \\
Band 8 (Cycle 5) & J0948+0022, J1028-0236 & J1058+0133, J1229+02023 & J1058+0133, J1229+02023  \\
\hline 
\bf{\citealt{bowler2018}}  \\ 
\hline 
Band 6 (Cycle 4) & J0948+0022 & J1058+0133 & Ganymede \\
\hline
\end{tabular}
}
\end{table*}

\begin{table*}
\tbl{Spatial Offsets of \name\ among the Two ALMA Band 6 Data. \label{tab:appendix2}}
{
\begin{tabular}{ccccc}
\hline
Data 1 & Data 2 & Spatial Offsets & Uncertainty & Significance \\ 
& & $\Delta_{\rm tot.}$, $\Delta_{\rm R.A.}$, $\Delta_{\rm Dec.}$  & Data 1, Data 2 & \\
(1) & (2) & (3) & (4)  & (5) \\ 
\hline
Band 6 (This Study) & Band 6 (\citealt{bowler2018})  & $0''.44$, $0''.05$, $0''.44$ & $0''.02$, $0''.17$ & $2.3\sigma$ \\
\hline
\end{tabular}
}
\tabnote{Note. (1) and (2) Two data sets to be compared; (3) Spatial offset in the units of arcsec, where $\Delta_{\rm R.A.}$ and $\Delta_{\rm Dec.}$ correspond to the offset toward the direction of R.A. and Dec., respectively, and the total offset is $\Delta_{\rm tot.}$; (4) Positional uncertainties of Data 1 and Data 2 in the units of arcsec; and (5) Statistical significance of the spatial offset. 
}
\end{table*}

\begin{thebibliography}{148}
\expandafter\ifx\csname natexlab\endcsname\relax\def\natexlab#1{#1}\fi

\bibitem[{{Behrens} {et~al.}(2018){Behrens}, {Pallottini}, {Ferrara},
  {Gallerani}, \& {Vallini}}]{behrens2018}
{Behrens}, C., {Pallottini}, A., {Ferrara}, A., {Gallerani}, S., \& {Vallini},
  L. 2018, \mnras, 477, 552

\bibitem[{{Binney} \& {Tremaine}(2008)}]{binney.tremaine2008}
{Binney}, J., \& {Tremaine}, S. 2008, {Galactic Dynamics: Second Edition}
  (Princeton University Press)

\bibitem[{{Bock} {et~al.}(1993){Bock}, {Hristov}, {Kawada}, {Matsuhara},
  {Matsumoto}, {Matsuura}, {Mauskopf}, {Richards}, {Tanaka}, \&
  {Lange}}]{bock1993}
{Bock}, J.~J., {et~al.} 1993, \apjl, 410, L115

\bibitem[{{Bouwens} {et~al.}(2009){Bouwens}, {Illingworth}, {Franx}, {Chary},
  {Meurer}, {Conselice}, {Ford}, {Giavalisco}, \& {van Dokkum}}]{bouwens2009}
{Bouwens}, R.~J., {et~al.} 2009, \apj, 705, 936

\bibitem[{{Bouwens} {et~al.}(2010){Bouwens}, {Illingworth}, {Oesch}, {Trenti},
  {Stiavelli}, {Carollo}, {Franx}, {van Dokkum}, {Labb{\'e}}, \&
  {Magee}}]{bouwens2010}
{Bouwens}, R.~J., {et~al.} 2010, \apjl, 708, L69

\bibitem[{{Bouwens} {et~al.}(2014){Bouwens}, {Bradley}, {Zitrin}, {Coe},
  {Franx}, {Zheng}, {Smit}, {Host}, {Postman}, {Moustakas}, {Labb{\'e}},
  {Carrasco}, {Molino}, {Donahue}, {Kelson}, {Meneghetti}, {Ben{\'{\i}}tez},
  {Lemze}, {Umetsu}, {Broadhurst}, {Moustakas}, {Rosati}, {Jouvel},
  {Bartelmann}, {Ford}, {Graves}, {Grillo}, {Infante}, {Jimenez-Teja}, {Lahav},
  {Maoz}, {Medezinski}, {Melchior}, {Merten}, {Nonino}, {Ogaz}, \&
  {Seitz}}]{bouwens2014.z9}
{Bouwens}, R.~J., {et~al.} 2014, \apj, 795, 126

\bibitem[{{Bouwens} {et~al.}(2016){Bouwens}, {Aravena}, {Decarli}, {Walter},
  {da Cunha}, {Labb{\'e}}, {Bauer}, {Bertoldi}, {Carilli}, {Chapman}, {Daddi},
  {Hodge}, {Ivison}, {Karim}, {Le Fevre}, {Magnelli}, {Ota}, {Riechers},
  {Smail}, {van der Werf}, {Weiss}, {Cox}, {Elbaz}, {Gonzalez-Lopez},
  {Infante}, {Oesch}, {Wagg}, \& {Wilkins}}]{bouwens2016}
{Bouwens}, R.~J., {et~al.} 2016, \apj, 833, 72


\bibitem[{{Bowler} {et~al.}(2014){Bowler}, {Dunlop}, {McLure}, {Rogers},
  {McCracken}, {Milvang-Jensen}, {Furusawa}, {Fynbo}, {Taniguchi}, {Afonso},
  {Bremer}, \& {Le F{\`e}vre}}]{bowler2014}
{Bowler}, R.~A.~A., {et~al.} 2014, \mnras, 440, 2810


\bibitem[{{Bowler} {et~al.}(2017){Bowler}, {Dunlop}, {McLure}, \&
  {McLeod}}]{bowler2017}
{Bowler}, R.~A.~A., {Dunlop}, J.~S., {McLure}, R.~J., \& {McLeod}, D.~J. 2017,
  \mnras, 466, 3612

\bibitem[{{Bowler} {et~al.}(2018){Bowler}, {Bourne}, {Dunlop}, {McLure}, \&
  {McLeod}}]{bowler2018}
{Bowler}, R.~A.~A., {Bourne}, N., {Dunlop}, J.~S., {McLure}, R.~J., \&
  {McLeod}, D.~J. 2018, \mnras, 481, 1631


\bibitem[{{Brada{\v c}} {et~al.}(2017){Brada{\v c}}, {Garcia-Appadoo}, {Huang},
  {Vallini}, {Quinn Finney}, {Hoag}, {Lemaux}, {Borello Schmidt}, {Treu},
  {Carilli}, {Dijkstra}, {Ferrara}, {Fontana}, {Jones}, {Ryan}, {Wagg}, \&
  {Gonzalez}}]{bradac2017}
{Brada{\v c}}, M., {et~al.} 2017, \apjl, 836, L2

\bibitem[{{Brauher} {et~al.}(2008){Brauher}, {Dale}, \& {Helou}}]{brauher2008}
{Brauher}, J.~R., {Dale}, D.~A., \& {Helou}, G. 2008, \apjs, 178, 280

\bibitem[{{Bruzual} \& {Charlot}(2003)}]{bc03}
{Bruzual}, G., \& {Charlot}, S. 2003, \mnras, 344, 1000

\bibitem[{{Calura} {et~al.}(2017){Calura}, {Pozzi}, {Cresci}, {Santini},
  {Gruppioni}, {Pozzetti}, {Gilli}, {Matteucci}, \& {Maiolino}}]{calura2017}
{Calura}, F., {et~al.} 2017, \mnras, 465, 54

\bibitem[{{Calzetti} {et~al.}(2000){Calzetti}, {Armus}, {Bohlin}, {Kinney},
  {Koornneef}, \& {Storchi-Bergmann}}]{calzetti2000}
{Calzetti}, D., {Armus}, L., {Bohlin}, R.~C., {Kinney}, A.~L., {Koornneef}, J.,
  \& {Storchi-Bergmann}, T. 2000, \apj, 533, 682

\bibitem[{{Capak} {et~al.}(2015){Capak}, {Carilli}, {Jones}, {Casey},
  {Riechers}, {Sheth}, {Carollo}, {Ilbert}, {Karim}, {Lefevre}, {Lilly},
  {Scoville}, {Smolcic}, \& {Yan}}]{capak2015}
{Capak}, P.~L., {et~al.} 2015, \nat, 522, 455

\bibitem[{{Carilli} \& {Walter}(2013)}]{carilli_walter2013}
{Carilli}, C.~L., \& {Walter}, F. 2013, \araa, 51, 105

\bibitem[{{Carniani} {et~al.}(2017){Carniani}, {Maiolino}, {Pallottini},
  {Vallini}, {Pentericci}, {Ferrara}, {Castellano}, {Vanzella}, {Grazian},
  {Gallerani}, {Santini}, {Wagg}, \& {Fontana}}]{carniani2017a}
{Carniani}, S., {et~al.} 2017, \aap, 605, A42

\bibitem[{{Carniani} {et~al.}(2018{\natexlab{a}}){Carniani}, {Maiolino},
  {Smit}, \& {Amor{\'{\i}}n}}]{carniani2018a}
{Carniani}, S., {Maiolino}, R., {Smit}, R., \& {Amor{\'{\i}}n}, R.
  2018{\natexlab{a}}, \apjl, 854, L7

\bibitem[{{Carniani} {et~al.}(2018{\natexlab{b}}){Carniani}, {Maiolino},
  {Amorin}, {Pentericci}, {Pallottini}, {Ferrara}, {Willott}, {Smit},
  {Matthee}, {Sobral}, {Santini}, {Castellano}, {De Barros}, {Fontana},
  {Grazian}, \& {Guaita}}]{carniani2018b}
{Carniani}, S., {et~al.} 2018{\natexlab{b}}, \mnras, 478, 1170

\bibitem[{{Cassata} {et~al.}(2015){Cassata}, {Tasca}, {Le F{\`e}vre}, {Lemaux},
  {Garilli}, {Le Brun}, {Maccagni}, {Pentericci}, {Thomas}, {Vanzella},
  {Zamorani}, {Zucca}, {Amorin}, {Bardelli}, {Capak}, {Cassar{\`a}},
  {Castellano}, {Cimatti}, {Cuby}, {Cucciati}, {de la Torre}, {Durkalec},
  {Fontana}, {Giavalisco}, {Grazian}, {Hathi}, {Ilbert}, {Moreau}, {Paltani},
  {Ribeiro}, {Salvato}, {Schaerer}, {Scodeggio}, {Sommariva}, {Talia},
  {Taniguchi}, {Tresse}, {Vergani}, {Wang}, {Charlot}, {Contini}, {Fotopoulou},
  {Koekemoer}, {L{\'o}pez-Sanjuan}, {Mellier}, \& {Scoville}}]{cassata2015}
{Cassata}, P., {et~al.} 2015, \aap, 573, A24

\bibitem[{{Chabrier}(2003)}]{chabrier2003}
{Chabrier}, G. 2003, \apjl, 586, L133

\bibitem[{{Chevallard} \& {Charlot}(2016)}]{chevallard2016}
{Chevallard}, J., \& {Charlot}, S. 2016, \mnras, 462, 1415

\bibitem[{{Choudhury} {et~al.}(2015){Choudhury}, {Puchwein}, {Haehnelt}, \&
  {Bolton}}]{choudhury2015}
{Choudhury}, T.~R., {Puchwein}, E., {Haehnelt}, M.~G., \& {Bolton}, J.~S. 2015,
  \mnras, 452, 261

\bibitem[{{Cormier} {et~al.}(2015){Cormier}, {Madden}, {Lebouteiller}, {Abel},
  {Hony}, {Galliano}, {R{\'e}my-Ruyer}, {Bigiel}, {Baes}, {Boselli},
  {Chevance}, {Cooray}, {De Looze}, {Doublier}, {Galametz}, {Hughes},
  {Karczewski}, {Lee}, {Lu}, \& {Spinoglio}}]{cormier2015}
{Cormier}, D., {et~al.} 2015, \aap, 578, A53

\bibitem[{{Cuby} {et~al.}(2003){Cuby}, {Le F{\`e}vre}, {McCracken},
  {Cuillandre}, {Magnier}, \& {Meneux}}]{cuby2003}
{Cuby}, J.-G., {Le F{\`e}vre}, O., {McCracken}, H., {Cuillandre}, J.-C.,
  {Magnier}, E., \& {Meneux}, B. 2003, \aap, 405, L19

\bibitem[{{da Cunha} {et~al.}(2013){da Cunha}, {Groves}, {Walter}, {Decarli},
  {Weiss}, {Bertoldi}, {Carilli}, {Daddi}, {Elbaz}, {Ivison}, {Maiolino},
  {Riechers}, {Rix}, {Sargent}, \& {Smail}}]{da_cunha2013}
{da Cunha}, E., {et~al.} 2013, \apj, 766, 13

\bibitem[{{De Looze} {et~al.}(2014){De Looze}, {Cormier}, {Lebouteiller},
  {Madden}, {Baes}, {Bendo}, {Boquien}, {Boselli}, {Clements}, {Cortese},
  {Cooray}, {Galametz}, {Galliano}, {Graci{\'a}-Carpio}, {Isaak}, {Karczewski},
  {Parkin}, {Pellegrini}, {R{\'e}my-Ruyer}, {Spinoglio}, {Smith}, \&
  {Sturm}}]{de.looze2014}
{De Looze}, I., {et~al.} 2014, \aap, 568, A62

\bibitem[{{Decarli} {et~al.}(2017){Decarli}, {Walter}, {Venemans},
  {Ba{\~n}ados}, {Bertoldi}, {Carilli}, {Fan}, {Farina}, {Mazzucchelli},
  {Riechers}, {Rix}, {Strauss}, {Wang}, \& {Yang}}]{decarli2017}
{Decarli}, R., {et~al.} 2017, \nat, 545, 457

\bibitem[{{Dijkstra} {et~al.}(2006){Dijkstra}, {Haiman}, \&
  {Spaans}}]{dijkstra2006}
{Dijkstra}, M., {Haiman}, Z., \& {Spaans}, M. 2006, \apj, 649, 14

\bibitem[{{Dunlop} {et~al.}(2017){Dunlop}, {McLure}, {Biggs}, {Geach},
  {Micha{\l}owski}, {Ivison}, {Rujopakarn}, {van Kampen}, {Kirkpatrick},
  {Pope}, {Scott}, {Swinbank}, {Targett}, {Aretxaga}, {Austermann}, {Best},
  {Bruce}, {Chapin}, {Charlot}, {Cirasuolo}, {Coppin}, {Ellis}, {Finkelstein},
  {Hayward}, {Hughes}, {Ibar}, {Jagannathan}, {Khochfar}, {Koprowski},
  {Narayanan}, {Nyland}, {Papovich}, {Peacock}, {Rieke}, {Robertson},
  {Vernstrom}, {Werf}, {Wilson}, \& {Yun}}]{dunlop2017}
{Dunlop}, J.~S., {et~al.} 2017, \mnras, 466, 861

\bibitem[{{Dunne} {et~al.}(2000){Dunne}, {Eales}, {Edmunds}, {Ivison},
  {Alexander}, \& {Clements}}]{dunne2000}
{Dunne}, L., {Eales}, S., {Edmunds}, M., {Ivison}, R., {Alexander}, P., \&
  {Clements}, D.~L. 2000, \mnras, 315, 115

\bibitem[{{Ellis} {et~al.}(2013){Ellis}, {McLure}, {Dunlop}, {Robertson},
  {Ono}, {Schenker}, {Koekemoer}, {Bowler}, {Ouchi}, {Rogers}, {Curtis-Lake},
  {Schneider}, {Charlot}, {Stark}, {Furlanetto}, \& {Cirasuolo}}]{ellis2013}
{Ellis}, R.~S., {et~al.} 2013, \apjl, 763, L7

\bibitem[{{Erb} {et~al.}(2006){Erb}, {Steidel}, {Shapley}, {Pettini}, {Reddy},
  \& {Adelberger}}]{erb2006b}
{Erb}, D.~K., {Steidel}, C.~C., {Shapley}, A.~E., {Pettini}, M., {Reddy},
  N.~A., \& {Adelberger}, K.~L. 2006, \apj, 646, 107

\bibitem[{{Erb} {et~al.}(2014){Erb}, {Steidel}, {Trainor}, {Bogosavljevi{\'c}},
  {Shapley}, {Nestor}, {Kulas}, {Law}, {Strom}, {Rudie}, {Reddy}, {Pettini},
  {Konidaris}, {Mace}, {Matthews}, \& {McLean}}]{erb2014}
{Erb}, D.~K., {et~al.} 2014, \apj, 795, 33

\bibitem[{{Faisst} {et~al.}(2017){Faisst}, {Capak}, {Yan}, {Pavesi},
  {Riechers}, {Bari{\v s}i{\'c}}, {Cooke}, {Kartaltepe}, \&
  {Masters}}]{faisst2017}
{Faisst}, A.~L., {et~al.} 2017, \apj, 847, 21

\bibitem[{{Ferland} {et~al.}(2013){Ferland}, {Porter}, {van Hoof}, {Williams},
  {Abel}, {Lykins}, {Shaw}, {Henney}, \& {Stancil}}]{ferland2013}
{Ferland}, G.~J., {et~al.} 2013, RMxAA, 49, 137

\bibitem[{{F{\"o}rster Schreiber} {et~al.}(2009){F{\"o}rster Schreiber},
  {Genzel}, {Bouch{\'e}}, {Cresci}, {Davies}, {Buschkamp}, {Shapiro},
  {Tacconi}, {Hicks}, {Genel}, {Shapley}, {Erb}, {Steidel}, {Lutz},
  {Eisenhauer}, {Gillessen}, {Sternberg}, {Renzini}, {Cimatti}, {Daddi},
  {Kurk}, {Lilly}, {Kong}, {Lehnert}, {Nesvadba}, {Verma}, {McCracken},
  {Arimoto}, {Mignoli}, \& {Onodera}}]{forster-schreiber2009}
{F{\"o}rster Schreiber}, N.~M., {et~al.} 2009, \apj, 706, 1364

\bibitem[{{Fudamoto} {et~al.}(2017){Fudamoto}, {Oesch}, {Schinnerer}, {Groves},
  {Karim}, {Magnelli}, {Sargent}, {Cassata}, {Lang}, {Liu}, {Le F{\`e}vre},
  {Leslie}, {Smol{\v c}i{\'c}}, \& {Tasca}}]{fudamoto2017}
{Fudamoto}, Y., {et~al.} 2017, \mnras, 472, 483

\bibitem[{{Furusawa} {et~al.}(2016){Furusawa}, {Kashikawa}, {Kobayashi},
  {Dunlop}, {Shimasaku}, {Takata}, {Sekiguchi}, {Naito}, {Furusawa}, {Ouchi},
  {Nakata}, {Yasuda}, {Okura}, {Taniguchi}, {Yamada}, {Kajisawa}, {Fynbo}, \&
  {Le F{\`e}vre}}]{furusawa2016}
{Furusawa}, H., {et~al.} 2016, \apj, 822, 46

\bibitem[{{Gaia Collaboration} {et~al.}(2016){Gaia Collaboration}, {Brown},
  {Vallenari}, {Prusti}, {de Bruijne}, {Mignard}, {Drimmel}, {Babusiaux},
  {Bailer-Jones}, {Bastian}, \& et~al.}]{gaia2016}
{Gaia Collaboration} {et~al.} 2016, \aap, 595, A2

\bibitem[{{Garel} {et~al.}(2012){Garel}, {Blaizot}, {Guiderdoni}, {Schaerer},
  {Verhamme}, \& {Hayes}}]{garel2012}
{Garel}, T., {Blaizot}, J., {Guiderdoni}, B., {Schaerer}, D., {Verhamme}, A.,
  \& {Hayes}, M. 2012, \mnras, 422, 310

\bibitem[{{Gnerucci} {et~al.}(2011){Gnerucci}, {Marconi}, {Cresci}, {Maiolino},
  {Mannucci}, {Calura}, {Cimatti}, {Cocchia}, {Grazian}, {Matteucci}, {Nagao},
  {Pozzetti}, \& {Troncoso}}]{gnerucci2011}
{Gnerucci}, A., {et~al.} 2011, \aap, 528, A88

\bibitem[{{Gonz{\'a}lez-L{\'o}pez} {et~al.}(2017){Gonz{\'a}lez-L{\'o}pez},
  {Bauer}, {Romero-Ca{\~n}izales}, {Kneissl}, {Villard}, {Carvajal}, {Kim},
  {Laporte}, {Anguita}, {Aravena}, {Bouwens}, {Bradley}, {Carrasco}, {Demarco},
  {Ford}, {Ibar}, {Infante}, {Messias}, {Mu{\~n}oz Arancibia}, {Nagar},
  {Padilla}, {Treister}, {Troncoso}, \& {Zitrin}}]{gonzalez_lopez2017}
{Gonz{\'a}lez-L{\'o}pez}, J., {et~al.} 2017, \aap, 597, A41

\bibitem[{{Gronke} {et~al.}(2015){Gronke}, {Bull}, \& {Dijkstra}}]{gronke2015}
{Gronke}, M., {Bull}, P., \& {Dijkstra}, M. 2015, \apj, 812, 123

\bibitem[{{Haiman}(2002)}]{haiman2002}
{Haiman}, Z. 2002, \apjl, 576, L1



\bibitem[{{Harikane} {et~al.}(2018){Harikane}, {Ouchi}, {Shibuya}, {Kojima},
  {Zhang}, {Itoh}, {Ono}, {Higuchi}, {Inoue}, {Chevallard}, {Capak}, {Nagao},
  {Onodera}, {Faisst}, {Martin}, {Rauch}, {Bruzual}, {Charlot}, {Davidzon},
  {Fujimoto}, {Hilmi}, {Ilbert}, {Lee}, {Matsuoka}, {Silverman}, \&
  {Toft}}]{harikane2018b}
{Harikane}, Y., {et~al.} 2018, \apj, 859, 84

\bibitem[{{Hashimoto} {et~al.}(2013){Hashimoto}, {Ouchi}, {Shimasaku}, {Ono},
  {Nakajima}, {Rauch}, {Lee}, \& {Okamura}}]{hashimoto2013}
{Hashimoto}, T., {Ouchi}, M., {Shimasaku}, K., {Ono}, Y., {Nakajima}, K.,
  {Rauch}, M., {Lee}, J., \& {Okamura}, S. 2013, \apj, 765, 70

\bibitem[{{Hashimoto} {et~al.}(2015){Hashimoto}, {Verhamme}, {Ouchi},
  {Shimasaku}, {Schaerer}, {Nakajima}, {Shibuya}, {Rauch}, {Ono}, \&
  {Goto}}]{hashimoto2015}
{Hashimoto}, T., {et~al.} 2015, \apj, 812, 157

\bibitem[{{Hashimoto} {et~al.}(2017{\natexlab{a}}){Hashimoto}, {Ouchi},
  {Shimasaku}, {Schaerer}, {Nakajima}, {Shibuya}, {Ono}, {Rauch}, \&
  {Goto}}]{hashimoto2017a}
{Hashimoto}, T., {et~al.} 2017{\natexlab{a}}, \mnras, 465, 1543

\bibitem[{{Hashimoto} {et~al.}(2017{\natexlab{b}}){Hashimoto}, {Garel},
  {Guiderdoni}, {Drake}, {Bacon}, {Blaizot}, {Richard}, {Leclercq}, {Inami},
  {Verhamme}, {Bouwens}, {Brinchmann}, {Cantalupo}, {Carollo}, {Caruana},
  {Herenz}, {Kerutt}, {Marino}, {Mitchell}, \& {Schaye}}]{hashimoto2017b}
{Hashimoto}, T., {et~al.} 2017{\natexlab{b}}, \aap, 608, A10


\bibitem[{{Hashimoto} {et~al.}(2018{\natexlab{a}}){Hashimoto}, {Laporte},
  {Mawatari}, {Ellis}, {Inoue}, {Zackrisson}, {Roberts-Borsani}, {Zheng},
  {Tamura}, {Bauer}, {Fletcher}, {Harikane}, {Hatsukade}, {Hayatsu}, {Matsuda},
  {Matsuo}, {Okamoto}, {Ouchi}, {Pell{\'o}}, {Rydberg}, {Shimizu}, {Taniguchi},
  {Umehata}, \& {Yoshida}}]{hashimoto2018a}
{Hashimoto}, T., {et~al.} 2018{\natexlab{a}}, \nat, 557, 392


\bibitem[{{Hashimoto} {et~al.}(2018{\natexlab{b}}){Hashimoto}, {Inoue},
  {Tamura}, {Matsuo}, {Mawatari}, \& {Yamaguchi}}]{hashimoto2018c}
{Hashimoto}, T., {Inoue}, A.~K., {Tamura}, Y., {Matsuo}, H., {Mawatari}, K., \&
  {Yamaguchi}, Y. 2018{\natexlab{b}}, submitted to PASJ (arXiv:1811.00030) 


\bibitem[{{Herenz} {et~al.}(2016){Herenz}, {Gruyters}, {Orlitova}, {Hayes},
  {{\"O}stlin}, {Cannon}, {Roth}, {Bik}, {Pardy}, {Ot{\'{\i}}-Floranes},
  {Mas-Hesse}, {Adamo}, {Atek}, {Duval}, {Guaita}, {Kunth}, {Laursen},
  {Melinder}, {Puschnig}, {Rivera-Thorsen}, {Schaerer}, \&
  {Verhamme}}]{herenz2015}
{Herenz}, E.~C., {et~al.} 2016, \aap, 587, A78

\bibitem[{{Herrera-Camus} {et~al.}(2018){Herrera-Camus}, {Sturm},
  {Graci{\'a}-Carpio}, {Lutz}, {Contursi}, {Veilleux}, {Fischer},
  {Gonz{\'a}lez-Alfonso}, {Poglitsch}, {Tacconi}, {Genzel}, {Maiolino},
  {Sternberg}, {Davies}, \& {Verma}}]{herrera-camus2018a}
{Herrera-Camus}, R., {et~al.} 2018, \apj, 861, 94

\bibitem[{{Hildebrand}(1983)}]{hildebrand1983}
{Hildebrand}, R.~H. 1983, \qjras, 24, 267

\bibitem[{{Inoue} \& {Kamaya}(2004)}]{inoue.kamaya2004}
{Inoue}, A.~K., \& {Kamaya}, H. 2004, \mnras, 350, 729

\bibitem[{{Inoue}(2011{\natexlab{a}})}]{inoue2011}
{Inoue}, A.~K. 2011{\natexlab{a}}, \mnras, 415, 2920

\bibitem[{{Inoue}(2011{\natexlab{b}})}]{inoue2011dust}
{Inoue}, A.~K. 2011{\natexlab{b}}, Earth, Planets, and Space, 63, 1027

\bibitem[{{Inoue} {et~al.}(2014{\natexlab{a}}){Inoue}, {Shimizu}, {Iwata}, \&
  {Tanaka}}]{inoue2014igm}
{Inoue}, A.~K., {Shimizu}, I., {Iwata}, I., \& {Tanaka}, M. 2014{\natexlab{a}},
  \mnras, 442, 1805

\bibitem[{{Inoue} {et~al.}(2014{\natexlab{b}}){Inoue}, {Shimizu}, {Tamura},
  {Matsuo}, {Okamoto}, \& {Yoshida}}]{inoue2014alma}
{Inoue}, A.~K., {Shimizu}, I., {Tamura}, Y., {Matsuo}, H., {Okamoto}, T., \&
  {Yoshida}, N. 2014{\natexlab{b}}, \apjl, 780, L18

\bibitem[{{Inoue} {et~al.}(2016){Inoue}, {Tamura}, {Matsuo}, {Mawatari},
  {Shimizu}, {Shibuya}, {Ota}, {Yoshida}, {Zackrisson}, {Kashikawa}, {Kohno},
  {Umehata}, {Hatsukade}, {Iye}, {Matsuda}, {Okamoto}, \&
  {Yamaguchi}}]{inoue2016}
{Inoue}, A.~K., {et~al.} 2016, Science, 352, 1559

\bibitem[{{Jiang} {et~al.}(2013){Jiang}, {Egami}, {Fan}, {Windhorst}, {Cohen},
  {Dav{\'e}}, {Finlator}, {Kashikawa}, {Mechtley}, {Ouchi}, \&
  {Shimasaku}}]{jiang2013}
{Jiang}, L., {et~al.} 2013, \apj, 773, 153

\bibitem[{{Jones} {et~al.}(2017{\natexlab{a}}){Jones}, {Willott}, {Carilli},
  {Ferrara}, {Wang}, \& {Wagg}}]{jones2017a}
{Jones}, G.~C., {Willott}, C.~J., {Carilli}, C.~L., {Ferrara}, A., {Wang}, R.,
  \& {Wagg}, J. 2017{\natexlab{a}}, \apj, 845, 175

\bibitem[{{Jones} {et~al.}(2017{\natexlab{b}}){Jones}, {Carilli}, {Shao},
  {Wang}, {Capak}, {Pavesi}, {Riechers}, {Karim}, {Neeleman}, \&
  {Walter}}]{jones2017b}
{Jones}, G.~C., {et~al.} 2017{\natexlab{b}}, \apj, 850, 180

\bibitem[{{Kashino} {et~al.}(2013){Kashino}, {Silverman}, {Rodighiero},
  {Renzini}, {Arimoto}, {Daddi}, {Lilly}, {Sanders}, {Kartaltepe}, {Zahid},
  {Nagao}, {Sugiyama}, {Capak}, {Carollo}, {Chu}, {Hasinger}, {Ilbert},
  {Kajisawa}, {Kewley}, {Koekemoer}, {Kova{\v c}}, {Le F{\`e}vre}, {Masters},
  {McCracken}, {Onodera}, {Scoville}, {Strazzullo}, {Symeonidis}, \&
  {Taniguchi}}]{kashino2013}
{Kashino}, D., {et~al.} 2013, \apjl, 777, L8

\bibitem[{{Katz} {et~al.}(2017){Katz}, {Kimm}, {Sijacki}, \&
  {Haehnelt}}]{katz2017}
{Katz}, H., {Kimm}, T., {Sijacki}, D., \& {Haehnelt}, M.~G. 2017, \mnras, 468,
  4831

\bibitem[{{Knudsen} {et~al.}(2016){Knudsen}, {Richard}, {Kneib}, {Jauzac},
  {Cl{\'e}ment}, {Drouart}, {Egami}, \& {Lindroos}}]{knudsen2016}
{Knudsen}, K.~K., {Richard}, J., {Kneib}, J.-P., {Jauzac}, M., {Cl{\'e}ment},
  B., {Drouart}, G., {Egami}, E., \& {Lindroos}, L. 2016, \mnras, 462, L6

\bibitem[{{Knudsen} {et~al.}(2017){Knudsen}, {Watson}, {Frayer}, {Christensen},
  {Gallazzi}, {Micha{\l}owski}, {Richard}, \& {Zavala}}]{knudsen2017}
{Knudsen}, K.~K., {Watson}, D., {Frayer}, D., {Christensen}, L., {Gallazzi},
  A., {Micha{\l}owski}, M.~J., {Richard}, J., \& {Zavala}, J. 2017, \mnras,
  466, 138

\bibitem[{{Komatsu} {et~al.}(2011){Komatsu}, {Smith}, {Dunkley}, {Bennett},
  {Gold}, {Hinshaw}, {Jarosik}, {Larson}, {Nolta}, {Page}, {Spergel},
  {Halpern}, {Hill}, {Kogut}, {Limon}, {Meyer}, {Odegard}, {Tucker}, {Weiland},
  {Wollack}, \& {Wright}}]{komatsu2011}
{Komatsu}, E., {et~al.} 2011, \apjs, 192, 18

\bibitem[{{Koprowski} {et~al.}(2018){Koprowski}, {Coppin}, {Geach}, {McLure},
  {Almaini}, {Blain}, {Bremer}, {Bourne}, {Chapman}, {Conselice}, {Dunlop},
  {Farrah}, {Hartley}, {Karim}, {Knudsen}, {Micha{\l}owski}, {Scott},
  {Simpson}, {Smith}, \& {van der Werf}}]{koprowski2018}
{Koprowski}, M.~P., {et~al.} 2018, \mnras, 479, 4355

\bibitem[{{Kusakabe} {et~al.}(2015){Kusakabe}, {Shimasaku}, {Nakajima}, \&
  {Ouchi}}]{kusakabe2015}
{Kusakabe}, H., {Shimasaku}, K., {Nakajima}, K., \& {Ouchi}, M. 2015, \apjl,
  800, L29

\bibitem[{{Lagache} {et~al.}(2018){Lagache}, {Cousin}, \&
  {Chatzikos}}]{lagache2018}
{Lagache}, G., {Cousin}, M., \& {Chatzikos}, M. 2018, \aap, 609, A130

\bibitem[{{Laporte} {et~al.}(2015){Laporte}, {Streblyanska}, {Kim},
  {Pell{\'o}}, {Bauer}, {Bina}, {Brammer}, {De Leo}, {Infante}, \&
  {P{\'e}rez-Fournon}}]{laporte2015}
{Laporte}, N., {et~al.} 2015, \aap, 575, A92

\bibitem[{{Laporte} {et~al.}(2017){Laporte}, {Ellis}, {Boone}, {Bauer},
  {Qu{\'e}nard}, {Roberts-Borsani}, {Pell{\'o}}, {P{\'e}rez-Fournon}, \&
  {Streblyanska}}]{laporte2017}
{Laporte}, N., {et~al.} 2017, \apjl, 837, L21

\bibitem[{{Laursen} {et~al.}(2013){Laursen}, {Duval}, \&
  {{\"O}stlin}}]{laursen2013}
{Laursen}, P., {Duval}, F., \& {{\"O}stlin}, G. 2013, \apj, 766, 124

\bibitem[{{Madden} {et~al.}(2012){Madden}, {R{\'e}my}, {Galliano}, {Galametz},
  {Bendo}, {Cormier}, {Lebouteiller}, \& {Hony}}]{madden2012}
{Madden}, S.~C., {R{\'e}my}, A., {Galliano}, F., {Galametz}, M., {Bendo}, G.,
  {Cormier}, D., {Lebouteiller}, V., \& {Hony}, S. 2012, in IAU Symposium, Vol.
  284, The Spectral Energy Distribution of Galaxies - SED 2011, ed. R.~J.
  {Tuffs} \& C.~C. {Popescu}, 141--148

\bibitem[{{Mainali} {et~al.}(2017){Mainali}, {Kollmeier}, {Stark}, {Simcoe},
  {Walth}, {Newman}, \& {Miller}}]{mainali2017}
{Mainali}, R., {Kollmeier}, J.~A., {Stark}, D.~P., {Simcoe}, R.~A., {Walth},
  G., {Newman}, A.~B., \& {Miller}, D.~R. 2017, \apjl, 836, L14

\bibitem[{{Maiolino} \& {Mannucci}(2019)}]{maiolino.manucci2019}
{Maiolino}, R., \& {Mannucci}, F. 2019, \aapr, 27, 3

\bibitem[{{Maiolino} {et~al.}(2015){Maiolino}, {Carniani}, {Fontana},
  {Vallini}, {Pentericci}, {Ferrara}, {Vanzella}, {Grazian}, {Gallerani},
  {Castellano}, {Cristiani}, {Brammer}, {Santini}, {Wagg}, \&
  {Williams}}]{maiolino2015}
{Maiolino}, R., {et~al.} 2015, \mnras, 452, 54

\bibitem[{{Malhotra} {et~al.}(1997){Malhotra}, {Helou}, {Stacey}, {Hollenbach},
  {Lord}, {Beichman}, {Dinerstein}, {Hunter}, {Lo}, {Lu}, {Rubin},
  {Silbermann}, {Thronson}, \& {Werner}}]{malhotra1997}
{Malhotra}, S., {et~al.} 1997, \apjl, 491, L27

\bibitem[{{Malhotra} {et~al.}(2001){Malhotra}, {Kaufman}, {Hollenbach},
  {Helou}, {Rubin}, {Brauher}, {Dale}, {Lu}, {Lord}, {Stacey}, {Contursi},
  {Hunter}, \& {Dinerstein}}]{malhotra2001}
{Malhotra}, S., {et~al.} 2001, \apj, 561, 766

\bibitem[{{Marrone} {et~al.}(2018){Marrone}, {Spilker}, {Hayward}, {Vieira},
  {Aravena}, {Ashby}, {Bayliss}, {B{\'e}thermin}, {Brodwin}, {Bothwell},
  {Carlstrom}, {Chapman}, {Chen}, {Crawford}, {Cunningham}, {De Breuck},
  {Fassnacht}, {Gonzalez}, {Greve}, {Hezaveh}, {Lacaille}, {Litke}, {Lower},
  {Ma}, {Malkan}, {Miller}, {Morningstar}, {Murphy}, {Narayanan}, {Phadke},
  {Rotermund}, {Sreevani}, {Stalder}, {Stark}, {Strandet}, {Tang}, \&
  {Wei{\ss}}}]{marrone2018}
{Marrone}, D.~P., {et~al.} 2018, \nat, 553, 51

\bibitem[{{Martin}(2005)}]{martin2005}
{Martin}, C.~L. 2005, \apj, 621, 227

\bibitem[{{Mason} {et~al.}(2018{\natexlab{a}}){Mason}, {Treu}, {Dijkstra},
  {Mesinger}, {Trenti}, {Pentericci}, {de Barros}, \& {Vanzella}}]{mason2017}
{Mason}, C.~A., {Treu}, T., {Dijkstra}, M., {Mesinger}, A., {Trenti}, M.,
  {Pentericci}, L., {de Barros}, S., \& {Vanzella}, E. 2018{\natexlab{a}},
  \apj, 856, 2

\bibitem[{{Mason} {et~al.}(2018{\natexlab{b}}){Mason}, {Treu}, {de Barros},
  {Dijkstra}, {Fontana}, {Mesinger}, {Pentericci}, {Trenti}, \&
  {Vanzella}}]{mason2018}
{Mason}, C.~A., {et~al.} 2018{\natexlab{b}}, \apjl, 857, L11

\bibitem[{{Matsuhara} {et~al.}(1997){Matsuhara}, {Tanaka}, {Yonekura}, {Fukui},
  {Kawada}, \& {Bock}}]{matsuhara1997}
{Matsuhara}, H., {Tanaka}, M., {Yonekura}, Y., {Fukui}, Y., {Kawada}, M., \&
  {Bock}, J.~J. 1997, \apj, 490, 744

\bibitem[{{Matthee} {et~al.}(2017){Matthee}, {Sobral}, {Boone},
  {R{\"o}ttgering}, {Schaerer}, {Girard}, {Pallottini}, {Vallini}, {Ferrara},
  {Darvish}, \& {Mobasher}}]{matthee2017}
{Matthee}, J., {et~al.} 2017, \apj, 851, 145

\bibitem[{{Mawatari} {et~al.}(2016){Mawatari}, {Yamada}, {Fazio}, {Huang}, \&
  {Ashby}}]{mawatari2016}
{Mawatari}, K., {Yamada}, T., {Fazio}, G.~G., {Huang}, J.-S., \& {Ashby},
  M.~L.~N. 2016, \pasj, 68, 46

\bibitem[{{McMullin} {et~al.}(2007){McMullin}, {Waters}, {Schiebel}, {Young},
  \& {Golap}}]{McMullin2007}
{McMullin}, J.~P., {Waters}, B., {Schiebel}, D., {Young}, W., \& {Golap}, K.
  2007, in Astronomical Society of the Pacific Conference Series, Vol. 376,
  Astronomical Data Analysis Software and Systems XVI, ed. R.~A. {Shaw},
  F.~{Hill}, \& D.~J. {Bell}, 127

\bibitem[{{Meurer} {et~al.}(1999){Meurer}, {Heckman}, \&
  {Calzetti}}]{meurer1999}
{Meurer}, G.~R., {Heckman}, T.~M., \& {Calzetti}, D. 1999, \apj, 521, 64

\bibitem[{{Micha{\l}owski}(2015)}]{michalowski2015}
{Micha{\l}owski}, M.~J. 2015, \aap, 577, A80

\bibitem[{{Nakajima} {et~al.}(2016){Nakajima}, {Ellis}, {Iwata}, {Inoue},
  {Kusakabe}, {Ouchi}, \& {Robertson}}]{nakajima2016}
{Nakajima}, K., {Ellis}, R.~S., {Iwata}, I., {Inoue}, A.~K., {Kusakabe}, H.,
  {Ouchi}, M., \& {Robertson}, B.~E. 2016, \apjl, 831, L9

\bibitem[{{Nakajima} {et~al.}(2018){Nakajima}, {Fletcher}, {Ellis},
  {Robertson}, \& {Iwata}}]{nakajima2018}
{Nakajima}, K., {Fletcher}, T., {Ellis}, R.~S., {Robertson}, B.~E., \& {Iwata},
  I. 2018, \mnras, 477, 2098

\bibitem[{{Oesch} {et~al.}(2018){Oesch}, {Bouwens}, {Illingworth}, {Labb{\'e}},
  \& {Stefanon}}]{oesch2018}
{Oesch}, P.~A., {Bouwens}, R.~J., {Illingworth}, G.~D., {Labb{\'e}}, I., \&
  {Stefanon}, M. 2018, \apj, 855, 105

\bibitem[{{Oesch} {et~al.}(2015){Oesch}, {van Dokkum}, {Illingworth},
  {Bouwens}, {Momcheva}, {Holden}, {Roberts-Borsani}, {Smit}, {Franx},
  {Labb{\'e}}, {Gonz{\'a}lez}, \& {Magee}}]{oesch2015}
{Oesch}, P.~A., {et~al.} 2015, \apjl, 804, L30

\bibitem[{{Oke} \& {Gunn}(1983)}]{oke1983}
{Oke}, J.~B., \& {Gunn}, J.~E. 1983, \apj, 266, 713

\bibitem[{{Olsen} {et~al.}(2017){Olsen}, {Greve}, {Narayanan}, {Thompson},
  {Dav{\'e}}, {Niebla Rios}, \& {Stawinski}}]{olsen2017}
{Olsen}, K., {Greve}, T.~R., {Narayanan}, D., {Thompson}, R., {Dav{\'e}}, R.,
  {Niebla Rios}, L., \& {Stawinski}, S. 2017, \apj, 846, 105

\bibitem[{{Ono} {et~al.}(2010){Ono}, {Ouchi}, {Shimasaku}, {Dunlop}, {Farrah},
  {McLure}, \& {Okamura}}]{ono2010b}
{Ono}, Y., {Ouchi}, M., {Shimasaku}, K., {Dunlop}, J., {Farrah}, D., {McLure},
  R., \& {Okamura}, S. 2010, \apj, 724, 1524

\bibitem[{{Ono} {et~al.}(2012){Ono}, {Ouchi}, {Mobasher}, {Dickinson},
  {Penner}, {Shimasaku}, {Weiner}, {Kartaltepe}, {Nakajima}, {Nayyeri},
  {Stern}, {Kashikawa}, \& {Spinrad}}]{ono2012}
{Ono}, Y., {et~al.} 2012, \apj, 744, 83

\bibitem[{{Ota} {et~al.}(2014){Ota}, {Walter}, {Ohta}, {Hatsukade}, {Carilli},
  {da Cunha}, {Gonz{\'a}lez-L{\'o}pez}, {Decarli}, {Hodge}, {Nagai}, {Egami},
  {Jiang}, {Iye}, {Kashikawa}, {Riechers}, {Bertoldi}, {Cox}, {Neri}, \&
  {Weiss}}]{ota2014}
{Ota}, K., {et~al.} 2014, \apj, 792, 34

\bibitem[{{Ouchi} {et~al.}(2013){Ouchi}, {Ellis}, {Ono}, {Nakanishi}, {Kohno},
  {Momose}, {Kurono}, {Ashby}, {Shimasaku}, {Willner}, {Fazio}, {Tamura}, \&
  {Iono}}]{ouchi2013}
{Ouchi}, M., {et~al.} 2013, \apj, 778, 102

\bibitem[{{Pentericci} {et~al.}(2011){Pentericci}, {Fontana}, {Vanzella},
  {Castellano}, {Grazian}, {Dijkstra}, {Boutsia}, {Cristiani}, {Dickinson},
  {Giallongo}, {Giavalisco}, {Maiolino}, {Moorwood}, {Paris}, \&
  {Santini}}]{pentericci2011}
{Pentericci}, L., {et~al.} 2011, \apj, 743, 132

\bibitem[{{Pentericci} {et~al.}(2016){Pentericci}, {Carniani}, {Castellano},
  {Fontana}, {Maiolino}, {Guaita}, {Vanzella}, {Grazian}, {Santini}, {Yan},
  {Cristiani}, {Conselice}, {Giavalisco}, {Hathi}, \&
  {Koekemoer}}]{pentericci2016}
{Pentericci}, L., {et~al.} 2016, \apjl, 829, L11

\bibitem[{{Popping} {et~al.}(2017){Popping}, {Somerville}, \&
  {Galametz}}]{popping2017}
{Popping}, G., {Somerville}, R.~S., \& {Galametz}, M. 2017, \mnras, 471, 3152

\bibitem[{{Reddy} {et~al.}(2006){Reddy}, {Steidel}, {Fadda}, {Yan}, {Pettini},
  {Shapley}, {Erb}, \& {Adelberger}}]{reddy2006}
{Reddy}, N.~A., {Steidel}, C.~C., {Fadda}, D., {Yan}, L., {Pettini}, M.,
  {Shapley}, A.~E., {Erb}, D.~K., \& {Adelberger}, K.~L. 2006, \apj, 644, 792

\bibitem[{{R{\'e}my-Ruyer} {et~al.}(2015){R{\'e}my-Ruyer}, {Madden},
  {Galliano}, {Lebouteiller}, {Baes}, {Bendo}, {Boselli}, {Ciesla}, {Cormier},
  {Cooray}, {Cortese}, {De Looze}, {Doublier-Pritchard}, {Galametz}, {Jones},
  {Karczewski}, {Lu}, \& {Spinoglio}}]{remy-ruyer2015}
{R{\'e}my-Ruyer}, A., {et~al.} 2015, \aap, 582, A121

\bibitem[{{Richard} {et~al.}(2011){Richard}, {Kneib}, {Ebeling}, {Stark},
  {Egami}, \& {Fiedler}}]{richard2011}
{Richard}, J., {Kneib}, J.-P., {Ebeling}, H., {Stark}, D.~P., {Egami}, E., \&
  {Fiedler}, A.~K. 2011, \mnras, 414, L31

\bibitem[{{Rieke} {et~al.}(2009){Rieke}, {Alonso-Herrero}, {Weiner},
  {P{\'e}rez-Gonz{\'a}lez}, {Blaylock}, {Donley}, \& {Marcillac}}]{rieke2009}
{Rieke}, G.~H., {Alonso-Herrero}, A., {Weiner}, B.~J.,
  {P{\'e}rez-Gonz{\'a}lez}, P.~G., {Blaylock}, M., {Donley}, J.~L., \&
  {Marcillac}, D. 2009, \apj, 692, 556

\bibitem[{{Roberts-Borsani} {et~al.}(2016){Roberts-Borsani}, {Bouwens},
  {Oesch}, {Labbe}, {Smit}, {Illingworth}, {van Dokkum}, {Holden}, {Gonzalez},
  {Stefanon}, {Holwerda}, \& {Wilkins}}]{roberts-borsani2016}
{Roberts-Borsani}, G.~W., {et~al.} 2016, \apj, 823, 143

\bibitem[{{Rodighiero} {et~al.}(2011){Rodighiero}, {Daddi}, {Baronchelli},
  {Cimatti}, {Renzini}, {Aussel}, {Popesso}, {Lutz}, {Andreani}, {Berta},
  {Cava}, {Elbaz}, {Feltre}, {Fontana}, {F{\"o}rster Schreiber},
  {Franceschini}, {Genzel}, {Grazian}, {Gruppioni}, {Ilbert}, {Le Floch},
  {Magdis}, {Magliocchetti}, {Magnelli}, {Maiolino}, {McCracken}, {Nordon},
  {Poglitsch}, {Santini}, {Pozzi}, {Riguccini}, {Tacconi}, {Wuyts}, \&
  {Zamorani}}]{rodighiero2011}
{Rodighiero}, G., {et~al.} 2011, \apjl, 739, L40

\bibitem[{{Santini} {et~al.}(2017){Santini}, {Fontana}, {Castellano}, {Di
  Criscienzo}, {Merlin}, {Amorin}, {Cullen}, {Daddi}, {Dickinson}, {Dunlop},
  {Grazian}, {Lamastra}, {McLure}, {Micha{\l}owski}, {Pentericci}, \&
  {Shu}}]{santini2017}
{Santini}, P., {et~al.} 2017, \apj, 847, 76

\bibitem[{{Sawicki}(2012)}]{sawicki2012}
{Sawicki}, M. 2012, \pasp, 124, 1208

\bibitem[{{Schaerer}(2003)}]{schaerer2003}
{Schaerer}, D. 2003, \aap, 397, 527

\bibitem[{{Schaerer} {et~al.}(2015){Schaerer}, {Boone}, {Zamojski}, {Staguhn},
  {Dessauges-Zavadsky}, {Finkelstein}, \& {Combes}}]{schaerer2015}
{Schaerer}, D., {Boone}, F., {Zamojski}, M., {Staguhn}, J.,
  {Dessauges-Zavadsky}, M., {Finkelstein}, S., \& {Combes}, F. 2015, \aap, 574,
  A19

\bibitem[{{Schenker} {et~al.}(2014){Schenker}, {Ellis}, {Konidaris}, \&
  {Stark}}]{schenker2014}
{Schenker}, M.~A., {Ellis}, R.~S., {Konidaris}, N.~P., \& {Stark}, D.~P. 2014,
  \apj, 795, 20

\bibitem[{{Schenker} {et~al.}(2012){Schenker}, {Stark}, {Ellis}, {Robertson},
  {Dunlop}, {McLure}, {Kneib}, \& {Richard}}]{schenker2012}
{Schenker}, M.~A., {Stark}, D.~P., {Ellis}, R.~S., {Robertson}, B.~E.,
  {Dunlop}, J.~S., {McLure}, R.~J., {Kneib}, J.-P., \& {Richard}, J. 2012,
  \apj, 744, 179

\bibitem[{{Schinnerer} {et~al.}(2007){Schinnerer}, {Smol{\v c}i{\'c}},
  {Carilli}, {Bondi}, {Ciliegi}, {Jahnke}, {Scoville}, {Aussel}, {Bertoldi},
  {Blain}, {Impey}, {Koekemoer}, {Le Fevre}, \& {Urry}}]{schinnerer2007}
{Schinnerer}, E., {et~al.} 2007, \apjs, 172, 46

\bibitem[{{Shibuya} {et~al.}(2012){Shibuya}, {Kashikawa}, {Ota}, {Iye},
  {Ouchi}, {Furusawa}, {Shimasaku}, \& {Hattori}}]{shibuya2012}
{Shibuya}, T., {Kashikawa}, N., {Ota}, K., {Iye}, M., {Ouchi}, M., {Furusawa},
  H., {Shimasaku}, K., \& {Hattori}, T. 2012, \apj, 752, 114

\bibitem[{{Shibuya} {et~al.}(2014){Shibuya}, {Ouchi}, {Nakajima}, {Hashimoto},
  {Ono}, {Rauch}, {Gauthier}, {Shimasaku}, {Goto}, {Mori}, \&
  {Umemura.}}]{shibuya2014b}
{Shibuya}, T., {et~al.} 2014, \apj, 788, 74

\bibitem[{{Smit} {et~al.}(2018){Smit}, {Bouwens}, {Carniani}, {Oesch},
  {Labb{\'e}}, {Illingworth}, {van der Werf}, {Bradley}, {Gonzalez}, {Hodge},
  {Holwerda}, {Maiolino}, \& {Zheng}}]{smit2018}
{Smit}, R., {et~al.} 2018, \nat, 553, 178

\bibitem[{{Sobral} {et~al.}(2015){Sobral}, {Matthee}, {Darvish}, {Schaerer},
  {Mobasher}, {R{\"o}ttgering}, {Santos}, \& {Hemmati}}]{sobral2015}
{Sobral}, D., {Matthee}, J., {Darvish}, B., {Schaerer}, D., {Mobasher}, B.,
  {R{\"o}ttgering}, H.~J.~A., {Santos}, S., \& {Hemmati}, S. 2015, \apj, 808,
  139

\bibitem[{{Speagle} {et~al.}(2014){Speagle}, {Steinhardt}, {Capak}, \&
  {Silverman}}]{speagle2014}
{Speagle}, J.~S., {Steinhardt}, C.~L., {Capak}, P.~L., \& {Silverman}, J.~D.
  2014, \apjs, 214, 15

\bibitem[{{Stark} {et~al.}(2010){Stark}, {Ellis}, {Chiu}, {Ouchi}, \&
  {Bunker}}]{stark2010}
{Stark}, D.~P., {Ellis}, R.~S., {Chiu}, K., {Ouchi}, M., \& {Bunker}, A. 2010,
  \mnras, 408, 1628

\bibitem[{{Stark} {et~al.}(2013){Stark}, {Schenker}, {Ellis}, {Robertson},
  {McLure}, \& {Dunlop}}]{stark2013}
{Stark}, D.~P., {Schenker}, M.~A., {Ellis}, R., {Robertson}, B., {McLure}, R.,
  \& {Dunlop}, J. 2013, \apj, 763, 129

\bibitem[{{Stark} {et~al.}(2015{\natexlab{a}}){Stark}, {Walth}, {Charlot},
  {Cl{\'e}ment}, {Feltre}, {Gutkin}, {Richard}, {Mainali}, {Robertson},
  {Siana}, {Tang}, \& {Schenker}}]{stark2015b}
{Stark}, D.~P., {et~al.} 2015{\natexlab{a}}, \mnras, 454, 1393

\bibitem[{{Stark} {et~al.}(2015{\natexlab{b}}){Stark}, {Richard}, {Charlot},
  {Cl{\'e}ment}, {Ellis}, {Siana}, {Robertson}, {Schenker}, {Gutkin}, \&
  {Wofford}}]{stark2015a}
{Stark}, D.~P., {et~al.} 2015{\natexlab{b}}, \mnras, 450, 1846

\bibitem[{{Stark} {et~al.}(2017){Stark}, {Ellis}, {Charlot}, {Chevallard},
  {Tang}, {Belli}, {Zitrin}, {Mainali}, {Gutkin}, {Vidal-Garc{\'{\i}}a},
  {Bouwens}, \& {Oesch}}]{stark2017}
{Stark}, D.~P., {et~al.} 2017, \mnras, 464, 469

\bibitem[{{Steidel} {et~al.}(2010){Steidel}, {Erb}, {Shapley}, {Pettini},
  {Reddy}, {Bogosavljevi{\'c}}, {Rudie}, \& {Rakic}}]{steidel2010}
{Steidel}, C.~C., {Erb}, D.~K., {Shapley}, A.~E., {Pettini}, M., {Reddy}, N.,
  {Bogosavljevi{\'c}}, M., {Rudie}, G.~C., \& {Rakic}, O. 2010, \apj, 717, 289

\bibitem[{{Sugahara} {et~al.}(2017){Sugahara}, {Ouchi}, {Lin}, {Martin}, {Ono},
  {Harikane}, {Shibuya}, \& {Yan}}]{sugahara2017}
{Sugahara}, Y., {Ouchi}, M., {Lin}, L., {Martin}, C.~L., {Ono}, Y., {Harikane},
  Y., {Shibuya}, T., \& {Yan}, R. 2017, \apj, 850, 51

\bibitem[{{Takeuchi} {et~al.}(2012){Takeuchi}, {Yuan}, {Ikeyama}, {Murata}, \&
  {Inoue}}]{takeuchi2012}
{Takeuchi}, T.~T., {Yuan}, F.-T., {Ikeyama}, A., {Murata}, K.~L., \& {Inoue},
  A.~K. 2012, \apj, 755, 144

\bibitem[{{Tamura} {et~al.}(2018){Tamura}, {Mawatari}, {Hashimoto}, {Inoue},
  {Zackrisson}, {Christensen}, {Binggeli}, {Matsuda}, {Matsuo}, {Takeuchi},
  {Asano}, {Shimizu}, {Okamoto}, {Yoshida}, {Lee}, {Shibuya}, {Taniguchi},
  {Umehata}, {Hatsukade}, {Kohno}, \& {Ota}}]{tamura2018}
{Tamura}, Y., {et~al.} 2019, \apj, 874, 27 

\bibitem[{{Trainor} {et~al.}(2015){Trainor}, {Steidel}, {Strom}, \&
  {Rudie}}]{trainor2015}
{Trainor}, R.~F., {Steidel}, C.~C., {Strom}, A.~L., \& {Rudie}, G.~C. 2015,
  \apj, 809, 89

\bibitem[{{Vallini} {et~al.}(2017){Vallini}, {Ferrara}, {Pallottini}, \&
  {Gallerani}}]{vallini2017}
{Vallini}, L., {Ferrara}, A., {Pallottini}, A., \& {Gallerani}, S. 2017,
  \mnras, 467, 1300

\bibitem[{{Vallini} {et~al.}(2015){Vallini}, {Gallerani}, {Ferrara},
  {Pallottini}, \& {Yue}}]{vallini2015}
{Vallini}, L., {Gallerani}, S., {Ferrara}, A., {Pallottini}, A., \& {Yue}, B.
  2015, \apj, 813, 36

\bibitem[{{Vanzella} {et~al.}(2011){Vanzella}, {Pentericci}, {Fontana},
  {Grazian}, {Castellano}, {Boutsia}, {Cristiani}, {Dickinson}, {Gallozzi},
  {Giallongo}, {Giavalisco}, {Maiolino}, {Moorwood}, {Paris}, \&
  {Santini}}]{vanzella2011}
{Vanzella}, E., {et~al.} 2011, \apjl, 730, L35

\bibitem[{{Verhamme} {et~al.}(2015){Verhamme}, {Orlitov{\'a}}, {Schaerer}, \&
  {Hayes}}]{verhamme2015}
{Verhamme}, A., {Orlitov{\'a}}, I., {Schaerer}, D., \& {Hayes}, M. 2015, \aap,
  578, A7

\bibitem[{{Verhamme} {et~al.}(2006){Verhamme}, {Schaerer}, \&
  {Maselli}}]{verhamme2006}
{Verhamme}, A., {Schaerer}, D., \& {Maselli}, A. 2006, \aap, 460, 397

\bibitem[{{Verhamme} {et~al.}(2018){Verhamme}, {Garel}, {Ventou}, {Contini},
  {Bouch{\'e}}, {Herenz}, {Richard}, {Bacon}, {Schmidt}, {Maseda}, {Marino},
  {Brinchmann}, {Cantalupo}, {Caruana}, {Cl{\'e}ment}, {Diener}, {Drake},
  {Hashimoto}, {Inami}, {Kerutt}, {Kollatschny}, {Leclercq}, {Patr{\'{\i}}cio},
  {Schaye}, {Wisotzki}, \& {Zabl}}]{verhamme2018}
{Verhamme}, A., {et~al.} 2018, \mnras, 478, L60

\bibitem[{{Walter} {et~al.}(2018){Walter}, {Riechers}, {Novak}, {Decarli},
  {Ferkinhoff}, {Venemans}, {Banados}, {Bertoldi}, {Carilli}, {Fan}, {Farina},
  {Mazzucchelli}, {Neeleman}, {Rix}, {Strauss}, {Uzgil}, \&
  {Wang}}]{walter2018}
{Walter}, F., {et~al.} 2018, \apjl, 869, L22

\bibitem[{{Watson} {et~al.}(2015){Watson}, {Christensen}, {Knudsen}, {Richard},
  {Gallazzi}, \& {Micha{\l}owski}}]{watson2015}
{Watson}, D., {Christensen}, L., {Knudsen}, K.~K., {Richard}, J., {Gallazzi},
  A., \& {Micha{\l}owski}, M.~J. 2015, \nat, 519, 327

\bibitem[{{Weiner} {et~al.}(2009){Weiner}, {Coil}, {Prochaska}, {Newman},
  {Cooper}, {Bundy}, {Conselice}, {Dutton}, {Faber}, {Koo}, {Lotz}, {Rieke}, \&
  {Rubin}}]{weiner2009}
{Weiner}, B.~J., {et~al.} 2009, \apj, 692, 187

\bibitem[{{Willott} {et~al.}(2015){Willott}, {Carilli}, {Wagg}, \&
  {Wang}}]{willott2015}
{Willott}, C.~J., {Carilli}, C.~L., {Wagg}, J., \& {Wang}, R. 2015, \apj, 807,
  180

\bibitem[{{Willott} {et~al.}(2013){Willott}, {Omont}, \&
  {Bergeron}}]{willott2013}
{Willott}, C.~J., {Omont}, A., \& {Bergeron}, J. 2013, \apj, 770, 13

\bibitem[{{Zheng} {et~al.}(2014){Zheng}, {Shu}, {Moustakas}, {Zitrin}, {Ford},
  {Huang}, {Broadhurst}, {Molino}, {Diego}, {Infante}, {Bauer}, {Kelson}, \&
  {Smit}}]{wei.zheng2014}
{Zheng}, W., {et~al.} 2014, \apj, 795, 93

\bibitem[{{Zheng} {et~al.}(2017){Zheng}, {Zitrin}, {Infante}, {Laporte},
  {Huang}, {Moustakas}, {Ford}, {Shu}, {Wang}, {Diego}, {Bauer}, {Troncoso
  Iribarren}, {Broadhurst}, \& {Molino}}]{wei.zheng2017}
{Zheng}, W., {et~al.} 2017, \apj, 836, 210

\bibitem[{{Zheng} \& {Wallace}(2014)}]{zheng2014}
{Zheng}, Z., \& {Wallace}, J. 2014, \apj, 794, 116

\bibitem[{{Zitrin} {et~al.}(2015){Zitrin}, {Labb{\'e}}, {Belli}, {Bouwens},
  {Ellis}, {Roberts-Borsani}, {Stark}, {Oesch}, \& {Smit}}]{zitrin2015}
{Zitrin}, A., {et~al.} 2015, \apjl, 810, L12

\end{thebibliography}

\enddocument